\documentclass[aps,pre,preprint,groupedaddress,draft,showkeys,amsfonts,amsmath]{revtex4-2}

\DeclareMathOperator\erf{erf}
\DeclareMathOperator\sech{sech}
\DeclareMathOperator\Res{Res}
\DeclareMathOperator\md{md}
\DeclareMathOperator\var{var}
\DeclareMathOperator\cov{cov}
\DeclareMathOperator\B{B}
\DeclareMathOperator\T{T}
\DeclareMathOperator{\sgn}{sgn}
\usepackage[final]{graphicx}
\usepackage[outdir=./img/]{epstopdf}
\usepackage{mathbbol}
\usepackage{amssymb}

\begin{document}


\title{
Discrete scaling and criticality in a chain of adaptive excitable integrators 
}


\author{Mario Martinez-Saito}
\email[]{mmartinezsaito@gmail.com}
\affiliation{Institute of Cognitive Neuroscience, HSE University, Russian Federation}
\affiliation{Department of Psychology, National University of Singapore, Singapore}


\date{\today}

\begin{abstract}
We describe a chain of unidirectionally coupled adaptive excitable elements slowly driven by a stochastic process from one end and open at the other end, as a minimal toy model of unresolved irreducible uncertainty in a system performing inference through a hierarchical model. Threshold potentials adapt slowly to ensure sensitivity without being wasteful. Activity and energy are released as intermittent avalanches of pulses with a discrete scaling distribution largely independent of the exogenous input form. Subthreshold activities and threshold potentials exhibit Lorentzian temporal spectra, with a power-law range determined by position in the chain. Subthreshold bistability closely resembles empirical measurements of intracellular membrane potential. We suggest that critical cortical cascades emerge from a trade-off between metabolic power consumption and performance requirements in a critical world, and that the temporal scaling patterns of brain electrophysiological recordings ensue from weighted linear combinations of subthreshold activities and pulses from different hierarchy levels.
\end{abstract}


\maketitle

\section{Introduction}

Nature proliferates with complex spontaneous structures that feed on a sustained energy flow \cite{Prigogine1985a}. For example, the Sun's radiant energy induces Earth's atmospheric circulation, and sustains life on its surface. Many such structures are open systems involving many nonlinear interacting subunits slowly driven by an exogenous force that evolve into a non-equilibrium steady state (NESS) characterized by self-organized criticality (SOC). Critical phenomena entail spatiotemporally correlated fluctuations over several orders of magnitude, power-law distributed quantities, and scaling invariance with exponents that are universal across a large class of systems. Critical phenomena are typically poised near phase transitions and occur for a narrow range of parameters, but SOC systems evolve naturally toward criticality or quasi-criticality without requiring parameter tuning. These phenomena are typically characterized by driven excitable media with a wide separation between slow driving and fast relaxation temporal scales \cite{Vespignani1998} and a conservation law \cite{Bonachela2009}. The concept of SOC was first demonstrated via a sandpile model \cite{Bak1987}. Since then, many studies have put forward SOC models to explain power-laws empirically found in nature. Some of these models describe earthquake seismic moments and stick-slip processes \cite{Gutenberg1956,Feder1991,Olami1992}, traffic jams \cite{Nagel1995}, forest fires \cite{Drossel1992}, epidemics \cite{Rhodes1997}, neuronal avalanches \cite{Corral1995,Herz1995,Chialvo1999}, chemical oscillators, and solar flares \cite{Lu1993}. 

The cerebral cortex seems to operate near a critical regime \cite{Beggs2003} that affords optimal computational properties, viz. optimal dynamic range (e.g.\ quick reconfiguration in response to inputs), information transmission and information capacity \cite{Langton1990,Shew2013}. 
In this article, we surmise that criticality in the brain ensues from three related preconditions: (1) any living creature, insofar as it comprises a regulator (nervous system) selected to enhance survival odds, must have a good enough model of its environment or external milieu \cite{Conant1970,Friston2006}, (2) the multilayered architecture of the brain is a reflection of the multiple spatio-temporal scales of its environment \cite{Chialvo2006}, and (3) a balance between performance in terms of representational accuracy and metabolic power consumption \cite{Hasenstaub2010,Roberts2014} is a characteristic of life indispensable not to overstep the limited homeostatic range compatible with survival. These preconditions would entail the empirically observed power-law distributed neuronal avalanches \cite{Beggs2003} and temporal long-range dependencies in behavior and neural oscillations \cite{Novikov1997,Linkenkaer-Hansen2001,Kello2010,Palva2013,Zhigalov2015}. 
There are few theories laying down organic principles of general cortical circuitry operation. The perhaps most promising ones posit that the brain incorporates a generative model of the world \cite{Dayan1995,Friston2006} instantiated as a hierarchy of layers where neural processing accomplishes inversion of the causes of sensory input \citep{Mumford1992,Friston2003,Lee2003,Friston2005}. The hierarchical nature of the brain \cite{Felleman1991} and its role in implementing contextual invariance \cite{Phillips1997} are well substantiated. Although it has been proposed that the hierarchical structure of the brain recapitulates the temporal hierarchy of its environment \cite{Kiebel2008} and that SOC is a necessary attribute of any (living) system that persists as a corollary of its proclivity to resolve uncertainty through exploration \cite{Friston2012a}, simple mechanistic accounts of how uncertainty suppression is expressed as neural dynamics are scant.

We model a hierarchical system that checks the estimated surprisal about the state of the environment at multiple temporal scales, without delving into the details of its implicit internal generative model, thus bringing to focus the flow of surprisal between levels and the tolerance to surprisal determined by the threshold adaptation rule. Unlike previous studies linking avalanche dynamics of threshold units to the long range dependencies of brain dynamics, which focused on networks of recurrent connections on a complete graph \cite{Eurich2002} with adaptive synapses \cite{Levina2007}, two-dimensional lattice \cite{Poil2012} with adaptive synapses \cite{DeArcangelis2006}, or partially connected random network configurations \cite{Bertschinger2004,Levina2007}, here we only use excitable nodes embedded in a single directed chain, characterized by a separation of time scales. Crucially, here the closest analogue in the brain to nodes would be cortical regions classified by their characteristic time scale, rather than single neurons. Our analysis endorses the notion that criticality in the brain is a manifestation of the idiosyncratic spatio-temporal scales of the world's causal structure, mimicked by the cortical hierarchy, under metabolic energy constraints.

\section{The perfusive cascade model \label{sec:perfcasc}}

The perfusive cascade model (PC, from ``diffuse through") simulates the behavior of a good regulator \cite{Conant1970} that incorporates a dynamical model of its environment. Let there be $l=1..n_l$ threshold integrator units or levels, coupled in a daisy chain with linear topology. The first and fastest unit, which we will denote as sensory unit, is driven by an exogenous input at discrete time steps $t=1,...,n_t$. The (square of) exogenous input and unit activities can be for now roughly interpreted as a measure of uncertainty about the state of the world, that emanates from the diffusive nature of the environment. We will assume the input to be distributed as a discrete white standard Gaussian noise process $I \sim \mathcal{N}(0,1)$. This is justified by the homeostatic equilibrium (in the NESS sense) in which living creatures coexist with their environment \cite{Friston2007a,Friston2012a}. This assumes that the system can track environmental states to the extent that the uncertainty of the exogenous input is on average just irreducible noise resulting from the added contribution of many sources ---modeled as Gaussian random effects. In other words, the system should entertain a reasonably accurate internal representation of its environment. 

The system operates in a stop-and-go manner: exogenous input drives the sensory unit only after the chain has reached quiescence. This corresponds to an infinite separation between the temporal scales of driving and relaxation \cite{Loreto1996}. At each time step or iteration $t$, the exogenous input drives the sensory unit ($l=1$), and in general any unit $l$ may receive a signal $\varepsilon_{l-1}$ from its subordinate neighbor $l-1$ and add to its sub-threshold activity $a_l \in \mathbf{R}$, yielding the post-pulse activity 
\begin{equation}
\tilde{a}_l^{(t)} = a_l^{(t)} + \varepsilon_{l-1}^{(t)}. \label{eq:a_tl}
\end{equation}
The unit then fires and resets (like the integrate-and-fire neuron \cite{Izhikevich2007}), or stays unchanged according to the firing rule
\begin{equation}
a_l^{(t+1)} = 
\left\{ \begin{array}{lll}
\tilde{a}_l^{(t)}        & \text{if} \quad |\tilde{a}_l^{(t)}| < \theta_l^{(t)} & \text{(rest)} \\
\alpha \tilde{a}_l^{(t)} & \text{else.} & \text{(fire, reset)}
\end{array} \right. \label{eq:a}
\end{equation}
where $\varepsilon_l \in \mathbf{R}$ is the error signal or pulse, which is propagated forward between consecutive units whenever activity reaches the threshold $\theta_l \in \mathbf{R}^+$(for the sensory unit the error signal is the exogenous input $\varepsilon_0 = I$). Error signals, pertaining to the subsequent iteration $t+1$, are computed after updating the activities
\begin{equation} 
\varepsilon_l^{(t+1)} = 
\left\{ \begin{array}{ll}
I      & \text{if} \quad l=0 \\
0      & \text{if} \quad l>0  \; \text{and} \; \tilde{a}_l^{(t)} < \theta_l^{(t)} \\ 
\tilde{a}_l^{(t)} & \text{if} \quad l>0  \; \text{and} \; \tilde{a}_l^{(t)} \geq \theta_l^{(t)}.
\end{array} \right. \label{eq:e}
\end{equation}

An above-threshold unit transfers all its activity to its supraordinate unit, in the same way an open sluice gate dumps stored water in a canal or a s\=ozu \footnote{The PC system is similar to a sequence of sluices or a stack of s\=ozus. A s\=ozu is a water-powered rocking device ---a bamboo pipe pivoted to one side of its center of gravity. At rest its mouth is tilted upwards while its bottom lies on a support. A trickle of water fills it till its tipping point (the center of gravity is displaced past the pivot), when the pipe rotates around its pivot and dumps out \emph{all} the water it contained. A s\=ozu operates intermittently in the seesaw fashion of a relaxation oscillator \cite{Ginoux2012} as long as the water flow persists. Thus, in a hierarchy of s\=ozus arranged so that each lies below the preceding, outflows and inflows are coupled sequentially, and the ensemble constitutes a cascade with level-wise intermittent falls. However, there are two important differences: water can only steadily build up, until it is discharged (there is no anti-water equivalent) and thresholds are fixed. Threshold adaptation could be accomplished by fastening the pivot shaft to the pipe, filling it with a dilatant (shear-thickening fluid), and making it rotate about another internal coaxial shaft in turn attached to the support. If the dilatant viscosity change had a long enough time constant, it would mimic approximately the behavior of adaptive thresholds.} tilts to dump water. This is also similar to the Olami-Feder-Christensen stick-slip (OFC) model of earthquake dynamics in the conservative regime, where plates undergo mechanical stress similar to how unit activities index surprisal \cite{Olami1992}. This choice is important, because minutes details about how surprisal is transferred at the local scale determine the emergent dynamics \cite{Corral1995,Herz1995,DeLosRios1999}. However, the OFC model differs in that slipping plates spill the stored stress energy through faults shared with four neighbors in a square lattice. 

The thresholds $\theta_l$ are gates that set the value of $|a_l|$ beyond which pulses are emitted. Crucially, this mimics the trade-off between metabolic power consumption and performance \cite{Hasenstaub2010,Roberts2014} by setting how much surprisal can be tolerated without (costly) updating the internal representation. Note this plasticity rule is fundamentally different from Hebbian learning \cite{DeArcangelis2006} and short-term synaptic plasticity \cite{Levina2007,Levina2009} rules used previously to model dynamic synaptic efficacies. Shifting thresholds has an additive, as opposed to multiplicative, effect on the excitability of units. $\theta_l^{(t)}$ is the time series of threshold fluctuations defined by:
\begin{equation} 
\theta_l^{(t+1)} = 
\left\{ \begin{array}{ll}
\theta_l^{(t)} (1-w^-) + |\tilde{a}_l^{(t)}| w^- & \text{if} \; |\tilde{a}_l^{(t)}|<\theta_l^{(t)} \\
\theta_l^{(t)} (1-w^+) + |\tilde{a}_l^{(t)}| w^+ & \text{if} \; |\tilde{a}_l^{(t)}| \geq \theta_l^{(t)}. \\
\end{array} \right. \label{eq:thetal}
\end{equation}
Thresholds shift as a function of the concurrent activities and error signals (delta rule), thus emulating perceptual inference and sensory adaptation with a learning rate $0<w<1$ (we will assume $w=w^+=w^-$).

Sensory adaptation is modeled via threshold fluctuations. Increasing the threshold upon discharging also emulates a metabolic energy saving scheme. Decreasing the threshold of stimulated yet quiescent units is
motivated by the tendency of living creatures to explore and learn --- an evolutionary imperative to prevent death in the long-term--- that pushes them toward uncertain situations or the edge of failure \cite{Carlson1999,Chialvo2008,Friston2012a}.
In neurons, sensory adaptation induces changes in responsiveness to input \cite{Fairhall2001a,Kohn2007}, which has the effect of whitening the steady-state distribution of incoming signals. At least for some modeled causal structures, synaptic plasticity can lead to approximate Bayesian inference through Expectation-Maximization \cite{Nessler2013}, and synaptic plasticity in conjunction with activity-depend changes in the responsiveness of neurons can induce the required empirical Bayesian priors to perform Bayesian inference \cite{Bernacchia2014}. Importantly, the adaptability of synaptic efficacies can be interpreted as perceptual inference by assuming the generative model has been shaped by selective pressure \cite{Friston2006} within the realm of complexity afforded by NESS structures \cite{Kauffman1992}. 

\begin{figure*}
\includegraphics[width=1\textwidth]{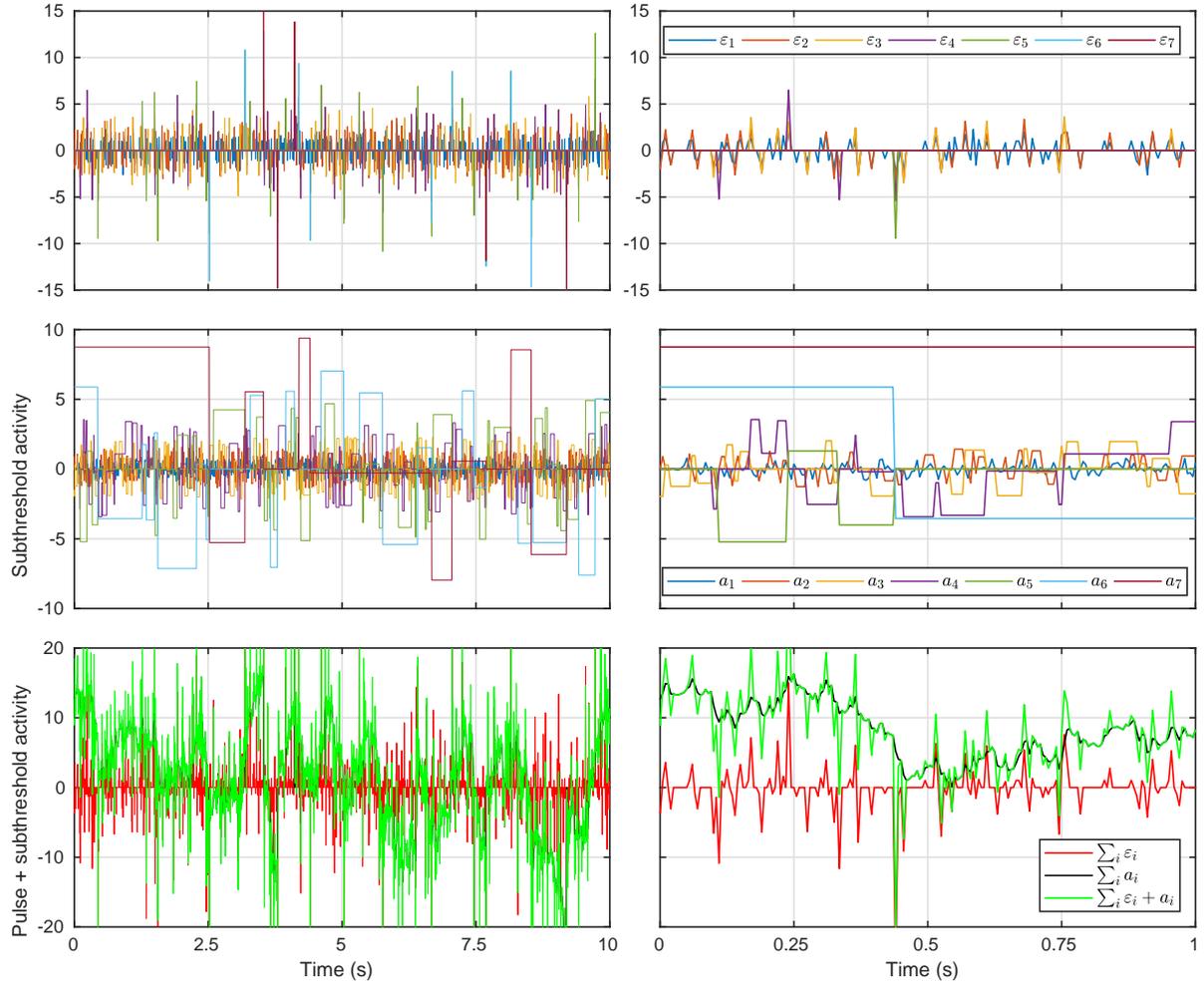}
\caption{Time series of pulses (top) and subthreshold activities (center) for each level $l=1..7$, and their sum (bottom), in steady state regime driven by Gaussian noise input. Left: interval of 2000 iterations (10 s). Right: as in left, but for 200 time steps (1 s). Exogenous driving input: white Gaussian noise. Refractory period (stepsize): $T_0=5$ ms. See main text for details. \label{fig:AEts}}
\end{figure*}

After the cascade stops somewhere in the chain, the activities of suprathreshold nodes are suppressed by a constant fraction $\alpha \in [0,1]$ (we will use full suppression, $\alpha=1$). Activity and threshold updates are triggered only by incoming signals ---subthreshold quiescent units remain unchanged. A time series instantiation of pulses, activities, and their sum are shown in Fig.\ \ref{fig:AEts}. The total absolute activity stored in the chain is $A = \Sigma_{l=1}^L |a_l|$, whereas $E = \Sigma_{l=1}^{L} |\varepsilon_l|$ represents the total absolute activity that is being transferred. The terminal node $L$ has a special status because it has no node to whom relay pulses. The resetting of $a_L$ upon relaxing or discharging is the only way to dissipate activity in the system; in other words, $L$ is an open boundary or sink of activity.

Relaxation oscillator neuron models, which comprise a fast-slow dynamical variable pair (e.g. FitzHugh-Nagumo), may exhibit intermittency and critical slowing of their dynamics when poised near a (Andronov-Hopf \cite{Izhikevich2007}) bifurcation between resting and firing regimes, especially when the bifurcation parameter is subject to noise \cite{Platt1993}. Although the PC may exhibit similar behavior, it (as we will see) is induced by the separation of timescales between its driving and relaxation processes, which implicitly determines a near critical steady-state \cite{Vespignani1998,Bonachela2009} (Section \ref{sec:toymodel_crit}), and by the combination of stochastic stretching (diffusion) and contraction (resetting and truncation, see Section \ref{sec:RRLC}), which ensures intermittent oscillations.

\section{Model properties \label{sec:properties}}

Eqs.\ \ref{eq:a_tl} -- \ref{eq:thetal} constitute a stochastic discrete-time nonlinear dynamical system or map. The walk step of the sensory unit $\mathcal{E}_0$ is a random variable sampled from $\mathcal{N}(0,1)$, unless otherwise stated. By symmetry, subthreshold activities evolve as one-dimensional reflected random walks. For $l>0$, walk steps are distributed as $\mathcal{E}_l  \sim | \tilde{A}_l |_{\geq \theta_l}$ restricted to $[-\infty, -\theta_l] \cup [\theta_l, \infty]$. For each level $l$, these random variables converge in distribution in the limit $t \to \infty$. The probability density function (density) of pulses is conditioned on activities hitting the threshold:
\begin{equation}
 f_{\mathcal{E}_l} \equiv f_{\tilde{A}_l \; | \; | \tilde{A}_l | > \Theta_l}.
\end{equation}
Hence, $\mathcal{E}_l$ depends jointly on $A_l$, $\mathcal{E}_{l-1}$, and $\Theta_l$, through the probability that unit $l$ fires after receiving a signal from $l-1$. The zero value is excluded from the image of $\mathcal{E}$, unlike for $A$.

\subsection{Simulations with white Gaussian noise as exogenous driving input \label{sec:gausim}}

\begin{figure}
\includegraphics[width=.5\textwidth]{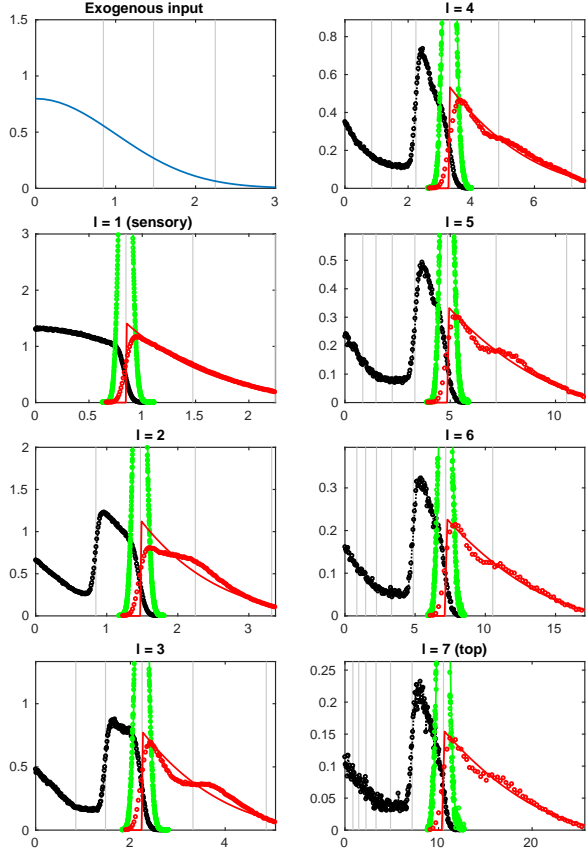}
\caption{Threshold (green dotted), subthreshold activity (black dotted), and pulse (red dotted) numerical distributions for $l=1..7$, and white Gaussian noise exogenous input (blue). All distributions are symmetric with respect to zero. Only the positive half of the (symmetric) densities is shown. The subthreshold activity $A_l^+$ is shown excluding the rest state $A_l=0$, instead of the original $A_l$; their densities are related by $f_{A_l} = g_l\delta(0) + (1-g_l) f_{A_l^+}$. $P(A_l = 0) = g_l$ are included in Table \ref{tab:a1}. Vertical gray lines indicate the location of $\langle \Theta_l \rangle$. Solid red line: generalized Pareto distribution fit (Appendix \ref{sec:appPulseGpd}). Number of iterations $n_t=9 \cdot 10^6$. \label{fig:pdfs}}
\end{figure}

Using white Gaussian noise as driving input, numerical evaluation with $L=7$, $n_t=9 \cdot 10^6$, and $w=.01$, we obtain the numerical distributions for $A_l$ and $\Theta_l$ in Fig.\ \ref{fig:pdfs}. As implied by Eq.\ \ref{eq:thetal}, $\langle \Theta_l \rangle \approx \langle |\tilde{A}_l| \rangle$. To see this, note that by definition the profiles of $f_{A_l}$ and $f_{\mathcal{E}_l}$ result from the truncation of $f_{\tilde{A}_l}$ (Eq.\ \ref{eq:a}, also Eqs.\ \ref{eq:Bae}, \ref{eq:Tae}) at $\Theta_l$, which is close to $\langle |\tilde{A}_l| \rangle$. The sensory unit threshold is $\langle \Theta_1 \rangle = .8464$, and its mean probability of firing is $g_1=P(\mathcal{E}_1 \neq 0) = .4251$ (see Fig.\ \ref{fig:pdfs} and Table \ref{tab:a1}). The gating or thresholding rule (Eq.\ \ref{eq:a}) sets the system behavior apart from random walks by dynamically conditioning the accumulation process on the location of the threshold.

The positive half of subthreshold activity $|A^+_l|$  densities is bimodal, with one peak at zero and other close to $\langle \Theta_{l-1} \rangle$. The pulse $\mathcal{E}_l$ densities are the smoothly truncated tails of $\tilde{A}_l$, roughly similar to the generalized Pareto distribution (Appendix \ref{sec:appPulseGpd}). In the vicinity of the origin, $A^+_l$ is well approximated by a constant slope graph, as in systems governed by continuous diffusion with resetting beyond a fixed threshold \cite{Evans2011b}. The stationary densities $f_{\Theta_l}$, $f_{\tilde{A}_l}$ and $f_{\mathcal{E}_l}$ for each unit $l>1$ are scaled by $k_l \approx 1.5$ with respect to the preceding unit (Fig.\ \ref{fig:pdfs}, Table \ref{tab:01}). The scaling factors $k_l$ are related to the properties of one-dimensional driftless random walks. By construction, consecutive unit thresholds cannot be scaled by a factor larger than 2: the fluctuating thresholds are attracted to the expected value of post-pulse activities $\langle \tilde{A}_l \rangle$, which cannot be larger than  $2\langle \Theta_{l-1} \rangle = 2\langle \tilde{A}_{l-1} \rangle$ (cf. Section \ref{sec:edgecases}). On the other hand, a unit's threshold is larger its preceding unit's threshold because the sequence of accumulating inputs is a (diffusive) random walk that eventually reaches any (finite) threshold $\theta_l > \theta_{l-1}$ almost surely. Hence, $k_l \in [1,2]$.

\begin{table}
\caption{Numerically estimated between-level scaling factors, gains and discharge rates. The gain is the probability of discharging after receiving a signal $g_l \equiv P(\mathcal{E}_l \neq 0 | \mathcal{E}_{l-1} \neq 0)$, with $P(\mathcal{E}_0 \neq 0)=1$ (only the sensory unit is driven at each iteration; note that unit $l$ discharging is tantamount to $\mathcal{E}_l \neq 0$). The last digit is the least significant; the expected variability in round brackets is in s.e. units. Probabilities are modeled as binomial proportions and their standard uncertainties are estimated using a normal approximation. Burn-in draws ($10^6$) were discarded; all values are averaged across $9 \cdot 10^6$ time steps. Only significant digits are displayed. *For convenience, $\langle \Theta_0\rangle$ is arbitrarily set to 1, so this cell value equals $\langle \Theta_1\rangle$. $g_l$: gain; $200\nu_l$: firing rate of unit $l$, thus roughly assuming a 200 Hz firing rate for the sensory unit. \label{tab:01}}
\begin{ruledtabular}
\begin{tabular}{cllllc}
$l$ & $k_l \equiv \frac{\langle \Theta_l \rangle}{\langle \Theta_{l-1} \rangle}$ & $k_{\theta,l} \equiv \frac{\langle \Theta_l \rangle}{\langle |\mathcal{E}_{l-1}| \rangle}$  & $k_{\varepsilon,l} \equiv  \frac{\langle |\mathcal{E}_l| \rangle}{\langle \Theta_l \rangle}$ & \multicolumn{1}{c}{$g_l$} & $200 \nu_l$ \\
\hline
1 & $0.8470^{*}$& 1.0615 & 1.7100 & .4252 & 85 \\
2 & 1.7404      & 1.0178 & 1.5044 & .4503 & 38 \\
3 & 1.5241      & 1.0131 & 1.4605 & .4602 & 17 \\
4 & 1.4748      & 1.0119 & 1.4482 & .4697 & 8.2 \\
5 & 1.4660      & 1.0123 & 1.4570 & .4620 & 3.7 \\
6 & 1.4749      & 1.0123 & 1.4494 & .4672 & 1.7 \\
7 & 1.4662      & 1.0116 & 1.4552 & .461  & 0.8
\end{tabular}
\end{ruledtabular}
\end{table}

Table \ref{tab:01} displays, for each level, numerical estimates of scaling factors between successive levels between average thresholds and error signals, the firing rate $\nu_l=P(\mathcal{E}_l \neq 0)$, and the gain or probability of firing after an input
\begin{equation}
g_l \equiv 1-\int_0^{\Theta_l} f_{\tilde{A}_l} = 1-F_{\tilde{A}_l}(\Theta_l) = P(\tilde{A}_l > \Theta_l), \label{eq:glint}
\end{equation}
where $F_{\tilde{A}_l}$ is the cumulative distribution function of $A_l$. The firing frequency $\nu_l$ is the accumulated product of the current and subordinate levels' gains 
\begin{equation}
\nu_l = \prod_{i=1}^l g_i.
\end{equation}
We assumed a relative refractory period of $T_0=5$ms (or equivalently a sensory input rate of $\nu_0=200$Hz), which is typical of neurons \cite{Holland1998a,Dayan2001,Izhikevich2007} and in neural network simulations \cite{Dehaene2005}; this is also a rough estimate of the firing rate expected at each level under continuous exogenous driving. See Table \ref{tab:a1} for estimates of average thresholds and subthreshold activities for each level.

For $\langle \Theta_1 \rangle = .8464$ (Table \ref{tab:a1}), the probability that the sensory unit discharges at any time  $g_1=P(\mathcal{E}_1 \neq 0)$ is lower bounded by the firing rate of a memoryless process where subthreshold activities are reset at every iteration (white Gaussian noise activity, $P(\tilde{A}_{\mathcal{N}(0,1)} = 0) = .3973$), and upper bounded by the firing rate of a process where the size of time step approaches zero (Wiener process, $P(\tilde{A}_{B^{(t)}} = 0) = .5158$). This is because, conditioned on not having reached the threshold, the subthreshold activity $a_1^{(t)}$ follows a Gaussian random walk with unit variance within $[-\theta_1,\theta_1]$, so the quadratic mean of its translation distance after $t$ steps is distributed normally with variance $t$. This means that the variance of the stochastic process $A_1^{(t)}$ is larger than for white Gaussian noise, but smaller than for a Wiener process (see appendix \ref{sec:appb}). Therefore, the refractory period explains that the gain is lower than if each unit behaved as a continuous-time accumulator ---a ``Brownian neuron'' (Appendix \ref{sec:appb}).

The addition of subthreshold activity $A_l$ to $\mathcal{E}_{l-1}$ causes a threshold expansion: for the sensory unit $\langle \Theta_1 \rangle > \langle | \mathcal{E}_0| \rangle = \langle |\mathcal{N}(0,1)| \rangle = \sqrt{2/\pi} = .7979$, which evinces that $\langle \Theta_1 \rangle$  is larger than what it would be if $A_1$ were memoryless (i.e.\ if it were reset to zero after each iteration). Leftover subthreshold activity leads to $k_{\theta,l} \equiv \frac{\langle \Theta_l \rangle}{\langle |\mathcal{E}_{l-1}| \rangle} > 1$ because the addition of random sign signals diffuses bilaterally toward the thresholds (eventually triggering a discharge almost surely, and thus perfusing or diffusing through to the following level). Numerical simulations show that $\frac{\langle \Theta_l \rangle}{\langle |\mathcal{E}_{l-1}| \rangle} = 1.000$ if activities are reset after each iteration.
But even for memoryless units, if we let the threshold adapt, the gain of an element driven by white Gaussian noise $P(\tilde{A}_{\mathcal{N}(0,1)} = 0)$ (or most stationary process) would be less than .5. The reason lies in the asymmetry of $f_{|\mathcal{E}_l|}$. The probability of firing is the proportion of post-pulse activities lying above threshold; therefore a gain of .5 corresponds to a threshold equal to the \textit{median} of the pulse $\md{(\mathcal{E}_l)}$. However, the statistic that governs threshold location according to Eq.\ \ref{eq:thetal} is the mean of the pulse $\langle |\mathcal{E}_l| \rangle$, and $\langle |\mathcal{E}_l| \rangle > \md{|\mathcal{E}_l|}$ because $|\mathcal{E}_l|$ is right-skewed. In summary, $\frac{\langle \Theta_l \rangle}{\langle |\mathcal{E}_{l-1}| \rangle}>1$ is caused by the diffusive effect of leftover subthreshold activity, and gains are less than $.5$ because $f_{|\mathcal{E}_l|}$ is right-skewed.

\subsection{Spectral densities in the steady state}
\begin{figure}
\includegraphics[width=.5\textwidth]{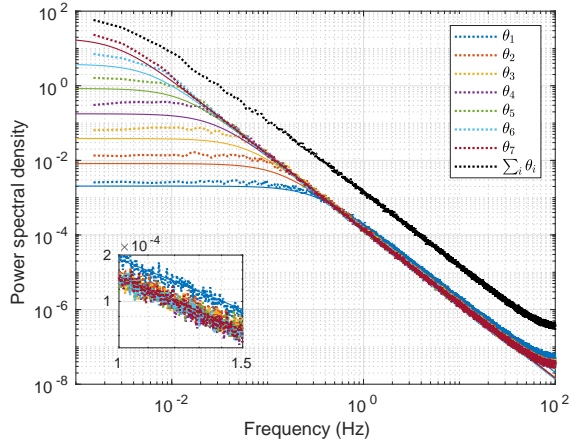}
\caption{Power spectral density estimate of threshold time series $\theta_l^{(t)}$ (dotted curves), and their corresponding AR(1) expressions (solid curves), for each level $l=1..7$, and their sum $\sum_{l=1}^7 \theta_l^{(t)}$. Sampling rate $T_0=5$ms; number of iterations $n_t=9 \cdot 10^6$. Estimation through Welch's method with Hamming windows of size $9 \cdot 10^4$ and $50\%$ overlap.\label{fig:psdT}}
\end{figure}

$\theta_l^{(t)}, a_l^{(t)}$ and $\varepsilon_l^{(t)}$ are stochastic processes indexed by $t$. $\theta_l^{(t)}$ can be construed (Eq.\ \ref{eq:thetal}) as an autoregressive model of order one or AR(1) with constant $w\langle \Theta_l \rangle$, regressor coefficient $1-w$ and white noise $w(|\tilde{a}_l^{(t)}| - \langle \Theta_l \rangle)$ with variance $w^2 \var{|\tilde{A}|}$. In the limit $T_0 \to 0$, it becomes a continuous-time Ornstein-Uhlenbeck process (OUP) with mean $\langle \Theta_l \rangle$, mean reversal term $-w$, and the same noise variance $w^2 \var{|\tilde{A}|}$ (Appendix \ref{sec:appc}). Its power spectral density (PSD) is
\begin{equation}
S_{\theta,\theta} (\nu | l) \approx \var{|\tilde{A}|} \frac{T_l\gamma_{\theta}^2}{\gamma_{\theta}^2+ \nu^2}, \label{eq:Stt}
\end{equation}
which is a Lorentzian function, with maximum $S_{\theta,\theta}(0) = T_l\var{|\tilde{A}|}$ at $\nu=0$ and corner frequency $\gamma_\theta = \frac{w}{2\pi T_l}$ (particular values for each unit are shown in Appendix \ref{sec:appc}, Table \ref{tab:c1}). For each level $l=1..7$, $T_l$ corresponds to the discharge rate of the preceding unit $T_l= \nu_{l-1}^{-1}$, with $\nu_0^{-1} = T_0$. The power spectral density estimate of numerically generated $\theta_l^{(t)}$ agrees with the spectral function of its corresponding AR(1) process (Fig.\ \ref{fig:psdT}). The discrepancies at the high and low frequency ends are caused by the estimation method bias (spectral leakage, see Appendix \ref{sec:appc}). PSD agree for frequencies above $\approx 0.5$ Hz (assuming a sampling rate of $T_0=5$ms) for all units except for the sensory unit (see blown up inset of Fig.\ \ref{fig:psdT}), which is slightly larger; this difference is due to the sensory unit being the only one that receives arbitrarily small input signals, as can be seen in Fig.\ \ref{fig:pdfs}. The corner frequencies $\gamma_{\theta,l}$ are all smaller than $\approx 0.3$ Hz (Table \ref{tab:c1}).

\begin{figure}
\includegraphics[width=.5\textwidth]{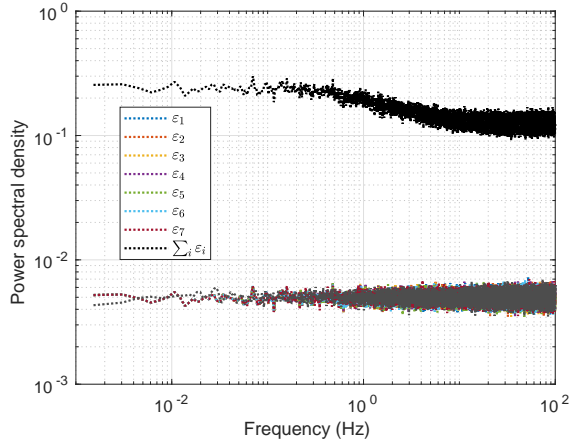}
\caption{Power spectral density estimate of pulses $\varepsilon_l^{(t)}$. Individual PSD's are identical to the PSD of  exogenous input (white Gaussian noise with unit variance) $S_{\varepsilon_0,\varepsilon_0}=T_0=5$ms. Note white noise signals are not affected by spectral leakage. Estimation parameters as in Fig.\ \ref{fig:psdT}.\label{fig:psdE}}
\end{figure}

Pulses derive their temporal structure from the exogenous input: the auto-correlation of discharges is directly inherited from the auto-correlation of its input. With white noise exogenous input, $\varepsilon_l^{(t)}$ also becomes white noise (albeit in general with a different density): its flat PSD (Fig.\ \ref{fig:psdE}) coincides with the PSD of the (standard Gaussian process) input distribution $\varepsilon_0^{(t)}$. Pulses from all units bear the same PSD because, as one ascends the hierarchy of units, increasing signal activity is offset by sparser firing. The sum of pulses from all units has larger PSD for low frequencies because the pulses are pairwise coherent and in phase only at low frequencies (see Fig.\ \ref{fig:epscoh}), which implies that the PSD of the sum of pulses is larger than the sum of their PSDs only at low frequencies.

The dynamics of the activities $a_l^{(t)}$ is defined (Eq.\ \ref{eq:a}) by a random walk with stationary and independent increments $\varepsilon_l^{(t)}$ (Fig.\ \ref{fig:pdfs}, in red) that resets to zero upon leaving $[-\theta_l, \theta_l]$. Crucially, the stochastic resetting implicit in Eq.\ \ref{eq:a} alone induces a NESS (with non-Gaussian fluctuations) because the combination of reset and diffusion constitutes a globally transition probability current-carrying loop between any activity value outside $[-\theta_l, \theta_l]$ and zero (the reset point), thus violating detailed balance \cite{Evans2011b}. This feature endows the system with complex dynamics.

Although at first sight subthreshold activities behave unlike thresholds, it turns out that their covariances and power spectra do. As for thresholds, we can use a continuous-time approximation, whereby in the limit $T_0 \to 0$ they are rendered an instance of OUP, where the mean-reverting force replaces the zero-resetting rule. Knowing that the probability that a unit resets at time $t$ since the last reset is its hitting time distribution $f_{\tau_{\pm\theta}}(t)$ (Appendix \ref{sec:appa}), we can estimate the covariance function of $a^{(t)}_l$ by taking the limit $T_0 \to 0$ , making some approximations, and applying the Fourier transform (see Appendix \ref{sec:appc}):
\begin{equation}
S_{a,a}(\nu|l) \approx \frac{\sigma_{\mathcal{E}_0}^2 \theta_l^2 (1-g_l)^2 \gamma_a}{\pi(\nu^2 + \gamma_a^2)}, \label{eq:Saa}
\end{equation}
where $\gamma_a = (2\pi\frac{3}{4}\theta_l^2T_0)^{-1}$ is the corner frequency and the maximum is $S_{a,a}(0) = 2\sigma_{\mathcal{E}_0}^2T_0(\frac{3}{4})^2\theta_l^4$. Importantly, the frequency spectrum of activities time series bears the same Lorentzian shape as thresholds do, so barring approximations activity resetting is effectively a restoring force similar to that of OUP.

\begin{figure}
\includegraphics[width=.5\textwidth]{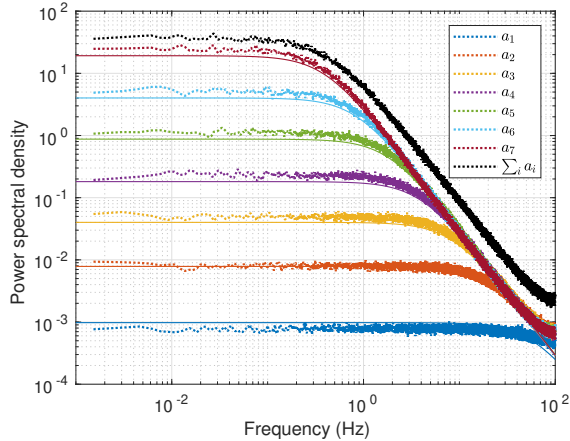}
\caption{Power spectral density estimate of activities $a_l^{(t)}$. Estimation parameters as in Fig.\ \ref{fig:psdT}.\label{fig:psdA}}
\end{figure}

\begin{figure*}
\includegraphics[width=1\textwidth]{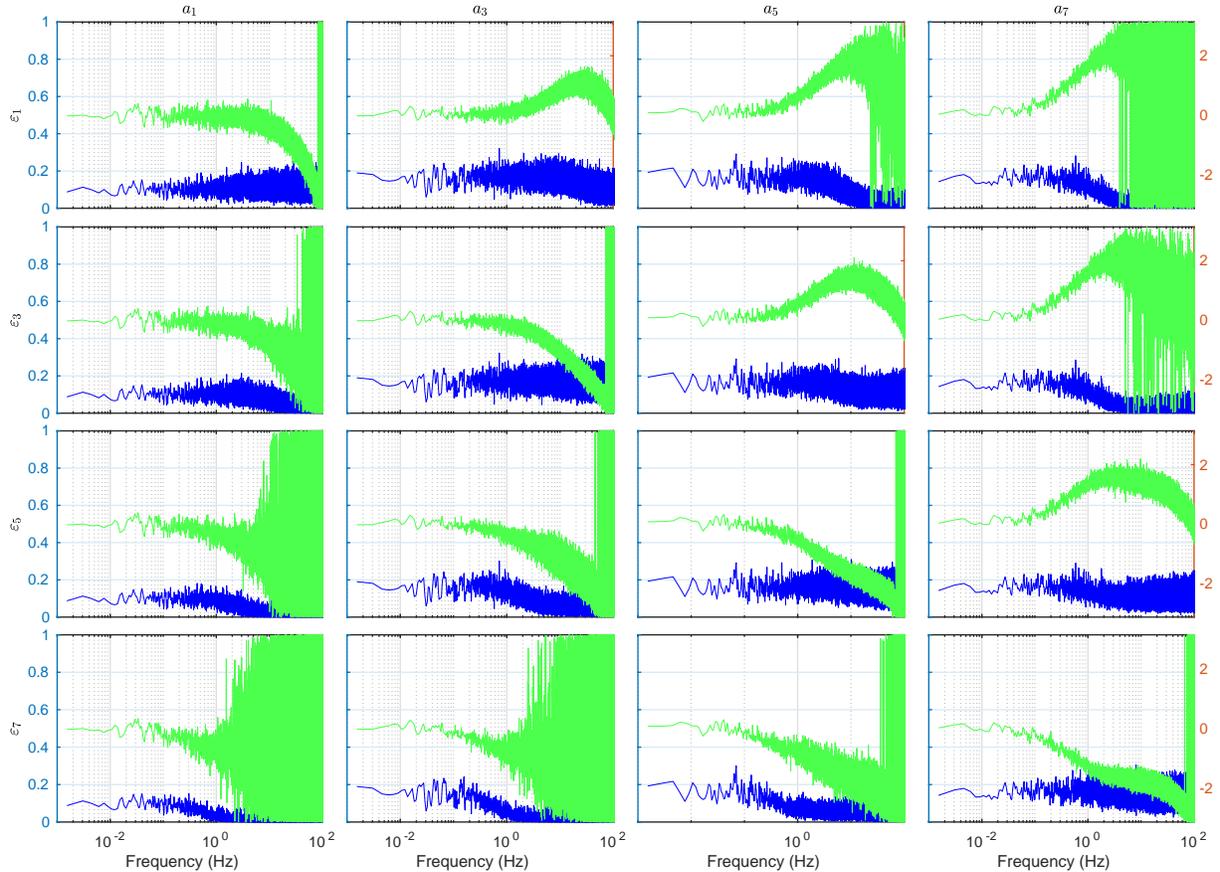}
\caption{Cross-spectral phase (green, in radians) and magnitude-squared coherence (blue) pair-wise estimates between pulses $\varepsilon_l^{(t)}$ (in rows) and subthreshold activities $a_l^{(t)}$ (in columns) for $l=1,3,5,7$. Cells above the main diagonal are populated by causal pulse-subthreshold activity interactions, i.e.\ combinations of $\varepsilon_m^{(t)}$ and $a_n^{(t)}$ with $m<n$, whereas cells in or below the diagonal display causality in the opposite direction. The smeared bands of some phase plots indicate that estimation is unreliable due to low or null coherence. Estimation parameters as in Fig.\ \ref{fig:psdT}.\label{fig:epsactcoh}}
\end{figure*}

The cross-power spectral density plot of pulses versus activities (Fig.\ \ref{fig:epsactcoh}) shows that some power is linearly transferred from pulses to subthreshold activities (cells above the diagonal, blue), and that the cross-spectral phase is positive, which evinces a delay of subthreshold activity phase with respect to preceding pulses (cells above the main diagonal, green). This ensues from the effect of pulses on supraordinate subthreshold activities. The within-level (outside the main diagonal) phase response is nonlinear. Negative cross-spectral phases in and below the main diagonal evince that subthreshold activities induce pulses in units of the same or higher levels. Unlike pulses, subthreshold activities are pair-wise incoherent over the whole frequency range (Appendix \ref{sec:appc}, Figs.\ \ref{fig:epscoh} and \ref{fig:actcoh}). This reflects that subthreshold activities exert a negligible effect on firing probabilities compared to pulses. In the steady state, the probability of firing given the preceding unit pulse $P(\mathcal{E}_l \neq 0 \; |\; \varepsilon_{l-1})$ is a monotonic increasing function of $|\varepsilon_{l-1}|$ because $P(\mathcal{E}_l \neq 0) = P(|A_l + \mathcal{E}_{l-1}| > \Theta_l)$ (cf. Fig.\ \ref{fig:TransMat}, middle left). In contrast, the probability of firing given the current unit subthreshold activity $P(\mathcal{E}_l \neq 0 \; |\; A_l)$ is approximately independent of $A_l$ (Fig.\ \ref{fig:TransMat}, right), which follows from the non-correlation of Brownian fluctuations.

In summary, subthreshold activity and threshold dynamics can be modeled as an AR(1) model or its continuous version OUP. This ensues from a weighted combination of the white noise injected at the sensory unit and a mean-reverting process effected by the adaptive threshold and subthreshold activity gating rules. Although activity power spectra plateau at low frequencies and thus display only short-range correlations \cite{Colaiori2004}, the corner frequencies marking the end of the plateau at different levels span two orders of magnitude (Table \ref{tab:c1}), so is their weighted linear combination. The temporal covariance functions are exponentially decaying, with a decay constant that at each level is proportional to the discharge period (Appendix \ref{sec:appc}). 
More generally, this behavior can be described with a Markov renewal process with discrete discharge times $t$, inter-event times $\nu_{l-1}$, subthreshold activity as a continuously valued state $A_l$, and jumps $\mathcal{E}_{l-1}$ (Appendix \ref{sec:appc}). 
Inter-event times are distributed as the first hitting time for reflected Brownian motion (of subthreshold activity with respect to its corresponding threshold), which has en exponentially decaying tail (Appendix \ref{sec:appc}, Eqs.\ \ref{eq:rbmtau1}, \ref{eq:rbmtau2}). The Lorentzian form of subthreshold activities (Eq.\ \ref{eq:Saa}) and thresholds (Eq.\ \ref{eq:Stt}) PSD is reminiscent of shot noise, a process that explains flicker noise in electronic devices where the sum of events in a Poisson process (with exponentially decaying inter-event times) has a Lorentzian PSD \cite{Schottky1926a}. Although it is also possible to construct fractal shot noise or fractal Markov renewal models that display a power law PSD with exponents in $[0,2]$ over an arbitrarily wide range of frequencies by setting power law decaying inter-event times \cite{Lowen1993}, in our system PSD are always Lorentzian-like, and power law behavior of subthreshold activities occurs for a restricted frequency span above a value $\gamma_{a,l}$ (corner frequency) dependent on the level $l$. Ultimately, the power-law frequency range of subthreshold activities stems from the intermittent integration of the driving input at each level. This is similar to how power law PSD noise can be generated by fractionally integrating white noise \cite{Barnes1966,Mandelbrot1968}.

\subsection{Pulse avalanche power laws \label{sec:properties_ava}}
As activity (and its square) percolates through the chain, pulses become rarer by a factor of $g_l$ and larger by a factor of $k_l$ at each level $l$. The shape of $f_{\mathcal{E}_l}$, $f_{\Theta_l}$, $f_{A_l}$ evinces some degree of self-similarity across levels (Fig.\ \ref{fig:pdfs}). If we assume self-similarity, we can approximate pulse densities at contiguous levels as 
\begin{equation}
f_{\mathcal{E}_{l-1}}(x) = k_l f_{\mathcal{E}_l}(k_lx), \label{eq:kfE}
\end{equation}
where $\mathcal{E}_l = k_l \mathcal{E}_{l-1}$ (the factor $k_l$  ensures that probabilities sum to 1). Likewise for frequency-weighted pulse size densities $\nu_{\mathcal{E}_{l-1}}(x) = \frac{k_l}{g_l} \nu_{\mathcal{E}_l}(k_lx)$ (where we defined $\nu_{\mathcal{E}_l} \equiv \nu_l f_{|\mathcal{E}_l|}$). The area under $\nu_{\mathcal{E}_l}$ for a range of pulse sizes can be interpreted as the frequency at which unit $l$ emits pulses within that range. Analogously to Eq.\ \ref{eq:kfE}, subthreshold activities and threshold size densities are scaled by $k_l$ under self-similarity:
\begin{eqnarray}
f_{\Theta_{l-1}}(x) &=& k_l f_{\Theta_l}(k_lx), \\
f_{A_{l-1}}(x) &=& k_l f_{A_l}(k_lx).
\end{eqnarray}

\begin{figure}
\includegraphics[width=.5\textwidth]{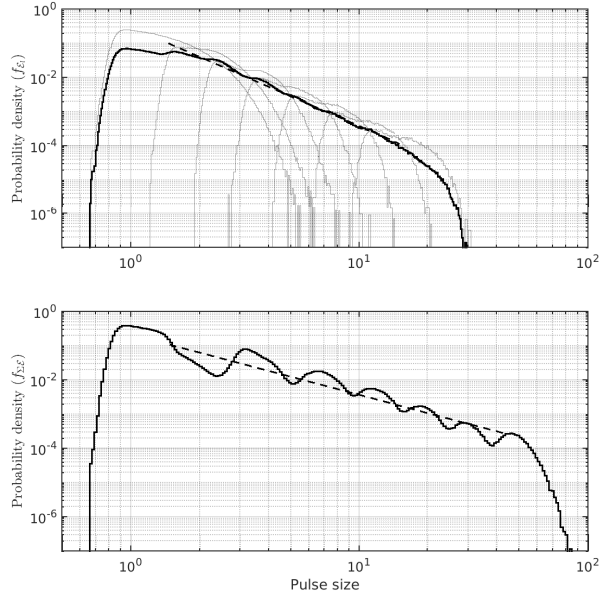}
\caption{Density estimates of pulse sizes (absolute values) for a 7-level chain. Upper: mixture density $f_{\mathcal{E}_{mix}}$ (black), and pulse size density $f_{\mathcal{E}_l}$ for each level $l=1\ldots 7$ (light grey). Lower: sum of pulse sizes density $f_{\Sigma\mathcal{E}}$. Slopes (dashed line) were fitted in the least squares sense in a range that  that excluded finite-size effects: $[\langle |\mathcal{E}_1| \rangle,\langle |\mathcal{E}_7| \rangle]$ for $f_{\mathcal{E}_{mix}}$ and $[\langle |\mathcal{E}_1| \rangle, \sum_{i=1}^7\langle |\mathcal{E}_i| \rangle]$ for $f_{\Sigma\mathcal{E}}$. \label{fig:Epdfhist}}
\end{figure}

The mixture density of pulses can be expressed as 
\begin{equation}
f_{\mathcal{E}_{mix}}(x) \propto \sum_{i=1}^{n_l} \nu_i f_{|\mathcal{E}_i|}(x) = \sum_{i=1}^{n_l} \nu_{\mathcal{E}_i}(x) \propto x^{-3} \label{eq:fEmixt},
\end{equation}
where the contribution of each level would be proportional to its discharge rate $\nu_i$, and the r.h.s.~becomes a density by dividing by $\sum_{i=1}^{n_l} \nu_i$. The envelope of the mixture density of pulse sizes (absolute values) rolls off as $f_{|\mathcal{E}|}(x) \propto x^{-3}$ (Fig.\ \ref{fig:Epdfhist}) because $\nu_l \propto l^{-2} $ holds in good approximation (Table \ref{tab:01}, the reason for this will be examined in the next section), and $\langle |\mathcal{E}_l| \rangle \propto l$, which entails that densities are spread over larger domains in inverse proportion to $l$, as expressed in Eq.\ \ref{eq:kfE}. Therefore, the exponent 3 comes from (in logarithmic scale) the exponents -2 and -1 that describe respectively  how $\nu_l$ and $f_{\mathcal{E}_l}$ are scaled down with respect to $l$ (Eq.\ \ref{eq:fEmixt}), and dividing by the expected pulse size scaling exponent 1.

If we picked up a signal constituted by the sum of pulses from different levels, then the sum of pulse densities (weighted by their frequencies) $f_{\Sigma\mathcal{E}}$ is the random variable that would describe it. Using the correspondence between the sum of random variables and convolution of their densities, we get
\begin{equation}
f_{\Sigma\mathcal{E}}(x) = \nu_{\mathcal{E}_1}(x) * \ldots * \nu_{\mathcal{E}_l}(x) \propto x^{-2} \label{eq:fEsum},
\end{equation}
where the decay exponent is obtained similarly to above by noting that $\langle |\mathcal{E}_1|\rangle + \ldots + \langle|\mathcal{E}_i|\rangle \propto \sum_l l \propto l^2$, using Eq.\ \ref{eq:kfE}. We have not specified  till now whether we are considering the sum of the absolute value of pulses or the absolute value of the sum of pulses because they are equivalent: $\langle | \sum_0^7 \mathcal{E}_l |\rangle= \sum_0^7 k^l\langle|\mathcal{E}_0|\rangle = \frac{k^8-1}{k-1}\langle|\mathcal{E}_0|\rangle$. This is because, by construction (cf. Eqs.\ \ref{eq:a}, \ref{eq:e}, Fig.\ \ref{fig:TransMat}), incoming pulses cannot trigger outgoing pulses of opposite sign. 

The mixture and sum density exponents in the scaling range are in rough agreement with the numerical estimates of -2.90 and -1.77 (Fig.\ \ref{fig:Epdfhist}). The log-periodic oscillations evinced by the densities in Fig.\ \ref{fig:Epdfhist} are reminiscent of discrete scale invariance (DSI) \cite{Sornette2000a}. For the mixture density, the logarithmic periods are $\Omega = k_l \approx 1.46$. As we will see later, $k_l \to k_*$ holds for large $l$; thus from the properties of geometric series $\Omega = \frac{k_*^l-1}{k_*^{l-1}-1} \approx k_*$ becomes independent of $l$ and $f_\mathcal{E}$ indeed displays DSI with the same logarithmic period.

\section{Steady state: flows and conservation}

\subsection{Diffusion and gating}
Activity dynamics can be construed as an iterative process consisting of two alternating stages: diffusion and gating. The presence of flows ---the signals transferring energy between units--- implies NESS. This precludes the existence of a Gibbs algorithm specifying a stationary distribution. Thus, averages are steady state averages, which are less informative than equilibrium averages. Nonetheless, in the steady state $\Theta_l$ converges (Fig.\ \ref{fig:pdfs}, green) in distribution to a Gaussian $ f_{\Theta_l} \overset{p}{\to} \mathcal{N}(\langle \Theta_l \rangle, \sigma_{\Theta_l})$ ; this was confirmed for a small enough learning rate ($w< .1$) in simulations. Because the dynamics of $a_l$ and $\theta_l$ are interlocked, working out their stationary distributions requires calculating their joint distribution. Note that $f_{\tilde{A}_l}$, $f_{A_l}$, and $f_{\mathcal{E}_l}$ are conditioned on the thresholds $\Theta_{\leq l} = \{\Theta_1,\ldots,\Theta_l\}$. The stationary distribution of activities $f_{A_l}$ has no closed-form expression, but it can be defined as the (infinite) sum over $t$ of the subthreshold activities resulting from $t$ iterations after the last discharge, weighted by the probability of \textit{not} having fired for $t$ time steps (which is a function of $\theta_l^{(t)}$; Eq.\ \ref{eq:Pal}). 

The stationary distributions of $A_l$ and $\mathcal{E}_l$ are interlocked and recursively defined by the composition of an addition (of $a_l$ to $\varepsilon_{l-1}$) and a truncation (gating of activity):
\begin{eqnarray}
f_{A_l}(x) & = &  g_l\delta(x) + (1-g_l)\B_{\Theta_l}f_{\tilde{A}_l}(x) = g_l\delta(x) + f_{\tilde{A}_l \: | \: \tilde{A}_l < \Theta_l}(x) \label{eq:Bae} \\
f_{\mathcal{E}_l}(x) & = & \frac{1}{g_l}\T_{\Theta_l}f_{\tilde{A}_l}(x) = \frac{1}{g_l}f_{\tilde{A}_l \:|\: \tilde{A}_l \geq \theta_l}(x), \label{eq:Tae}
\end{eqnarray}
where $\delta$ is the Dirac delta, and $\B_{\Theta}f$ and $\T_{\Theta}f$ are respectively the normalized body and tails of a two-sided truncated $f$ density, with support $[-\theta, \theta]$ and $[-\infty, -\theta] \cup [\theta, \infty]$ (notice $\Theta$ is a random variable). The term $g_l\delta(x)$ indicates that the probability of the subthreshold activity $A_l$ being zero is $f_{A_l}(0) = g_l$. The learning rate $w$ governs the variance of $\Theta$ (as $w \to 0$, the variance vanishes), thereby determining the smoothness of $f_{A_l}$ and $f_{\mathcal{E}_l}$. Stationarity mandates that $f_{\tilde{A}_l}$ be invariant under the composition of convolution with $f_{\mathcal{E}_{l-1}}$ and truncation (Eq.\ \ref{eq:Bae}).

\subsection{Conservation of squared activity (energy) \label{sec:ConsEn}}

A sequence of exogenous inputs constitutes a one-dimensional activity random walk. In a continuum approximation, and in the absence of gating, the diffusing activity $\phi$ density is defined by the heat equation $\frac{\partial \phi}{\partial t} = D \frac{\partial^2 \phi}{\partial x^2}$ with diffusion constant $D=\frac{\sigma_{\mathcal{E}_0}^2}{2T_0}$ (with $\sigma_{\mathcal{E}_0}^2 \equiv \var{\mathcal{E}_0} = \langle \mathcal{E}_0^2 \rangle$). Its fundamental solution is the heat kernel, which describes the evolution of a ``particle" of activity placed at time $t=0$ at $x=0$ diffusing in $(-\infty,\infty)$ without boundary conditions (Green's function number X00): a time-varying Gaussian function with zero mean and variance $2Dt$ (which is equivalent to the mean squared displacement $\text{MSD}=\langle(x(t)-0)^2\rangle$ in one dimension).
The entropy of an exogenous input with Gaussian density is $\frac{1}{2}(\log{2\pi\sigma^2}+1)$ (with variance $\sigma^2$), so the entropy of the heat kernel with MSD = $2Dt$ that describes diffusion of activity resulting from accumulating input is $\frac{1}{2}(\log{4\pi Dt}+1) \sim \frac{1}{2}\log{t}$, and its derivative or entropy production rate is $\dot{S} \sim \frac{1}{2t}$. Since the inflow of exogenous input is proportional to time $\sim t$, and entropy is an extensive variable (proportional to input activity) it ensues that the entropy production rate of the system is a constant, as is characteristic of a steady state regime. In the steady state, with appropriate boundary conditions specifying threshold location, the heat equation becomes the Poisson equation. Then, exogenous input at the sensory unit are the only source term of activity, appropriate source terms are needed to model incoming pulses for non-sensory units, and activity leaves the system at the end unit, which is the only sink. Because of symmetry, the solution at any level could also be obtained by solving for positive activities with a reflecting wall at zero. 

We have been sloppy in the wording because the quantity that is conserved is not activity but its square. This is because the Wiener process has independent increments, which implies that its quadratic variation or mean squared displacement equals the elapsed time; in other words, the variance of a sequence of inputs equals the sum of the variances of inputs. This is analogous to how, in classical mechanics, kinetic energy ---which is proportional to squared velocity--- is conserved, but not velocity. Under this view, the driving input is periodically (with period $T_0$) injecting into the system energy chunks that are sequentially and stochastically relayed by each unit until reaching the end, whence they exit the system. This suggests that we refer to squared activity as ``energy". Energy is locally conserved under addition and transfer of activity. Since the PC system has a single input (with variance or energy $\sigma_{\mathcal{E}_0}^2$), if we assume it devoid of activity at time $t=0$ with a time step of 1, the total energy injected after time $t$ is $t\sigma_{\mathcal{E}_0}^2$. By conservation, in the steady state the power or  energy flow entering the system must equal the energy flow leaving the system: over long enough time spans, the average rate of energy injected by the exogenous input equals the average rate of energy transferred between any two successive levels. Accordingly, the continuous limit solution in one dimension, which is given by the Poisson equation (after accounting for boundary conditions), is the constant $\sigma_{\mathcal{E}_0}^2$. Thus, in the steady state and in one dimension, the average energy flux traversing the chain is a constant (independent of location) and equal to the driving input energy flux.

Next we turn to the effect of gating, or how discharges affect subthreshold activity. Although these are all-or-none events of activity transfer between units that result in jumps in the subthreshold activity, as above mentioned the average energy flow over discharges throughout the chain is constant $\langle \mathcal{E}_0^2 \rangle = \sigma_{\mathcal{E}_0}^2$, and is driven solely via diffusion, as in Zhang's sandpile model \cite{Zhang1989}). Although the dynamic profile of spreading activity is Gaussian, the steady state profile of average activity is flat. This is expressed by the continuity equation
\begin{equation}
\Phi = \nu_l \langle \mathcal{E}_l^2 \rangle \quad \forall l = 1 \ldots n_l, \label{eq:conserv}
\end{equation}
where for simplicity we set $\nu_0=P(\mathcal{E}_0 \neq 0)=1$, so $\Phi = \sigma_{\mathcal{E}_0}^2$. Hence, at each level the product of the discharge frequency and the energy relayed per discharge is constant on average, and equal to the energy flow $\Phi$ injected into the sensory unit; this was confirmed in simulations. It follows that 
\begin{equation}
k_l = g_l^{-1/2} \quad \forall l = 1 \ldots n_l.  \label{eq:kgBrownian}
\end{equation}

With standard Gaussian driving input (Section \ref{sec:gausim}), we get $\Phi = \sigma_{\mathcal{E}_0}^2 = 1$, which implies $D=1/2$. Despite the intractability of subthreshold activity and pulse densities (Fig.\ \ref{fig:pdfs}, black and red) energy conservation entails that variances can be readily calculated. By symmetry $\langle \mathcal{E}_l\rangle = 0$ so we can decompose pulse variance into its absolute value raw and central components as $\sigma_{\mathcal{E}_l}^2 \equiv \var{\mathcal{E}_l} = \langle \mathcal{E}_l^2 \rangle - \langle \mathcal{E}_l \rangle^2 = \langle \mathcal{E}_l^2 \rangle = \langle |\mathcal{E}_l|^2 \rangle = \var{|\mathcal{E}_l|} + \langle |\mathcal{E}_l|\rangle^2$. Combining this with Eq.\ \ref{eq:conserv}, we get
\begin{eqnarray}
\sigma_{\mathcal{E}_l}^2 = \langle |\mathcal{E}_l|^2 \rangle = \var{|\mathcal{E}_l|} + \langle |\mathcal{E}_l| \rangle^2 = \nu_l^{-1}.
\end{eqnarray}
Due to diffusion, subthreshold activities, pulses, and thresholds from coupled units are scaled by a factor equal to the inverse gain: 
\begin{equation}
g_l^{-1} = \frac{\langle \mathcal{E}_l^2 \rangle}{\langle \mathcal{E}_{l-1}^2 \rangle} = \frac{\langle A_l^2 \rangle}{\langle A_{l-1}^2 \rangle} \quad \forall l = 1 \ldots n_l; \label{eq:interlevsc}
\end{equation}
this follows from conservation (Eq.\ \ref{eq:conserv}) because pulse energy is inversely proportional to firing rate, and subthreshold activity is in turn proportional to incoming pulse energy.

Since the discharge rate determines the variance of $\mathcal{E}$, it also determines indirectly some statistics of $A$ and $\Theta$. Using the law of total variance to split $\tilde{A}_l$ into $A_l^+$ and $\mathcal{E}_l$ at the threshold value $\Theta_l$ (Fig.\ \ref{fig:SFL}) yields $\var{\tilde{A}_l} = \langle \tilde{A}_l^2 \rangle = (1-g_l)\var{A_l^+} + g_l \var{\mathcal{E}_l}$, so
\begin{equation}
\langle \tilde{A}_l^2 \rangle = \langle A_l^2 \rangle + g_l\langle \mathcal{E}_l^2 \rangle, \label{eq:A2+gE2}
\end{equation}
where $A_l^+$ is the subthreshold activity excluding the rest state ($A_l=0$), with $f_{A_l} = g_l\delta(0) + (1-g_l) f_{A_l^+}$, and we used $\langle A_l^2 \rangle \equiv \sigma_{A_l}^2 = (1-g_l)\sigma_{A_l^+}^2$. This is an approximation valid only for small $w$ because this decomposition of $\tilde{A}$ assumes a constant $\Theta_l$. 

Next we seek the link between thresholds and activities. The absolute value of activity comes forth as a fundamental variable due to the symmetry of the threshold updating rule (Eq.\ \ref{eq:thetal}). Using the same approximation and the law of total expectation for $|\tilde{A}|$ at the threshold value we obtain
\begin{equation}
\langle |\tilde{A}_l| \rangle = \langle \Theta_l \rangle = \langle |A_l| \rangle + g_l \langle |\mathcal{E}_l| \rangle, \label{eq:mnAEmnT}
\end{equation}
where we used again the law of total expectation (on $|A_l|$) to derive $\langle |A_l| \rangle = (1-g_l)\langle |A_l^+| \rangle$ by noting that the pulse frequency $\nu_l$ (Table \ref{tab:01}) is also the fraction of firing trials, that resets subthreshold activity to zero. Similarly, applying the law of total variance to $|\tilde{A}_l|$ yields 
\[
\var{|\tilde{A}_l|} = g_l\var{\mathcal{E}_l} + (1-g_l)\var{|A^+_l|} + g_l(1-g_l)(\langle |\mathcal{E}_l| \rangle - \langle |A^+_l| \rangle)^2.
\]

Despite the expressions deduced above, the relationship between expected thresholds and pulse energies remains underdetermined. Thus, we now seek relationships between threshold expectation $\langle \Theta_l \rangle = \langle |\tilde{A}_l| \rangle$ and pulse size expectation $\langle |\mathcal{E}_l| \rangle$. If the subthreshold activity were reset after each iteration, $\langle |\tilde{A}_l| \rangle$ would equal $\langle | \mathcal{E}_{l-1} | \rangle$ because $\tilde{A}_l = A_l + \mathcal{E}_{l-1}$. But the presence of subthreshold activity has the effect of increasing $\langle |\tilde{A}_l| \rangle$ by a factor of $k_\varepsilon$:
\begin{eqnarray}
\langle \Theta_l \rangle = \langle |\tilde{A}_l| \rangle & = & k_{\theta,l} \langle |\mathcal{E}_{l-1}| \rangle  \\
\langle |\mathcal{E}_{l-1}| \rangle & = & k_{\varepsilon,l} \langle \Theta_{l-1} \rangle
\end{eqnarray}
for $l=2..L$, where $k_l = k_{\theta,l} k_{\varepsilon,l} \in [1,2]$ and $k_{\theta,l}, k_{\varepsilon,l} > 1$. Although all these variables are functions of $l$, they converge (separately) as the chain becomes infinitely long ($l \to \infty$), as we will see in the next section.

\subsection{Edge cases \label{sec:edgecases}}
To better understand our setup, it is instructive to probe a few limiting cases. For clarity, we will drop the subscript $l$ whenever an expression involves variables pertaining to the same level (e.g.\ Fig.\ \ref{fig:SFL}). A smaller $g$ entails lower discharge frequency, more energy accumulated, and a larger scaling factor $k$ (and vice versa). In the limit $g \to 0$, we get $\langle|\mathcal{E}|\rangle^2 = \langle\mathcal{E}^2\rangle = \Theta^2$ and $k\to\infty$. In the limit $g \to 1$, subthreshold activity $A$ vanishes along with scaling ($k=1$). However, these limit expressions are incompatible with the threshold setting rule (Eq.\ \ref{eq:thetal}): $\langle|\tilde{A}|\rangle\equiv\langle|A+\mathcal{E}/k|\rangle \approx \Theta$ cannot be located at an extreme, except in the degenerate case $\var{|\tilde{A}|}=0$. It can be observed (Table \ref{tab:01}) that the threshold expands with respect to pulse size mean while satisfying $\langle|\mathcal{E}|\rangle/k < \Theta$. Now we ask what would be the implications of $\langle|\mathcal{E}|\rangle/k=\Theta=\langle|\tilde{A}|\rangle$. This would hold if $|\mathcal{E}/k| + A \geq 0$ because the sum of the expectation of random variables equals the expectation of their sum, so $\langle |\mathcal{E}/k + A| \rangle = \langle |\mathcal{E}/k| + A \rangle = \langle|\mathcal{E}|\rangle/k + \langle A \rangle = \langle|\mathcal{E}|\rangle/k$, since $A$ is symmetric about zero. But $|\mathcal{E}/k| + A \geq 0$ is unfeasible because it would entail $\max{|A|} = \Theta \leq \Theta/k = \min{|\mathcal{E}/k|}$ (see Fig.\ \ref{fig:SFL}), which is achievable only in the limit $g\to 1$, where $A$ vanishes. However, by briefly indulging in this unfeasibility we can find an upper bound for the discharge rate: if $\langle|\mathcal{E}|\rangle/k=\Theta$ were true while somehow retaining multiplicative scaling, we can assume that (with frequency $g_u$) diffusive scaling should be counterpoised by the probability mass $g_u$ released by each discharge (cf. Fig.\ \ref{fig:SFL}, area below $f_{\mathcal{E}}$) as
\begin{equation}
g_u^{-1/2} = 1+g_u, \label{eq:gu}
\end{equation}
which after reshuffling becomes $g_u^3 + 2g_u^2 + g_u - 1 = 0$, which can be solved using Cardan's cumbersome formula or numerically, yielding a single real root $g_u \approx .46557$.

\subsection{Fixed threshold approximation \label{sec:FTA}}
For symmetric exogenous input, the expected value for $A$ and $\mathcal{E}$ can be worked out by noting that squared activity or energy is conserved (Eq.\ \ref{eq:conserv}) in perfusing the chain, and the threshold location is derived indirectly from their energies. To simplify the subsequent analysis, we fix the threshold to $\theta_* \equiv \langle \Theta \rangle$, corresponding to the limit of vanishing threshold fluctuations $w \to 0$ (so $\var{\Theta}=0$). Swapping the square of the expectation with the expectation of the square  $\langle\Theta^2\rangle = \langle\Theta\rangle^2$ is a form of mean-field approximation: it neglects correlations. This is justified because $w$ should be typically a small value for an adaptive system that learns the structure of its environment via iterative exposure to stimuli, so the timescales of threshold and activity dynamics are widely separated \cite{Vespignani1998}. Since $A$ and $\mathcal{E}$ are necessarily linked to $\Theta = \langle \tilde{A}^2 \rangle$ and $g = k^{-2}$, we need to write down four equations or constraints.

The steady state discharge frequency $g_* = k_*^{-2}$ must be such that the energy scaling induced by expanding diffusive forces $k^2 \sim g^{-1}$ matches pulse energy scaling $\sigma_\mathcal{E}^2 \equiv \langle\mathcal{E}^2\rangle$; this is expressed by Eq.\ \ref{eq:interlevsc}. Hence, a constraint among subthreshold activity, pulse, and threshold energies can be derived from a continuity equation integrated over one cycle of input, such that discharge frequency is $g$. Taking the interval $[\theta_*, \max{\mathcal{E}}]$ corresponding to the support of $f_{|\mathcal{E}|}$ (Fig.\ \ref{fig:SFL}, red), in the steady state the inward and outward flows of energy must cancel out by local conservation, so
\begin{equation}
g\sigma_\mathcal{E}^2 = g(g\sigma_\mathcal{E}^2 + \sigma_A^2) + \theta_*^4, \label{eq:sdE_sdA_T}
\end{equation}
where the left hand side is the averaged pulse energy as outward energy flow integrated over one cycle and $\sigma_A^2 \equiv \langle A^2 \rangle$. In the right hand side, the first term corresponds to the direct inflow of energy from post-pulse activity ---$\langle \tilde{A}\rangle = g\langle\mathcal{E}^2\rangle + \langle A^2\rangle$ weighted by the probability $P(\tilde{A} > \theta_*)=g$--- and the second term quantifies the diffusive transfer of energy from $A$ to $\mathcal{E}$. The rationale for the latter rests on Brownian motion properties: the energy flux through $\theta_*$ for a ``packet'' of activity concentrated at the threshold (the ``energy of the threshold'') is $\theta_*^2$ per time unit, and the duration of a discharge cycle is the expected hitting time for reflected Brownian motion (which embodies the symmetry of the densities) is also $\tau_{\pm \theta_*} = \theta_*^2$ (Appendix \ref{sec:appb}, Eq.\ \ref{eq:rbmhte}). Thus, the expected total energy transported across the threshold over one cycle with a continuous exogenous driving input energy flux $\Phi=1$ (Section \ref{sec:ConsEn}) by purely diffusive effects would be $\theta_*^2 \theta_*^2 = \theta_*^4$.

Another equation can be obtained by considering that discharge frequency is inversely proportional to the between-level scaling of energy. The average energy remaining in the form of (squared) subthreshold activity immediately after a discharge ($A^+$) divided by the squared threshold should equal the discharge frequency $\frac{\sigma_{A^+}^2}{\theta_*^2} = g$. This is because, by conservation, the accumulated squared subthreshold activity as a fraction of the squared threshold establishes the discharge frequency. Hence 
\begin{equation}
\frac{\sigma_A^2}{\theta_*^2} = g(1-g), \label{eq:sdA_T}
\end{equation}
which is in good agreement with the simulations (Table \ref{tab:01}). Note that $g(1-g) \approx 1/4$ holds for a relatively broad range of $g \approx 0.5$ values.

To spot a relationship between threshold and pulse energies,  an essential ingredient is the effect of diffusion on the density of absolute post-pulse activity $|\tilde{A}|$. Heuristically, the average pulse energy transmitted, over one cycle, as a multiple of squared threshold $\frac{g\langle \mathcal{E}^2\rangle}{\theta_*^2}$, should follow from an analogous ratio between the threshold energy and the incoming pulse size energy $\left( \frac{\langle |\tilde{A}| \rangle}{\sqrt{g}\langle|\mathcal{E}|\rangle}\right)^2=\frac{\theta_*^2}{g\langle|\mathcal{E}|\rangle^2}$. This ratio is larger than 1 and reflects that the addition of incoming pulses to subthreshold activity has the diffusive effect of ``bumping up'' the mean absolute value. By design, the threshold is set to the post-pulse \textit{absolute} activity mean (Eq.\ \ref{eq:thetal}), so increasing the mean absolute value from $\sqrt{g}\langle|\mathcal{E}|\rangle$ to $\theta_*$ (which equals $\langle |\tilde{A}| \rangle$ and the infimum of $|\mathcal{E}|$) necessarily affects the ratio of pulse energy discharged to threshold energy, such that energy conservation is satisfied. Bringing these expressions together, $\frac{\theta_*^2}{g\langle |\mathcal{E}| \rangle^2} = \frac{g\langle \mathcal{E}^2 \rangle}{\theta_*^2}$, which rearranged is
\begin{equation}
\theta_*^2 = g \langle|\mathcal{E}|\rangle \sigma_\mathcal{E},  \label{eq:T_aE}
\end{equation}
which evinces that larger scaling factors ($k=g^{-1/2}$) call for longer tails. In the limit of zero dispersion $\var{|\mathcal{E}|} \to 0$ and $\langle|\mathcal{E}|\rangle^2\to\langle\mathcal{E}^2\rangle\equiv \sigma_\mathcal{E}^2$, it becomes $\frac{\theta_*^2}{\langle\mathcal{E}^2\rangle} = g$, so the ratio of threshold and pulse energies equals discharge frequency.

A fourth (and last) equation stems from the properties of diffusion, applied to the absolute values of the system variables (Fig.\ \ref{fig:SFL}). By design, $\theta_*= \langle |\tilde{A}| \rangle$ minimizes the mean squared error of $|\tilde{A}|$, so $\var{|\tilde{A}|}=g\langle (|\mathcal{E}|-\theta_*)^2 \rangle + (1-g)\langle (\theta_*-|A|)^2 \rangle$ (the same relation is embodied by Eq.\ref{eq:mnAEmnT}). Noting that $\var{|\tilde{A}|}$ is the mean squared displacement from $\theta_*$, the contributions from the lower $|A^+|$ and upper $|\mathcal{E}|$ partial second order moments can be deduced to be inversely proportional to their respective probability masses, which coincide with their respective renewal frequencies ($g$ for $\mathcal{E}$ and $1-g$ for $A^+$):
\[\frac{\langle (|\tilde{A}|-\theta_*)^2 \rangle_{>\theta_*}}{\langle (\theta_* - |\tilde{A}|)^2 \rangle_{\leq\theta_*}} = \frac{g\langle (|\mathcal{E}|-\theta_*)^2 \rangle}{(1-g)\langle (\theta_*-|A^+|)^2 \rangle}=\left(\frac{g}{1-g}\right)^{-1}.
\]
After rearranging, 
\begin{equation}
\frac{\langle (|\mathcal{E}|-\theta_*)^2 \rangle}{\langle (\theta_*-|A^+|)^2 \rangle} = \left( \frac{1-g}{g} \right)^2, \label{eq:aE_T_aA}
\end{equation}
wherewith we can derive expressions for mean absolute values:
\begin{eqnarray}
\langle |\mathcal{E}| \rangle &=& \frac{\theta_*^2+g^2-\langle A^2 \rangle(1-g)}{2g\theta_*}  \\
\langle |A|\rangle &=& \frac{\theta_*^2 - g^2 + \langle A^2 \rangle(1-g)}{2\theta_*}  
\end{eqnarray}

\begin{figure}
\includegraphics[width=.5\textwidth]{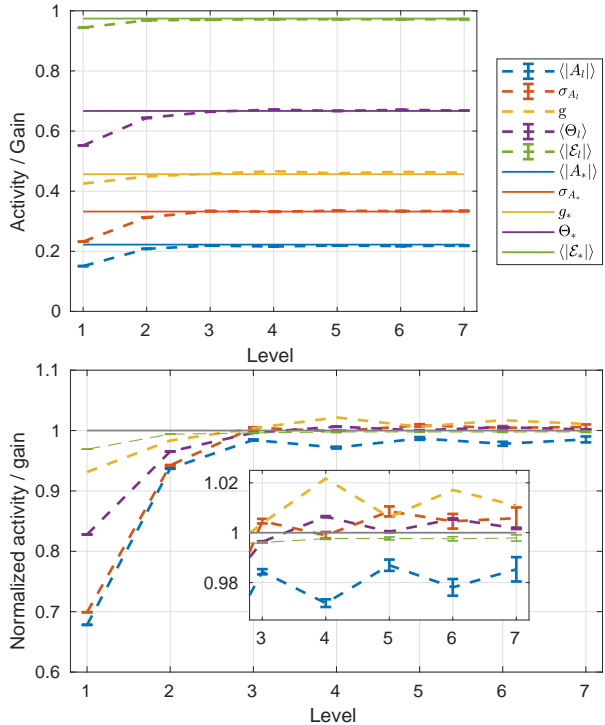}
\caption{Estimates of $A, g, \Theta, \mathcal{E}$ statistics (top plot legend) for levels $l=1 \ldots 7$. Top: dashed lines indicate numerical simulation estimates (number of iterations $n_t=9 \cdot 10^6$), whereas solid lines indicate values found with the fixed threshold approximation. Bottom: same as in top, but normalized by fixed threshold estimate homologues. Errorbars indicate s.e.m.\@ corrected for autocorrelation with effective sample sizes. \label{fig:ErrorbarGAET}}
\end{figure}

Equations \ref{eq:sdE_sdA_T}--\ref{eq:aE_T_aA} constitute a system of four interlocked nonlinear equations, which can be solved iteratively with a numerical routine; we used \texttt{fsolve} from MATLAB 9.2 (The MathWorks, Inc.; Natick, Massachusetts). In all, the fixed threshold approximation (FTA) yields
\begin{eqnarray}
g_* &\approx & 0.4565, \nonumber \\
\sigma_{A_*} &\approx & 0.3322, \nonumber \\
\theta_* &\approx & 0.6668, \nonumber \\
\sigma_{|\mathcal{E}_*|} &\approx & 0.9742, \nonumber \\
\sigma_{|A_*|} & \approx & 0.2222, \nonumber \\
k_* & \approx & 1.4801, \nonumber
\end{eqnarray}
whence $k_{\varepsilon,*} \approx 1.4609$, $k_{\theta,*}=1.0132$ also follows. These values are in reasonable agreement with numerical simulation estimates ($n_t=10^7$ iterations): Fig.\ \ref{fig:ErrorbarGAET} shows a reasonable convergence toward FTA estimates over the first 7 levels. To account for one-step autocorrelations (Eq.\ \ref{eq:Ctt}) in calculating error estimates, we used effective sample sizes based on the the autocorrelation time \cite{Straatsma1986,Kass1989}: $\tau = 1 + \sum_{i=1}^\infty \rho^i$ (with $\rho^i$ autocorrelation at lag $i$), which is roughly twice the decay time constant of the covariance function ($1/w$ and $\frac{3}{4}\theta^2$ for $\theta^{(t)}$ and $a^{(t)}$ respectively, see Appendix \ref{sec:appc}). 

In deriving these solutions we have implicitly assumed (Eqs.\ \ref{eq:sdE_sdA_T}, \ref{eq:T_aE}) that incoming pulses are identical to outgoing pulses, up to a scaling factor $k$ ---which implies that the same scaling relationship holds for any variable with respect to its subordinate level homologous. This assumption deserves a justification.

\section{Rescaled recurrent loop configuration \label{sec:RRLC}}
\begin{figure}
\includegraphics[width=.5\textwidth]{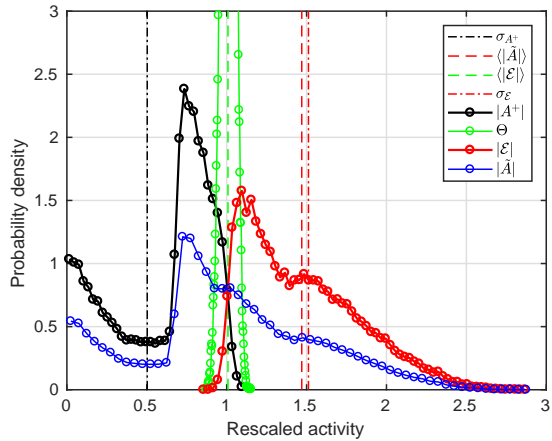}
\caption{Self-feeding configuration density estimates for $|A^+|$, $|\tilde{A}|$, $|\mathcal{E}|$, and $\Theta$. Recall that (signed) variable densities are symmetric. The incoming pulse density is equated to the outgoing pulse density divided by $\langle|\mathcal{E}|\rangle$, so that incoming pulse size mean equals 1. Learning rate $w=0.01$. A bootstrapped pool of $N=10^5$ data points for each variable was resampled and replaced one sample at a time for $10^6$ iterations. Vertical lines indicate the location of relevant statistics (see legend).\label{fig:SFL}}
\end{figure}

The rules defining PC (Eqs.\ \ref{eq:a_tl}--\ref{eq:thetal}) prescribe a summation of the incoming pulse $\mathcal{E}_{l-1}$ and truncated subthreshold activity $A_l$ random variables. Clearly, $f_{\tilde{A}_l}$ depends on $f_{\mathcal{E}_{l-1}}$, and both in turn depend on $f_{\tilde{A}_{l-1}}$ from the subordinate level. But what if all levels were iteratively coupled as chain links? Fig.\ \ref{fig:ErrorbarGAET} suggests that as the chain approaches an infinite number of links ($l \to \infty$), all variables converge respectively to specific densities, and that at level $l$ any variable becomes a larger ($k$-scaled) self-similar replica of its precursor in level $l-1$. Hence, we can set outgoing pulses to be self-similar versions of incoming pulses, with the recurrence relation
\begin{eqnarray}
\varepsilon_{t+1} &=& \left\{ \begin{array}{lll}
  a_t + \varepsilon_i/k_t \;\; \text{if}\; a_t + \varepsilon_i/k_t  \geq \theta_t \\
  \varepsilon_j  \qquad\qquad\;  \text{else}  
\end{array} \right. \label{eq:hc1} \\ 
a_{t+1} &=& \left\{ \begin{array}{lll}
  0  \qquad\qquad\; \text{if}\; a_t + \varepsilon_i/k_t  \geq \theta_t \\
  a_t + \varepsilon_i/k_t  \;\; \text{else}\;
\end{array} \right. \label{eq:hc2}\\
\theta_{t+1} &=& (1-w)\theta_t + w|a_t + \varepsilon_i/k_t| \label{eq:hc3} \\
k_{t+1} &=& \langle |\varepsilon_{1:N}| \rangle \label{eq:hc4},
\end{eqnarray} 
where subscripts denote time instead of level indices and 
$\varepsilon_i, \varepsilon_j$ are randomly and independently (bootstrap) resampled from a pool $\varepsilon_{1:N} = \{\varepsilon_1, \ldots, \varepsilon_N \}$ that in turn is replenished in a first in, first out manner with outgoing pulses. Resampling the incoming pulse $\varepsilon_i$ instead of picking the last pulse $\varepsilon_t$ is required because the actual configuration we are trying to mimic is essentially feedforward (unidirectional, toward higher levels $l$) and stochastically driven, so we need to avoid putting in spurious autocorrelations. The ``else'' assignment $\varepsilon_j$ in Eq.\ \ref{eq:hc1} picks out a past value to become $\varepsilon_{t+1}$, instead of zero as in Eq.\ \ref{eq:e}; this is consistent with the definition of $\mathcal{E}$ (Section \ref{sec:properties}) by ensuring that zero values are excluded. Eqs.\ \ref{eq:hc1} -- \ref{eq:hc4} constitute a form of bootstrap resampling (Section \ref{sec:RecLoopMcmc}); after resampling and replacing a pool of $N=10^5$ data points for $10^6$ iterations (see Fig.\ \ref{fig:3dBmATE}), we obtain densities invariant under rescaling (Fig.\ \ref{fig:SFL}), which can be seen as the limiting densities of Fig.\ \ref{fig:pdfs} for $l \to \infty$. This is analogous to how universal scaling functions emerge on large scales in critical systems \cite{Sethna2001}. 
Eq.\ \ref{eq:hc4} estimates the scaling factor through the time average of the outgoing pulse size ---note that the incoming pulse sizes are rescaled by the inverse of the scaling factor (Eqs.\ \ref{eq:hc1}, \ref{eq:hc2}). This time average is taken over a time span $[1, N]$; by the law of large numbers the larger is $N$, the scaling factor estimate becomes more accurate. 
Eq.\ \ref{eq:hc3} is the delta learning rule, an algorithm that converges in probability to the expectation of $|\tilde{A}|$ with a speed and variance dependent on the parameter $w$. Eq.\ \ref{eq:hc3} can be expanded as $\theta_t = w\sum_{i=1}^t(1-w)^{i-1}|\tilde{a}_{n-i}| + (1-w)^t\theta_0$, which determines the balance at which past values are forgotten and new ones are incorporated into the estimate of $\Theta$. Thus, Eq.\ \ref{eq:hc3} could be replaced by performing explicit threshold averaging as
\begin{equation}
\theta_{t+1} = \langle |\tilde{a}_{1:N}|\rangle, \label{eq:hc3mean}
\end{equation}
which, similarly to Eq.\ \ref{eq:hc4}, estimates $\langle \Theta \rangle$ with higher accuracy if $N$ is large enough.

Eqs.\ \ref{eq:hc1} -- \ref{eq:hc4} constitute a 4-dimensional nonlinear recurrence relation, that simulates the feedforward configuration defined in Eqs.\ \ref{eq:a_tl}--\ref{eq:thetal} in the limit $l \to \infty$. Borrowing the theory of Markov chains on continuous state space \cite{Durrett2019}, it can be shown that both this recurrence relation and the original feedforward configurations are equivalent to an aperiodic and recurrent Harris chain, which has a unique stationary distribution describing the steady state regime. This stationary distribution is the 4-tuple $[f_\mathcal{E}, f_A, f_k, f_\Theta]$ of the ``eigenfunctions'' invariant under iteration of the map composed of sum, truncation, and rescaling operations ($k$ also plays the role of eigenvalue, determining the scaling factor between successive levels). Assuming that $k_t$ converges for large $t$, then $g_t = k_t^{-2}$ also holds from conservation (Eq.\ \ref{eq:conserv}), with $g_t = \frac{1}{N} \sum_{t=1}^{N} \mathbb{1}_{\tilde{a}_t > \theta_t}$ the discrete-time version of Eq.\ \ref{eq:glint}.

Conventional linear stability analysis is here infeasible because both recurrent and feedforward configuration equations are non-differentiable (in fact discontinuous), and stochastic input precludes the existence of fixed points, except in terms of averages (Section \ref{sec:FTA}). The stationary density that emerges from the interplay of stochastic input and dynamical rules has no deterministic counterpart: without input, the system would remain forever quiescent. This is similar to how the persistence of noise-induced memory in excitable media \cite{Chialvo2000} is contingent on a continuous input of exogenous stochastic fluctuations.

Like the local stretching and global folding of orbits typical of deterministic nonlinear (chaotic) maps (e.g. Baker\'s map \cite{Strogatz2000}), the recurrent configuration map displays stretching (through the convolution operation) and folding or contraction (truncation and resetting) while preserving its density (measure-preserving), but with the difference that noise is injected by randomly sampling the pulse density. Noise is an indispensable ingredient to PC: the recurrence relation is effectively acting upon densities, rather than on ``point-states''. Note however that in general for a system, whether chaotic deterministic or stochastic and whether discrete or continuous, to display intermittent behavior it suffices to have an unstable invariant manifold that encompasses an attractor \cite{Geisel1984,Platt1993,Ding1995}; in PC, this is achieved via iterated diffusion (stretching) and resetting.

\begin{figure*}
\includegraphics[width=\textwidth]{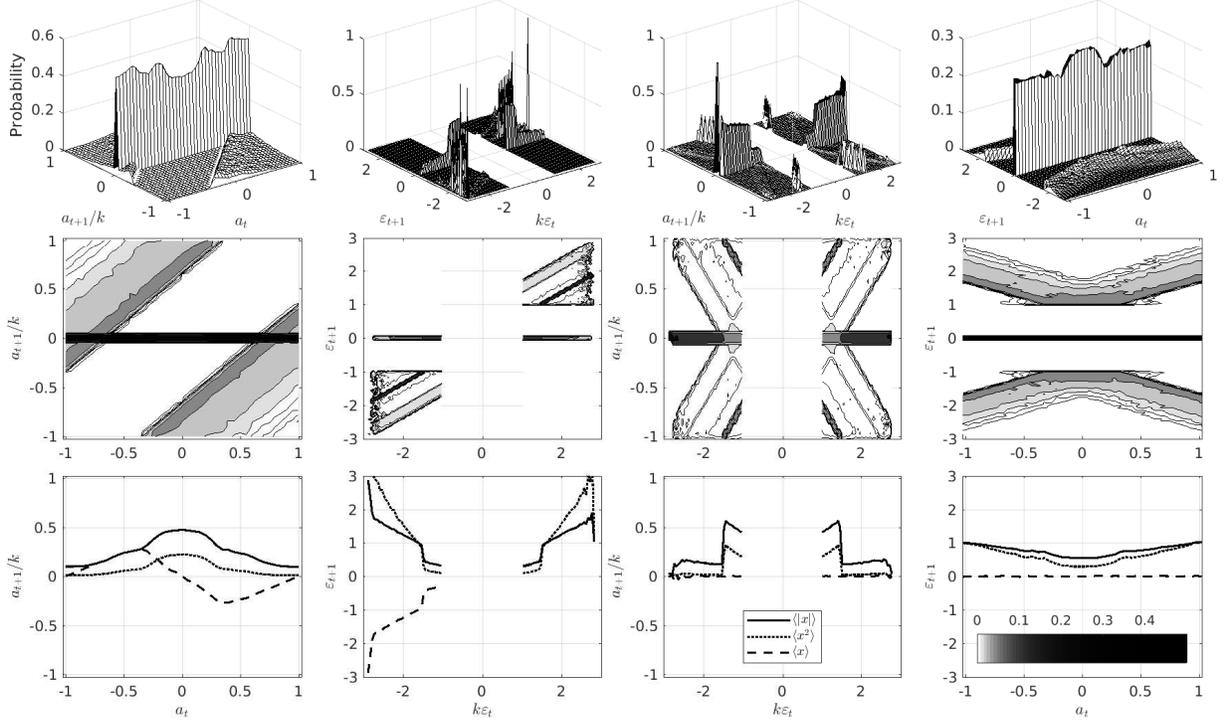}
\caption{Surface (top) and contour (middle) plots of the transition functions $P(a_{t+1}/k|a_t)$ in left column, $P(\varepsilon_{t+1}|k\varepsilon_t)$ in middle left, $P(a_{t+1}/k | k\varepsilon_t)$ in middle right, and $P(\varepsilon_{t+1}|a_t)$ in right. Some variables have been scaled by $k$ or $1/k$, so that $\langle|\tilde{A}|\rangle = \langle\Theta\rangle=1$, to facilitate comparison. The bottom row shows the conditional expectation of the next state given the current state $\langle x\rangle$, where $x$ is given by the figure column index, and the absolute $\langle |x|\rangle$ and squared $\langle x^2\rangle$ conditional expectations. The color bar in the bottom right plot is the color legend of the middle row; it represents probability. The abrupt fluctuations at the boundaries in the middle left column are artifacts due to binning of samples in low density regions. Computed with the same data as Fig.\ \ref{fig:SFL}. \label{fig:TransMat}}
\end{figure*}

To understand how states evolve locally, we can look at the transition function, which specifies the probability, given the current state, for all possible next states, i.e.\ the conditional density of the next state given the current state. By definition, the system satisfies global balance in the steady state  (the transition rate among all states is such that the distribution of states is stationary). The transition function was estimated numerically. First, consecutive duplets from the time series $a_t$ and $\varepsilon_t$ in recurrent configuration were sampled, and the resulting 2-D domain was split up into square bins of length 0.05, yielding bivariate histograms of the joint densities (Fig.\ \ref{fig:TransMatJoint}). The conditional densities were then calculated by dividing the values of each column by the column total sum; the plots are shown in Fig.\ \ref{fig:TransMat}. Resting for the next state is indicated by $\varepsilon_{t+1}=0$.
In the left column of Fig.\ \ref{fig:TransMat}, the low density in the bottom-right and top-left corners of $P(a_{t+1}|_t)$ confirm the intuition that a state of large subthreshold activity $a_t$ (of either sign) tends to be followed by a small $a_{t+1}$. The left middle column evinces that incoming pulses can only trigger outgoing pulses of the same sign. Outgoing pulse size increases monotonically with respect to both incoming pulse size and current subthreshold activity (middle left and right columns), with $\min{\langle |\varepsilon_{t+1}| \rangle_{|a_t}} \approx 0.54$ at $|a_t|=0$. However, the next state subthreshold activity reaches a maximum for non-boundary values of incoming pulse size ($\max{\langle |a_{t+1}/k|\rangle_{|\varepsilon_t}} \approx 0.57 $ for $|\varepsilon_t| \approx 1.43$, middle right column) and current subthreshold activity ($\max{\langle |a_{t+1}/k|\rangle_{|a_t}} \approx 0.47$ for $|a_t| = 0$, left column). The shape of discharge rate as a function of current subthreshold activity can be gleaned from $P(a_{t+1}=0|a_t)$ in the left column; here also we observe that discharge rate extrema are reached at non-boundary values (e.g., there is a minimum of $P(|a_{t+1}|=0)$ at $|a_t| \approx 0.35$). Note also that any incoming pulse of size larger than $2\theta_t$ (which rescaled equals $2k_*\approx 2.96$ if $\theta_t=1$ as in Fig.\ \ref{fig:TransMat}) would trigger a discharge, regardless of its sign and $a_t$. Such pulse sizes would be counteracted by a expansion of the threshold, so we expect them to lie close to $\max{|\mathcal{E}|}$; this is supported by the estimated densities of $\mathcal{E}$ in Fig.\ \ref{fig:TransMat}.

The transition function represents the probability $P(x_{t+1}=j|x_t=i)$ of moving to state $i$ given current state $i$. Fig.\ \ref{fig:TransMat} also shows that under repeated iteration, most trajectories of $x_t=[a_t,\varepsilon_t]$ would contract or decay to zero. This reflects that, at each level, the dynamics are dissipative and dissipation occurs at the same rate that energy arrives (although the cascade of levels as a whole conserves energy, except perhaps at the top level as a design feature). However, under a constant flow of energy with activity distributed as $P(x_t=i)$, the trajectory of the system is attracted to a stationary distribution defined by the ``eigendensity'' of the joint distribution of incoming energy and excitable element dynamics $P(x_{t+1}=j,x_t=i)=P(x_{t+1}=j|x_t=i)P(x_t=i)$ (Fig.\ \ref{fig:TransMatJoint} in Appendix \ref{sec:RecLoopMcmc}). The eigendensity of the system is precisely the stationary distributions $f_\mathcal{E}$ and $f_A$ (Fig.\@ \ref{fig:SFL}): if we arrayed the four transition matrices in Fig.\ \ref{fig:TransMat} as a the single system transition matrix ($a_t,a_{t+1}/k$ (left bottom) and $k\varepsilon_t,\varepsilon_{t+1}$ (top right) in the main diagonal and $k\varepsilon_t,a_{t+1}/k$  at the bottom right and $a_t,\varepsilon_{t+1}$ at the top left), diagonalizing this single matrix, the unique eigenvalue 1 would have eigendensity $f_A, f_\mathcal{E}$. The asymmetric form of the system transition matrix reflects that detailed balance is not satisfied: in general forward and backward transitions between any two states are not equiprobable.

\subsection{Fixed threshold approximation for the recurrent configuration}
We are chiefly interested in the case where the timescale of threshold and activities are separated, which leads to FTA (Section \ref{sec:FTA}). 
For $w \to 0$, $k$ and $\Theta$ stop being stochastic variables and become fixed to $k_*$ and $\theta_*$, which entails that the discharge probability $g_*$ is also constant. Notice that this also bears on the shape of the stationary densities of $\mathcal{E}$ and $A$. Removing threshold fluctuations reduces Eqs.\ \ref{eq:hc1}--\ref{eq:hc4} to a 2-D system. Further, the state of each unit is now defined by a single variable $a_t$ (as opposed to by $a_t$ and $\theta_t$). In symbols (similarly to Eqs.\ \ref{eq:Bae}, \ref{eq:Tae} but in terms of random variables):
\begin{eqnarray}
\mathcal{E} &=&  A + \mathcal{E}/k_*  \qquad \text{if} \; A + \mathcal{E}/k_* > \theta_*  \label{eq:shc1}\\
A &=& \left\{ \begin{array}{lll}
  0  \qquad\qquad\;\; \text{if}\; A + \mathcal{E}/k_*  > \theta_* \\ A + \mathcal{E}/k_*  \quad \text{else}\;
\end{array} \right. \label{eq:shc2} \\
\theta_* &=& \langle |A + \mathcal{E}/k_*|\rangle \label{eq:shc3}\\
k^{-2}_* &=& g_* = P(|A + \mathcal{E}/k_*| > \theta_*)
 \label{eq:shc4}
\end{eqnarray} 
These equations describe variable trajectories under the rescaled recurrent loop configuration, with FTA. This approximation is useful because it renders the system amenable to analytic approximations, while preserving the threshold-activity separation of timescales. Although we cannot implement a recurrence relation for the fixed threshold approximation as in Eqs.\ \ref{eq:hc1}--\ref{eq:hc4}, we can approximate it by choosing small $w$ (Appendix \ref{sec:RecLoopMcmc}).

\begin{figure}
\includegraphics[width=0.5\textwidth]{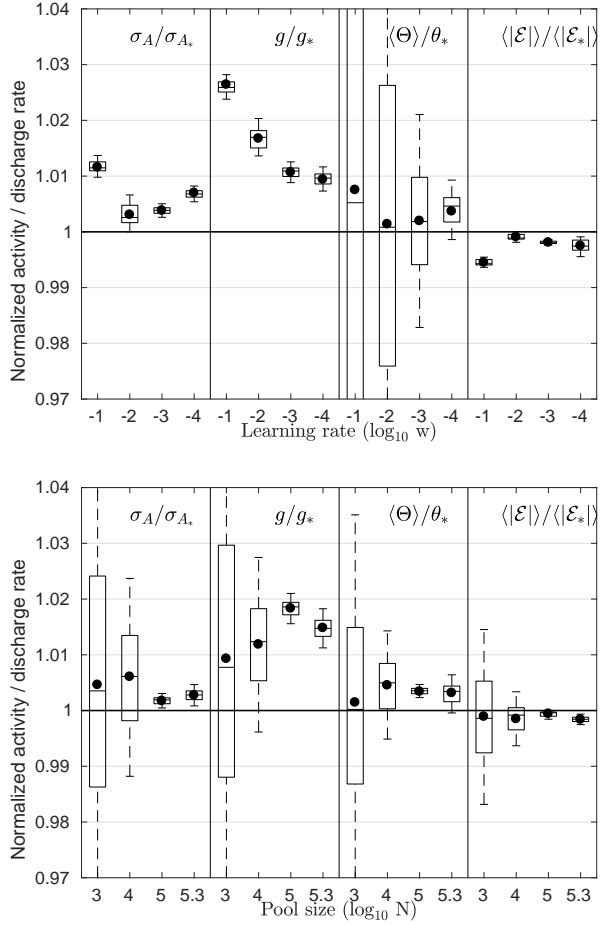} \caption{Numerical simulation for uncorrelated exogenous input in rescaled recurrent configuration. Variable estimates are normalized with respect to the fixed threshold approximation values. Top: estimates calculated with Eqs.\ \ref{eq:hc1}--\ref{eq:hc4} and $N=10^5$. Bottom: same with Eqs.\ \ref{eq:hc1}, \ref{eq:hc2}, \ref{eq:hc3mean}, \ref{eq:hc4}. Box edges: 1st and 4th quartiles; box central segment: median; whiskers: 5th and 95th percentiles; solid circles: mean. \label{fig:ErrorbarGaetResc}}
\end{figure}

Statistics estimates under the rescaled recurrent loop configuration are shown in Fig.\ \ref{fig:ErrorbarGaetResc}. That discharge rates are positively biased with respect to the fixed threshold approximation value, with bias being larger for noisier estimation procedures.
As seen in Fig.\ \ref{fig:SFL} (area under $f_{\tilde{A}}$ to the right of $\theta$) and Eq.\ \ref{eq:glint}, $g$ is a monotonic decreasing nonlinear function of $\theta$. Since the slope of $f_{\tilde{A}}$ is negative for any relevant value of $\tilde{a} = \theta$ (Figs.\ \ref{fig:pdfs}, \ref{fig:SFL}; Appendix \ref{sec:appPulseGpd}), if we assume $f_\Theta$ symmetric about its mean, we can venture that $g \geq g_*$, with equality when $\Theta$ has zero variance fluctuations (fixed threshold). This is because $f_{\tilde{A}}$ is larger to the left of $\theta$ than to the right, so the rate of increase of $g$ due to leftward fluctuations outweighs the rate of decrease due to rightward fluctuations, so the larger is the variance of $f_\Theta$, the larger becomes the expectation of $g$. This explains the positive bias of noisy $g$ estimates (which implies a negative bias for $k$).
Further, threshold fluctuations entail a larger variance of $A$ and $\mathcal{E}$, which also leads to positive biases for $\sigma_A$ and $\sigma_\mathcal{E}$ (Fig.\ \ref{fig:SFL}, where $f_{A^*}$ and $f_\mathcal{E}$ overlap due to the threshold fluctuations); the latter implies a negative bias for $\langle |\mathcal{E}| \rangle$ because $\langle |\mathcal{E}| \rangle = 1 - \sigma_\mathcal{E}$. We do not have simple explanation for $\langle \Theta \rangle$ estimates being positively biased, but we surmise that the hard boundary at the origin ``reflects'' threshold fluctuations.

\subsection{Autocorrelated inputs \label{sec:RRLC_autocorr}}
The PC system can be generalized by allowing autocorrelated or non-zero mean driving input. Assuming the driving input is Gaussian distributed, this can be suitably modeled with fractional Gaussian noise \cite{Mandelbrot1968} with Hurst exponent or parameter $H \in [0,1]$ or with a non-zero mean normal distribution, which corresponds to $H=1$.

Picking up the brief discussion in Section \ref{sec:gausim}, if the exogenous input has strictly positive (or in general non-zero) mean, then for $l \to \infty$ the pulse density $f_\mathcal{E}$ approaches a Gaussian density with mean twice the mean of the preceding unit: $\langle \mathcal{E}_l \rangle = 2\langle \mathcal{E}_{l-1} \rangle$ (note that for positive inputs $\langle \mathcal{E} \rangle = \langle |\mathcal{E}| \rangle$). In this case, $g=0.5$ because firing occurs as soon as exactly two pulses arrive, the first one leading to $A_l = \mathcal{E}_{l-1}$ and its summation with the second one to $\langle \mathcal{E}_{l-1} + \mathcal{E}_{l-1} \rangle = 2\langle \mathcal{E}_{l-1} \rangle = \langle \mathcal{E}_l \rangle$. Consequently, $\langle \Theta_l \rangle = \frac{3}{2}\sigma_{\mathcal{E}_{l-1}} = \frac{3}{4}\sigma_{\mathcal{E}_l}$ holds because $\tilde{A}_l$ is distributed as two lumps with the same probability mass corresponding to the random variables $A_l=\mathcal{E}_{l-1}$ and $\mathcal{E}_l$, which have expected value $\langle \mathcal{E}_{l-1} \rangle$ and $\langle \mathcal{E}_l \rangle$ respectively. 
For finite $l$, we get $\langle|\mathcal{E}|\rangle < \sigma_\mathcal{E} $ and $\sigma_A < \frac{1}{2} \sigma_\mathcal{E}$, which become equalities for $l \to \infty$. 
This is because for each level step-up, average pulse energy scales as $\langle \mathcal{E}_l^2 \rangle = 4\langle \mathcal{E}_{l-1}^2 \rangle$ by conservation (Eq.\ \ref{eq:conserv}), and squared mean pulse size scales as $\langle \mathcal{E}_l \rangle^2 = 2^2\langle \mathcal{E}_{l-1} \rangle^2$ by linearity of expectation. As $l \to \infty$, the contribution to energy due to variance vanishes: $\lim_{l\to\infty}\frac{\var{\mathcal{E}_l}}{\sigma_{\mathcal{E}_l}} = 0$ (idem for $A_l$). This is why the resulting densities are the same as for $H=1$, as we will see soon. These results were verified in numerical simulations (not shown). Note that both mean squared and mean absolute activities are conserved.

For an arbitrary $H \in [0,1]$, Eq.\ \ref{eq:kgBrownian} is not correct anymore. Instead
\begin{equation}
k_l = g_l^{-H} \quad \forall l = 1 \ldots n_l  \label{eq:kgH}
\end{equation}
holds. Eq.\ \ref{eq:kgBrownian} is a particular case of Eq.\ \ref{eq:kgH}, which accommodates autocorrelated Gaussian input with Hurst parameter $H$. Although not uncorrelated, fractional Gaussian noise preserves self-similarity \cite{Mandelbrot1968}.
Energy, as defined so far (squared activity, Section \ref{sec:ConsEn}), is in general not conserved for $H \neq 0.5$ because neither the sum of squared activities nor the mean squared displacement of ``activity particles'' (with respect to time) would be guaranteed to be linear. However, if we redefine energy as a function of activity that is conserved in time, then from Eq.\ \ref{eq:kgH} we can deduce that it would become $k^{1/H} \sim g^{-1}$, where the right hand side has units of time and the left hand side has (equivalent) units of activity raised to the power of $1/H$. Thus, PC with autocorrelated driving input can also be viewed as an ``anomalous cascade'', by analogy with anomalous diffusion: mean squared activity (mean energy) is not necessarily linear in time $\langle \mathcal{E}^2 \rangle \equiv \langle E \rangle \sim t^\alpha$, and the anomalous diffusion coefficient would be related to the Hurst exponent as $\alpha = 2H$. Similarly, first hitting time densities under fractional driving input generalize to $\sim t^{2-H}$ for fractional Brownian motion \cite{Ding1995} but remain exponentially decaying when flanked by two thresholds (Appendix \ref{sec:appb_HT}).

In this spirit, we can restate Eqs.\ \ref{eq:sdE_sdA_T}--\ref{eq:aE_T_aA}, in keeping with the heuristic arguments of Section \ref{sec:FTA}, by substituting $1/2$ for $H$ where suitable. After taking $\sigma_\mathcal{E} \equiv \langle \mathcal{E}^2 \rangle = 1$ for simplicity and without loss of generality, we get
\begin{eqnarray}
g &=& g(g+\sigma_A^2)+\theta_*^{2+H^{-1}} \label{eq:Hc1}, \\
\frac{\sigma_A^2}{\theta_*^2} &=&  g^{2H}(1-g) \label{eq:Hc2}, \\
\theta_*^2 &=& g^{H+1/2} \langle|\mathcal{E}|\rangle \label{eq:Hc3},\\
\frac{1+\theta_*^2-2\theta_*\langle|\mathcal{E}|\rangle}{\frac{\sigma_{A}^2}{1-g}+\theta_*^2-2\theta_*\frac{\langle|A|\rangle}{1-g}} &=& \left(\frac{1-g}{g}\right)^{2H+1}, \label{eq:Hc4}
\end{eqnarray}
which together with $\langle|A|\rangle = g\theta_*\langle|\mathcal{E}|\rangle$ constitutes a nonlinear system of five equations and five variables, solvable via an iterative algorithm. However, both the equations above and Eqs.\ \ref{eq:hc1}--\ref{eq:hc4} rest on FTA assumptions and on heuristics based on energy conservation in (anomalous) diffusion. Eq.\ \ref{eq:aE_T_aA} also makes the implicit assumption that the density of $\tilde{A}$ has a connected support, which breaks down for high enough $H$ (when $f_A$ and $f_\mathcal{E}$ are no longer connected through $\Theta$).
To extend the domain of gain-related variables to autocorrelated input, we substitute $H$ for $1/2$ as above (Eqs.\ \ref{eq:Hc1}--\ref{eq:Hc4}):
\begin{eqnarray}
k_*(H) &=& g_*^{-H}  \label{eq:Hc5} \\
k_{\theta,*}(H) &=& \frac{\theta_*}{g_*^H\langle|\mathcal{E}_*|\rangle} \label{eq:Hc6} \\
k_{\varepsilon,*}(H) &=& k_*(H) / k_{\theta,*}(H). \label{eq:Hc7}
\end{eqnarray}

\begin{figure}
\includegraphics[width=0.5\textwidth]{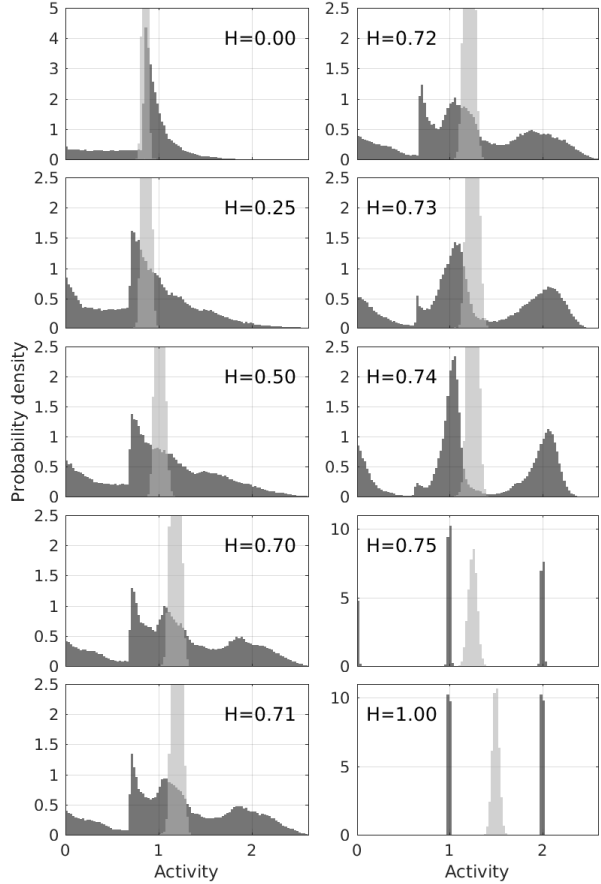}
\caption{Histogram of positive half densities of post-pulse activity $\tilde{A}$ (dark) and threshold $\Theta$ (light) for different degrees of Gaussian input autocorrelation (indexed by the Hurst parameter $H$) in rescaled recurrent loop configuration with explicitly averaged thresholds (Eqs.\ \ref{eq:hc1}, \ref{eq:hc2}, \ref{eq:hc3mean}, \ref{eq:hc4}). Simulation parameters: $5\cdot 10^6$ burn-in iterations; pool size $N=10^5$. As in Fig.\ \ref{fig:SFL}, all plots have been scaled such that activity 1 corresponds to the average incoming pulse size $\langle|\mathcal{E}|/k\rangle$. \label{fig:AncsHists}}
\end{figure}

\begin{figure}
\includegraphics[width=0.5\textwidth]{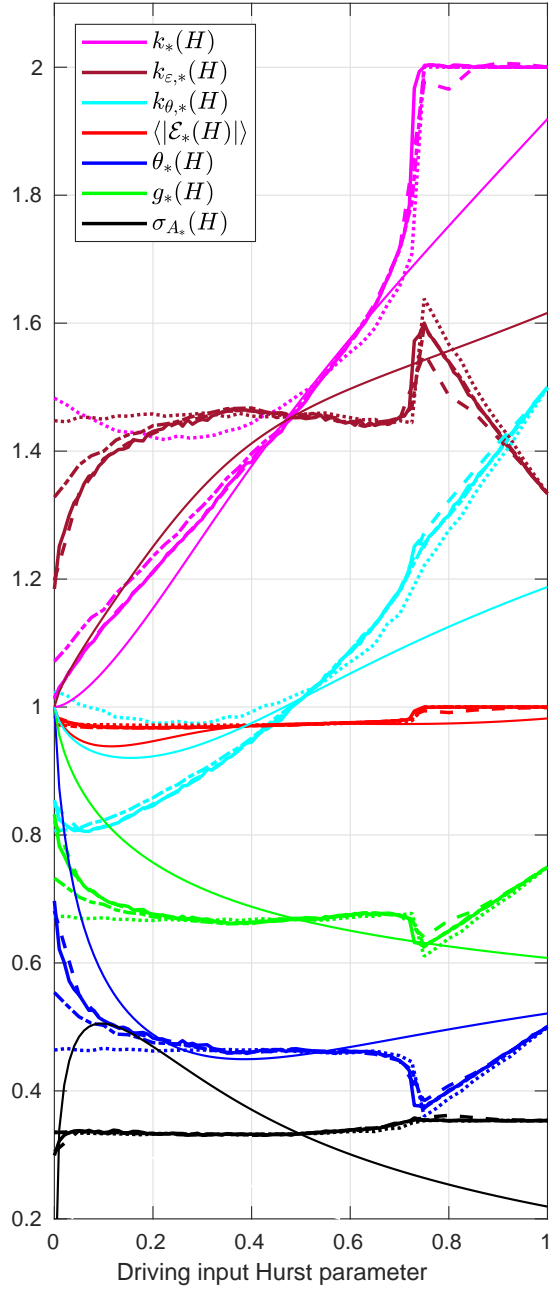}
\caption{Perfusive cascade variables' estimates (see legend) with explicit threshold averaging, as a function of driving input autocorrelation, in recurrent rescaled loop configuration. $A$ and $\mathcal{E}$ are scaled consistent with $\langle \mathcal{E}^2 \rangle = 1$. Line style indicates the estimation procedure: closed form expression (Eqs.\ \ref{eq:Hc1}--\ref{eq:Hc4}) derived through the fixed threshold approximation (solid thin); simulation with pool sizes $N=10^3$ (dotted), $N=10^4$ (dash-dotted), $N=10^5$ (solid thick), $N=2\cdot 10^5$ (dashed). \label{fig:AncsH2}}
\end{figure}

In recurrent configuration, with super-diffusive (anomalous diffusion with $H>0.5$) input, there is a value of $H$ beyond which incoming pulses never reach the threshold. This occurs because incoming pulses are so autocorrelated that any two consecutive pulses are very likely to have the same sign, triggering a discharge and thereby expanding the threshold enough that the incoming pulse maximum is less than the threshold minimum $\max{|\mathcal{E}|/k} < \min{\Theta}$ (where $f_\Theta$ is well approximated by a Gaussian; see Section \ref{sec:appc}). This starts occurring for some $H \in [0.74, 0.75]$ (see Fig.\ \ref{fig:AncsHists}), where a unimodal post-pulse activity density breaks up into three blobs (actually, into five because all densities are symmetrical about the origin). Hence, FTA is invalid beyond this value. Under repeated iteration, these blobs degenerate into increasingly narrow sticks, as described at the beginning of this subsection for $H=1$. This transition is marked by a discontinuity in the derivative of all the PC variables with respect to $H$ (Fig.\ \ref{fig:AncsH2}).

Plotting the map iteration defined by Eqs.\ \ref{eq:hc1}, \ref{eq:hc2}, \ref{eq:hc3mean}, \ref{eq:hc4}, which uses explicit averaging to determine the threshold with autocorrelated Gaussian noise input with parameter $H$, yields variables estimates as a function of $H$ (Fig.\ \ref{fig:AncsH2}). The homologous plot with thresholds calculated through the delta rule (Eqs.\ \ref{eq:hc1}--\ref{eq:hc4}), which are noisier, can be found in Fig.\ \ref{fig:AncsH1} in Appendix \ref{sec:RecLoopMcmc}. These curves display a few remarkable features. First, the sharp concavity (or convexity) with peak at $H \approx 0.75$, surging from $H \approx 0.71$, on the otherwise smooth curves. Second, FTA curves (solid thin in Fig.\ \ref{fig:AncsHists}, using Eqs.\ \ref{eq:Hc1} -- \ref{eq:Hc7}) are a good fit only in the vicinity of $H=1/2$; elsewhere they widely deviate from the numerical estimates. This was expected for $H$ high enough for $f_{\tilde{A}}$ to split up (Fig.\@ \ref{fig:AncsHists}); we surmise that the mismatch occurring elsewhere could be due either to the heuristics used in Eqs.\ \ref{eq:hc1}--\ref{eq:hc4} or to the difficulty of adjusting simulation parameters that expedite fast and reliable convergence to the FTA. This is because there is a trade-off between the variance and convergence speed (bias) of the numerical estimates: increasing larger pool sizes ($N$) or smaller learning rates ($w$) leads to estimates approaching FTA, but this also entails slower convergence rates (see also Appendix \ref{sec:RecLoopMcmc}). In either way, Fig.\ \ref{fig:AncsHists} shows that as expected curves with simulation parameters closest to FTA (largest $N$'s or smallest $w$'s) are also the closest to the closed-form FTA curves. Yet, only $\langle|\mathcal{E}| \rangle$, $g$ and $k$ are a reasonably  good fit in a neighborhood of 1/2. Third, most curves have non-trivial extrema (not at the boundaries of $[0, 1]$ or at 1/2).

\section{A minimal toy model for irreducible uncertainty in the cortical hierarchy \label{sec:toymodel}}

The perfusive cascade model is a chain of excitable elements with a finite refractory period, akin to neurons, that propagate pulses unidirectionally when an internal state variable (subthreshold activity, in analogy to membrane potential) exceeds a range of values defined by another state variable (threshold, in analogy to threshold potential). This arrangement echoes the electrophysiological properties of neurons and a taxonomy of the brain whereby temporal scales define the categories \cite{Kiebel2008}; its key motivation is the operation of good regulators \cite{Conant1970} compelled to trade off metabolic power consumption \footnote{We use the term metabolic energy (or metabolic power, for metabolic energy rate) when referring to the energy required by living organisms to maintain homeostasis, to distinguish it from squared activity.} and representational accuracy in environments comprising multiple spatio-temporal scales.
We envisage the PC as a toy model of a good regulator or living system that implicitly incorporates an internal generative model of its environment \cite{Friston2006} and conforms with the Bayesian brain hypothesis \cite{Dayan1995,Lee2003,Friston2003} ---whereby the brain performs Bayesian inference on its sensory inputs to uncover the hidden causes generating them--- and accomplishes inference through a hierarchical Bayesian scaffolding \cite{Mumford1992,Rao1999,Hochstein2002,Friston2005}. To accomplish inference in real time on a continuous flow of sensory input, a Bayesian filtering scheme ---which allows inverting dynamic non-linear state-space models by assuming that the conditional densities on the system's states and parameters are Gaussian--- can be used. To make inference tractable, the system can be assumed to rely on a mean-field partition of a variational density \cite{Dayan1995,Friston2003} to approximate its recognition model, which computes the posterior density over its internal states given the sensory input. In particular, dynamic expectation maximization (DEM) is a Bayesian filtering scheme that accommodates random fluctuations on sensory inputs and (hierarchical) internal states that is rendered tractable through its use of a mean-field partition of its variational density, and is thus a plausible candidate scheme for general brain computation \cite{Friston2008}. 
In contrast, the PC can be seen as a minimal model that captures salient dynamical properties common to systems that perform real time inference about the state of the environment, while overlooking the details of the inferential machinery. In particular, it emulates the dynamics of surprisal flowing through a hierarchical model.

In the PC system, surprisal denotes irreducible uncertainty, i.e.\ the fraction of uncertainty (information content) that the system cannot resolve by reconfiguring its generative model; instead the system can only resolve or suppress it by exerting action in the environment. We assume that the system can only act on (and thus suppress) uncertainty that reaches the top level. In this setup, pulses $\mathcal{E}$ represent prediction errors relayed to a higher area, thresholds $\Theta$ are excitability or synaptic efficacy, and subthreshold activity $A$ are subthreshold membrane potential. In turn, their associated squares represent surprisal relayed between-levels, surprisal threshold or tolerance, and accumulated surprisal. Importantly, these and their associated variables ---e.g.\ the discharge rates in Table \ref{tab:a1}'s rightmost column--- are not representative of single neurons, but of cortical areas assorted by their rank in a model hierarchy of increasingly larger spatio-temporal scales.

Perfect representational accuracy is infeasible in a stochastic world. In Bayesian inference, only in the large data limit and given some regularity conditions, the posterior density approaches a docile Gaussian around the mode \footnote{Bernstein-Von Mises theorem} \cite{Beal2003}. In practice, each biological species embodies a generative model molded by stochastic selective pressure, that only needs to be good enough to avoid extinction. Although so far these generative models have been good enough to preserve a multiplicity of life forms, they cannot foresee some of the hazardous and rare events the environment is capable of engendering; the effect of past rare events may have faded in the genetic memory, or some of them may simply have never occurred \cite{Kauffman1991,Taleb2010}. These outlier events are part of the complex NESS (heavy-tailed) structure that exerts selective pressure, but have remained hidden. Although these events carry catastrophic effects \cite{Raup1986,McKinney2003} for highly optimized living systems \cite{Carlson1999} and self-organized ecological systems \cite{Friston2006}, in this article we focus in the typical environmental conditions that the generative models have been selected to deal with, which justifies the use of mildly random inputs (Gaussian noise).

\subsection{The trade-off between metabolic power consumption and performance \label{sec:powperto}}

Living creatures are highly optimized systems \cite{Carlson1999,Newman2000a} that delicately balance performance, metabolic energy consumption, and robustness or surprisal tolerance requirements. We assume environmental uncertainty is suppressed fast enough to avoid death, but slow enough to make use of available resources efficiently. This implies maintaining both a good enough model of the environment and a good enough behavior, where good enough is understood in the Darwinian fitness sense. In animal nervous systems, synaptic scaling and firing rate homeostasis \cite{Beggs2019} can be construed as mechanisms to suppress uncertainty. Notably, under the parametric empirical Bayes formulation of the free energy principle \cite{Friston2006} (Section \ref{sec:toymodel_FE}), this trade-off is implicit in the phylogenetic priors that oversee the form of the internal generative model. In vertebrates, the hypothalamus is likely to make up a fraction these priors via its regulation of metabolic processes. Neuronal spikes (discharges) are costly: in humans, possibly fewer than 1\% of neurons on average are active at any given moment \cite{Lennie2003}, which means the typical activation pattern is sparse \cite{Olshausen1996a}. An upper bound estimate, assuming precise timing, for the information encoded by single neurons is $\approx 6$ bit/spike \cite{Stevens1995}. We assume that discharge cost is proportional to the pulse size: $C = \sum_l |\mathcal{E}_l|$, as in rate coding, so PC metabolic power consumption is $P = \sum_l \nu_l|\mathcal{E}_l|$. A reasonable alternative would be that all discharges were equally costly ($P' = \sum_l \nu_l$).

Surprisal increases linearly with time (Section \ref{sec:toymodel_FE}), but not so its square root (absolute activity). This motivates the existence of a finite refractory period: at small timescales, activity fluctuates wildly ---to the extent that it diverges in the limit of size zero fluctuations--- so allowing pulses to be triggered by small fluctuations would be exceedingly wasteful. Indeed, the ratio of pulse size to discharge period at level $l$ is proportional to the square root of discharge frequency: 
\begin{equation}
\frac{\langle |\mathcal{E}_l| \rangle}{T_l} \approx \frac{.97 \sqrt{\langle \mathcal{E}_l^2 \rangle}}{T_l} = \frac{.97 \nu_l^{-1/2}}{T_l} \sim \nu_l^{1/2},
\end{equation}
so metabolic \textit{power} cost grows as a square root law with firing rate, thus diverging for a vanishing refractory period (which entails a zero threshold, see Appendix \ref{sec:appb_HT}). For autocorrelated input with Hurst exponent $H$, this generalizes to $\nu_l^{1-H}$.
In opposition, the metabolic \textit{energy} cost (per spike) decreases with discharge frequency:
\begin{equation}
\langle |\mathcal{E}_l| \rangle \approx \nu_l^{-1/2}.
\end{equation} 
While surprisal accrues linearly, the ebb and flow of sensation-driven activity follows a $1/2$ exponent power law. This suggests that, in view to save metabolic energy, it is expedient to let surprisal accumulate as much as the attendant decrease in performance can be afforded.

The PC implicitly achieves representational accuracy by transmitting pulses to supraordinate levels that correct or suppress the received error signals through the reconfiguration of internal states (not actually modeled). At the sensory level, discharges are implicitly elicited by the discrepancy between the input expected under the system's generative model, and the actual sensory input; at non-sensory levels, this discrepancy is expressed by non-zero incoming pulses. This scheme emulates salient features of predictive coding, an iterative approach to inference related to the EM algorithm, that assumes random effects can be modeled as additive Gaussian fluctuations \cite{Rao1999,Friston2003}. 

Excitability defines the trade-off between metabolic power and representational accuracy, and  is regulated through two ``knobs'': threshold adaptation (Eq.\ \ref{eq:thetal}) and refractory period ($T_0$). The refractory period discretizes the allowed firing times; hence we cannot model our excitable elements as ``Brownian neurons'' to derive hitting times (Appendix \ref{sec:appb}). The refractory period is a physiological limitation in neurons: after a spike, neurons require some time to restore the electrochemical gradient that enables further spiking. But at the same time, a refractory period checks metabolic power consumption, at the expense of limiting perceptual accuracy and reaction time.
Threshold adaptation balances metabolic power consumption and performance: it enables adjusting stored surprisal to the sensory input variance, while keeping a balance between representation update frequency, which is directly related to performance, and consumption, both of which are proportional to discharge rates. Given a fixed driving input and number of levels $n_l$, setting more liberal thresholds entails higher discharge rate and gain (faster reaction time) and higher metabolic power consumption $P \sim \sum_l \nu_l^{1/2}$, but less stored subthreshold energy $\sum_l \langle A_l^2 \rangle$ (higher performance, Section \ref{sec:toymodel_FE}). However the average flow of surprisal or energy remains unchanged in the steady state (Section \ref{sec:toymodel_FE}).

\subsection{Complexity of the brain and its environment \label{sec:toymodel_EarthSurf}}

Any living being (good regulator \cite{Conant1970}), must incorporate a model of the world that is good enough to warrant its persistence. Hence, we can in principle estimate an upper bound for the complexity of this model by assessing the local complexity of the local environment where evolution took place. The relevant measure in this calculation is the minimum average information content \footnote{Although information content and surprisal are usually synonyms to Shannon information, here information content denotes the entropy of the living system's internal model, whereas surprisal denotes the fraction of that entropy attributable to surprisal, corresponding to mismatches between the sensory input and the expected sensory input.} per unit time of the local environment, that bears on Darwinian fitness.
Humans evolved on Earth's surface, an open and dissipative NESS system sustained by a continuous flow of energy. Its relevant information rate is loosely upper bounded by the entropy production rate. This is  because reversible processes and subsystems at equilibrium do not contribute to the generation of complexity: by definition, all microscopic states consistent with equilibrium maximize unpredictability (are equiprobable), which implies that they constitute irreducible noise (that cannot be suppressed via modeling). The Sun's radiant flux or power is (directly or indirectly) the main source of metabolic energy for most known forms of life. For each ecosystem, the energy flux dissipated while traversing it determines its entropy production (together with its temperature) and its complexity. This view resonates with synergetics \cite{Haken1983}, a theoretical framework that explains the formation of self-organized structures far from equilibrium, fed by a gradient of energy. Here, the radiant flux would be the gradient that feeds life, the self-organized structure born in it. 

The following is a back of envelope estimation of the average information content about the environment modeled, and the average surprisal tolerated, by the brain. Much of the complexity of the environment may be implicitly modeled by other organs; developmental processes, physico-chemical properties, and metabolic processes are encoded in the genotype and typically expressed with no involvement of the brain. A reference value for the complexity of the human \textit{organism} generative model $C_G$ (of which the brain is just a fraction) is the information encoded in the genome, which is estimated to be around $6 \text{Gb} \approx 10^{10}$b. This information is combined with the complexity generated by environmental through ontogenetic processes \cite{Kauffman1992} to yield the human organism generative model.
The Earth's surface entropy production is roughly $\sigma_\alpha \approx 0.6 \; \text{W} \text{K}^{-1} \text{m}^{-2}$, where absorption of the Sun's incoming irradiance makes by far the largest contribution \cite{Kleidon2009}. Thus, the information rate needed to encode the spatio-temporal structure of the average Earth's surface $\text{m}^2$-patch is approximately $\sigma_\alpha$.
Before the Neolithic revolution, a typical human foraged in a territory of area $\approx 10^{10} \text{m}^2$ (coarsely, assuming a home range of 100 km \cite{Gamble2013}) over a life span of $\approx 10^9$s. Hence, the human generative model must comprise a dynamical model of at least a fraction of this spatio-temporal ``home'' domain, which after assuming steady state conditions throughout the lifespan yields an entropy production of $\dot{S}_\alpha \approx 10^{10} \text{JK}^{-1}\text{s}^{-1}$ or $\dot{S}_\alpha / k_B \approx 10^{33}$ b/s ($k_B$: Boltzmann constant), which entails $C_{U,1} \equiv S_\alpha / k_B \approx 10^{31}$b within each refractory period of $\approx 10^{-2}$s. This is an upper bound to the generative model complexity $C_{U,1} > C_G$ that accounts for all sources of information that encode the environment of a typical creature on Earth's surface, but it is a gross overestimate because it includes many non-relevant degrees of freedom.
We can hone this bound by looking into the anatomy of the brain: it contains more than $\approx 10^{14}$ synapses and each synapse efficacy holds $\approx 4$b \cite{Bromer2018}, which yields a total storage capacity of $\approx 10^{15}$b. This is a tighter upper bound to the information content of the generative model, which we can approximate to its complexity $C_G \approx C_{U,2} \approx 10^{15}$b. This suggests that human brains only need to model a fraction $C_G / C_{U,1} \approx 10^{-16}$ of the ``home'' environment's information content to compose good enough models.

Looking into the benefits that brains afford enables to roughly estimate the irreducible information rate ---the fraction of uncertainty that cannot be resolved without acting on the environment, and that aggregates random fluctuations across all levels of the structure of the environment. The motivation is that the generative model complexity can be crudely thought of as the complexity of the dynamical trajectory that it induces in phase space \cite{Friston1997a}.
The fluctuation theorem \cite{Evans2002} relates the relative probability of average irreversible entropy $\langle S \rangle_t$ production and destruction over a time interval $t$ as $\frac{P(\langle S \rangle_t = B)}{P(\langle S \rangle_t = -B)} = e^{Bt}$ so it gives a lower bound to the surprisal bitrate $B$ by asking how much time could a living system operate without its brain without significantly jeopardizing its fitness. How long would a brainless human survive in its milieu? One day or $\approx 10^5$ s is a reasonable guess nowadays, but perhaps $\approx 10^4$ s was more pertinent before the Neolithic revolution. Substituting above $t=10^4$ and assuming a small probability ratio of 2 (beyond which the absence of brain is consequential), we get $B \approx 10^{-4}$ b/s, which is a very lax lower bound.
The aggregate afferent sensory bitrate is estimated to be $\approx 10^7$ b/s. Sensory information is compressed downstream the brain hierarchy by maximizing the mutual information between input and increasingly lower-dimensional neural representations \cite{Linsker1990}. Thus the information capacity of higher-order functions such as visual attention \cite{Verghese1992} and cognitive control is $B \approx 50$ b/s (roughly $2\text{--}100$ b/s \cite{Wu2016}). \footnote{Note that the width of the conscious integration kernel (the interval over which events are integrated into a single percept) is $\approx 100\text{--}400$ ms \cite{Efron1970,Burr1980,Herzog2016}, which suggests an information content for each ``conscious snapshot'' of $\approx 10$ b.} Assuming $B$ stands for the surprisal that is preserved throughout the hierarchy (and might account for a significant fraction of the stream of consciousness), the exogenous input surprisal that accrues each refractory period $T_0 = 5$ms (Section \ref{sec:gausim}) is $\sigma_{\mathcal{E}_0}^2 = B T_0 \approx .25$ b, whence the diffusion constant is (Section \ref{sec:ConsEn}) $D=\frac{\sigma_{\mathcal{E}_0}^2}{2T_0} \approx 25$ b/s.

\subsection{Energy, free energy, and surprisal tolerance \label{sec:toymodel_FE}}

Let a system be a good enough model of its environment that its  sensory input error is unbiased Gaussian distributed noise \cite{Friston2006}. If it incorporates a generative model and performs inference about environmental states through a variational density \cite{Friston2003}, then its analogous to the PC energy is denoted free energy in the context of variational Bayesian learning and hierarchical models of the brain \cite{Dayan1995,Friston2003}. The free energy principle states that by (locally) minimizing free energy with respect to the variational parameters (the sufficient statistics of the variational density, encoded in brain states), variational learning prescribes how a living system reacts to stimuli. This can be achieved by extending variational Bayesian inference methods to keep surprisal at check along time trajectories through DEM \cite{Friston2008}, which thereby furnishes a principled simulation of brain dynamics.

In contrast, PC's generative model of the environment is not explicitly modeled. However, its derived error energy or surprisal is locally conserved, relayed downstream the hierarchy, and suppressed at the top level. At each level $l$, surprisal is accumulated as $\tilde{A}_l^2$ until it triggers a discharge at a level set by $\Theta_l^2$ (threshold energy), with pulse and subthreshold densities determined by the hierarchy level (Fig.\ \ref{fig:pdfs}). For $l \to \infty$, they converge to the rescaled recurrent loop configuration density (Fig.\ \ref{fig:SFL}).
Another explanation for the observed scale invariance in PC (Section \ref{sec:toymodel_crit}) ---where squared activity or energy is a conserved quantity--- is that in systems with random noise in a locally conserved quantity, small fluctuations preserve their influence across increasingly larger scales, because fluctuations can die away only by diffusion \cite{Hwa1989,Sethna2001}. Accordingly, in SOC models conservation laws are usually associated with criticality \cite{Manna1990,Dickman1998,Bonachela2009}.

A parametric empirical Bayes model can be replicated as a chain of conditionally independent subsystems where the stochastic terms are independent at each level \cite{Efron1973}. This architecture has been incorporated explicitly as a hierarchical model of the cortex, where its stochastic terms at each level are typically assumed to be Gaussian distributed \cite{Friston2003}. 
In contrast, PC's stochastic terms are Gaussian only at the sensory level. This is not only because its generative model is not explicitly modeled; more essentially, uncertainty suppressed at a given level is not resolved and instead is relayed as a ``pending issue'' to its supraordinate level. Thus aggregate subthreshold activity is a form of ``tolerated'' surprisal stored across the hierarchy as $\sum_l A_l^2$, that measures surprisal tolerance. This is evinced in the high frequency content of subthreshold activity surprisal, which is relatively small and similar for all levels; however low frequencies contain increasingly large quantities of surprisal for higher levels (Fig.\ \ref{fig:psdA}). Conversely, pulse surprisal has a rather flat spectrum for all levels (Fig.\ \ref{fig:psdE}). This is a fundamental feature distinguishing PC from hierarchical Bayesian models of the cortex, and the reason that non-sensory levels have non-Gaussian stochastic terms.

\subsection{Separation of timescales \label{sec:toymodel_septs}}

An essential feature of systems displaying SOC is the timescale separation between slow driving and fast relaxation terms \cite{Vespignani1998}. In our setup, the exogenous input is injected only after all links of the chain to ``decide'' whether to discharge or not, which implicitly defines an infinite separation of timescales between forcing (input) and relaxation (discharges).

There are no free parameters in PC. Timescale ratios between hierarchical levels emerge naturally from the requirement to strike a balance between metabolic power consumption and representational accuracy; threshold and discharge rate are properties derived from the PC defining rules. Unlike the cascade of bifurcations propagated from fast to slow elements in a chain of R\"ossler systems studied by \cite{Fujimoto2003}, which requires strong pairwise correlations and adjusting timescale ratios to a specific range and parameters to the vicinity of bifurcations, PC exhibits long-range temporal correlations without parameter tuning. 

A separation of timescales is inherently incorporated in the physics of the world in the form of characteristic timescales. This can be expounded from the vantage point of synergetics \cite{Haken1983}, which rests on the adiabatic approximation or enslaving principle, whereby systems of coupled elements bearing separated timescales self-organize such that only a few slow and typically unstable order parameters become naturally relevant to explain the global structure, and fast oscillations become enslaved to slow modes. Synergetics has been demonstrated in behavioral dynamics (bimanual coordination) \cite{Jirsa1998}. 
Separation of (time) scales is also illustrated by the manifestation of discrete scale invariance (DSI) in nature. DSI is associated with log-periodic oscillations (cf. \ref{fig:Epdfhist}) and complex critical exponents, and has been studied e.g.\ in hierarchical geometrical systems, and rupture and growth phenomena \cite{Sornette2000a}. There is also some evidence of DSI in human waking electroencephalogram (EEG) large voltage deflections \cite{Zorick2015}, and in event-related changes of spectral power \cite{Jurgens1995}. The center and ranges of behaviorally relevant brain oscillations categories with successively faster frequencies form an arithmetic progression on logarithmic scale \cite{Penttonen2003}. It is plausible that the separation of timescales found in nature, through the evolution of a hierarchical architecture in the brain, resulted in neural variables exhibiting DSI.

\subsection{Diffusion and gating of surprisal in the cortical hierarchy \label{sec:toymodel_ciu}}

Crucially, we have assumed that our system is good enough not only to predict future environmental states with high enough accuracy to persist, but also to make unbiased predictions on average. This will be typically true, as long as the system remains in the environment that exerted selective pressure on it. However, this also means that the system is overly vulnerable to perturbations of exceptional nature \cite{Carlson1999}, i.e.\ those that were absent during its evolutionary history.
Setting aside extreme events, the complex structure of the system's environment must be well enough mimicked by its internal model. Yet not all the entropy of the environment has a recognizable structure that is reducible to a light parametric model: the remaining unexplained noise constitutes sensory error. This is the irreducible uncertainty (surprisal), as opposed to the reducible uncertainty accounted for by the internal model, which we assume to be caused by many independent events that average to a sort of thermal noise, whose time integral is Brownian motion (Appendix \ref{sec:appb}). The existence of an irreducible component to sensory uncertainty that is preserved through the cascade up to the top level is a PC assumption, that is not justified in general.

The main virtue of PC is simplicity: aside from the threshold adaptation rule and number of levels, it has no parameters. This comes at the cost of forgoing the internal configuration of the generative model. 
In DEM, bottom-up error signals modify their (supraordinate state) prior to suppress uncertainty. In turn, the prior constrains the kind of input that the its subordinate level expects. Each state carries Gaussian random fluctuations, parameterized by a precision variable. This scheme is remarkably comprehensive and can be used to model virtually any environmental or neurophysiological phenomenon, from perception to responsiveness to stimuli (attention) and tolerance to uncertainty (e.g.\ by tinkering with DEM's D-step \cite{Friston2008}). Thus, although DEM could be used to regulate the metabolic power-performance trade-off, through the judicious use of metabolic priors or through the regulation of the D-step duration, this would require ad hoc tuning.
In PC, activities represent conserved irreducible uncertainty that reaches the top level through a combination of diffusion, gating, and pulse transfer, so pulses are in general not Gaussian. Since activity behaves as Brownian motion, higher units store increasingly larger quantities of surprisal.

\begin{figure}
\includegraphics[width=.5\textwidth]{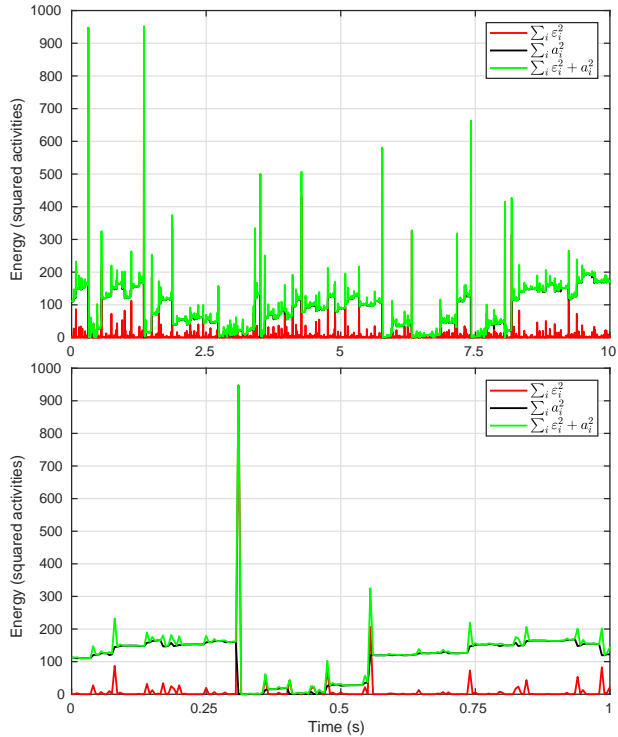}
\caption{Time series of pulse energies, subthreshold activity energies, for each $l=1..7$ level and their sum, in steady state regime. Energy or surprisal is emitted as pulses and stored as subthreshold activity at each level (Section \ref{sec:toymodel_ciu}). Interval length: 2000 (top) and 200 (bottom) time steps. Parameters as in Fig.\ \ref{fig:AEts}. \label{fig:sumsqAEts}}
\end{figure}

\begin{figure}
\includegraphics[width=.5\textwidth]{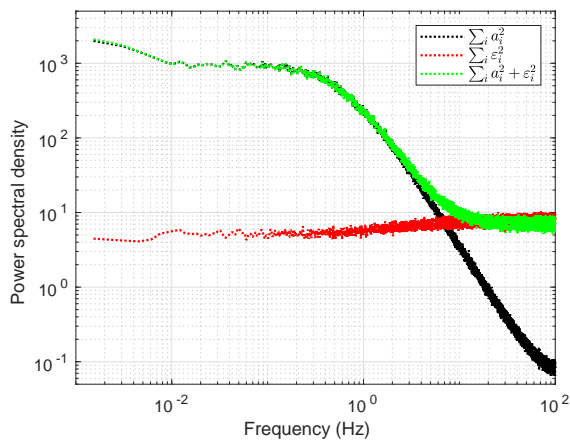}
\caption{Power spectral density estimate of the time series in Fig.\ 
\ref{fig:sumsqAEts}. Estimation parameters as in Fig.\ \ref{fig:psdT}.\label{fig:sumsqAEpsd}}
\end{figure}

$A_l$ undergoes a random walk with step size distribution $f_{\mathcal{E}_l}$, where activity steps occur at multiples of $T_0$. At each level, activity is stochastically reset to some value closer to zero ---depending on the level--- as a function of its distance to zero (Fig. \ref{fig:TransMat}, left)  \cite{Evans2011b}. This is one of many types of reversible Markov chain (random walk) that display power law dynamics \cite{Erland2007}.
Since exogenous input steadily accumulates and hops forward in the hierarchy, the set of all PC variables can be construed as a projection of the one-dimensional exogenous input time series onto the $n_l$-dimensional $a^{(t)}_l$ for $l=1 \ldots n_l$ random walk. However the subsetting of variables, activity MSD or energy is conserved (Section \ref{sec:ConsEn}): it travels from the sensory to the top level, alternating between emitted pulses and stored subthreshold activity energies (Fig.\ \ref{fig:sumsqAEts}), with a power spectrum similar to that of activities (Fig.\ \ref{fig:sumsqAEpsd}).

The extension from one- to $d$-dimensional input is straightforward; the fundamental solution to the diffusion equation (Section \ref{sec:ConsEn}) can be readily generalized \cite{Sornette2000a} to $d$ dimensions as $\text{MSD} = 2dDt$. Thus, the dimensionality of the state space is manifested as just a linear expansion factor in activity MSD.
This is important because it enables to expand the PC modeling scope to comprise high dimensional input time series, by only including a factor $d$ for energies (or $\sqrt{d}$ for activities) and letting the bilateral threshold generalize to a $d$-dimensional ball. Thus, surprisal remains a linear function of time, but if the multidimensional activity space is partitioned into submanifolds ---corresponding to its subelements, just as cortical columns constitute one brain area--- then the surprisal of each subspace $i$ will be weighted by its corresponding dimensionality $d_i$ such that $d = \sum_i d_i$. 

Although PC does not model the internal states that simulate the world (under the DEM scheme, this is the conditional expectations of states of the world \cite{Friston2008}), we can envisage what the dynamics of its internal states would have been like if the generative model had been explicitly modeled. 
Each level $l$ emulates an analogous level in the hierarchy of spatio-temporal scales that characterizes its environment, where $l$ indicates the cortical region rank in sensorifugal order. Levels are conceptually closer to cortical regions than to individual neurons, so the discharge rates of Table \ref{tab:01} indicate transmission of activity between sets of brain regions functioning at contiguous timescales. 
Pulse activity would be a measure of the prediction error or mismatch between the sensory input actually received and the sensory input predicted by the internal state configuration at each level. Likewise, pulse energy would reflect surprisal transferred to the supraordinate level. 
The system would comprise a continuously actualized representation of the environment, coupled to exogenous dynamics, and its state would be encoded in the spatial pattern of activation of its excitable elements. The system would try to minimize uncertainty about the environment ---match predictions with inputs--- by dynamically reconfiguring its connectivity within and between levels (which in brains involves a subtle balance of excitation and inhibition \cite{VanVreeswijk1996,Poil2012}). Hence, its dynamics would be largely defined by within-level \textit{spatial} patterns of activation, which would not be detectable by their amplitude, but only by the activation topography. 
In summary, as in PC, pulses would display a mostly flat temporal spectrum (Fig.\ \ref{fig:psdE}), whereas subthreshold activities would exhibit larger temporal correlations with a timescale congruent with the associated level (Fig.\ \ref{fig:psdA}), but in addition both would display a rich spatial structure.

What happens at the top level? By design, in PC surprisal is quenched only at the top level. This follows from the definition of surprisal: since it can only be suppressed by the top level, it is conserved throughout the hierarchy. 
One inspiration for this feature is the purported existence of surprisal that cannot be resolved through perceptual inferential processes, but instead requires acting on the environment to suppress surprisal. This would recruit top or near-top level mechanisms that elaborate a mapping between surprisal or expectation-input errors to actions (the inverse model \cite{Friston2010}). Such mechanisms typically require a model of the self-other distinction, which entails conscious processing \cite{Blakemore2002,Friston2018,Martinez-Saito2021}.

What is the best rule for setting thresholds? The threshold delta rule of Eq.\ \ref{eq:thetal} is equivalent to stochastic gradient descent on a surprisal quadratic loss (squared error) function \cite{Friston2003} with learning rate $w$. Because the derivative of surprisal has activity units, the minimum mean squared error (MSE) of $\tilde{A}$ is its mean. This is analogous to how the local derivative of free energy is proportional to prediction errors weighted by precision \cite{Friston2003}. 
However, there are other plausible adaptation rules. Minimum mean absolute error (MAE), perhaps the most relevant, converges to the the median  through a judiciously small learning rate $w$:
\begin{equation}
\theta_l^{(t+1)} = 
\left\{ \begin{array}{ll}
\theta_l^{(t)} + w & \text{if} \; |\tilde{a}_l^{(t)}|<\theta_l^{(t)} \\
\theta_l^{(t)} + w & \text{if} \; |\tilde{a}_l^{(t)}| \geq \theta_l^{(t)}. \\
\end{array} \right. \label{eq:thetal_median}
\end{equation}
Note this ignores the magnitude of the prediction error; instead it only acts on the sign of the prediction error by a fixed amount $w$. 

Is there an optimal threshold setting rule? MSE- and MAE-based rules are different ways to arbitrate the trade-off between metabolic energy and performance. Error or surprisal cost proportional to the error magnitude motivates MSE. MAE assumes that all non-zero errors are equally costly. Although so far we have only used MSE, another plausible rule would be: thresholds contract minimizing MSE, and expand minimizing MAE. The rationale is that a firing shortage should be corrected (by contracting) in proportion to its associated error size (thus MSE) because error size indicates susceptibility to fatal outcomes, whereas correcting a firing excess (by expanding) independently of error size (thus MAE) can be motivated by metabolic power saving (if metabolic power $\sim \sum_l \nu_l$, Section \ref{sec:powperto}).

\subsection{A hierarchy of timescales in the brain \label{sec:toymodel_hie}}

Activity power spectra (Fig.\ \ref{fig:psdA}) exhibit a $\sim \nu^{-2}$ scaling range, and pulse avalanches (Fig.\ \ref{fig:Epdfhist}) are also scale invariant across up to three orders of magnitude, for $n_l=7$. The corner frequencies of activity spectra (Figs.\ \ref{fig:psdE}, \ref{fig:psdA}) lie in the midrange $10 \approx 100$Hz and are associated with thresholds, which in turn are related to the timescale of the corresponding level (cf. Eq.\ \ref{eq:conserv}). Conversely, pulses can be approximated as homogeneous Poisson processes (discharge times are memoryless or independent of each other) which entails a flat spectrum \cite{Lombardi2014} (cf. Fig.\ \ref{fig:psdE}). This is because the random time at which a continuous-time stochastic process hits some target (hitting time) has a density that depends on the form of the process; in particular, the hitting time density for Brownian motion in a flanking threshold configuration (as in our system) has an exponentially decaying tail (Appendix \ref{sec:appb_HT}).

Linear superposition of these power spectra \cite{Srinivasan1996} might underpin the power law spectrum of EEG traces and LFPs of mammalian cortex \cite{Freeman2000}. Electrophysiological and modeling evidence indicate that LFPs may display Lorentzian function corner frequencies ---characterized by e.g.\ postsynaptic current and dendritic current efflux decay times--- that split the frequency range \cite{Bedard2006} into a flat ($\nu<20$Hz), $\sim \nu^{-2}$ scaling exponent ($20\text{Hz}<\nu<75$Hz), and $\nu^{-4}$ scaling exponent ($75\text{Hz}<\nu$) bands \cite{Miller2009}. Temporal integration in dendrites or soma or exponentially decaying membrane currents are some known neural processes that can be modeled with the same Ornstein-Uhlenbeck process that describes activity and pulse dynamics (Appendix \ref{sec:appc}).
This contrasts with magneto- and electroencephalographic trace temporal Hurst scaling exponents, which are typically $H \approx .8$ \cite{Novikov1997,Linkenkaer-Hansen2001}, but with a large intersubject variability of $H \approx .55 \text{--} 1.05$ \cite{Smit2011,Poil2012} (corresponding to a power spectrum scaling exponent \cite{Peng1994} $\beta = 2H-1$ of .1--1.1). Their correlations with behavioral scaling exponents are both more prominent in higher brain areas than in sensory areas and more widespread in resting state than during task performance \cite{Palva2013}, consistent with PC dynamics, where bottom-up signals have flat spectra (Fig.\ \ref{fig:psdE}) and higher levels display a wider frequency range of power-law form (Fig.\ \ref{fig:psdA}).

Most hierarchical accounts of brain architecture, like PC, prescribe that sensory input drives the fastest (lowest) areas, whose signals traverse the hierarchy sequentially activating areas with increasingly slower rhythms; likewise, slower rhythms modulate via top-down effects larger and faster subordinate domains \cite{Engel2001}. This is supported by much evidence. Higher frequency oscillations are confined to a small regions, whereas slow oscillations recruit large networks \cite{Buzsaki2004}. Beta oscillations are readily observable immediately after evoked gamma oscillations in sensory evoked potential recordings and can synchronize over long conduction delays that gamma rhythms cannot withstand \cite{Kopell2000}. Gamma rhythms display a smaller correlation length than beta rhythms \cite{VonStein1999}, and its coherence decreases with distance to neighboring cells \cite{Varela2001}. Nested frequencies, where the phase of slower rhythms are correlated with the amplitude of faster rhythms \cite{Hahn2006,Gireesh2008,Lombardi2014}, are pervasive across the whole spectrum of brain activity \cite{He2010}.

In PC's stylized architecture, each hierarchy level has associated a spatio-temporal scale, which emulates the brain hierarchy \cite{Yu1996,Hasson2008,Hasson2015}, where the number of levels, seven, is grounded on brain anatomical connectivity \cite{Mesulam1998,Kiebel2008}. The sensory level is a pre-cortical element such as a thalamic nucleus relay ---e.g.\ the lateral geniculate nucleus \cite{Sherman1998} is the main input that projects to the primary visual cortex. Hence, it would be located at the lowest level with a characteristic timescale of $\approx T_0$ or few milliseconds. Levels 2--5 would constitute the sensory-fugal processing streams that transform sensation into cognition, while the sixth and seventh levels contain transmodal gateways that compose a coherent multimodal representation of the state of the world \cite{Mesulam1998}. The hippocampus \cite{Felleman1991, Mesulam1998} and the hypothalamus \cite{Mesulam1998} are likely to be regions sitting at the top level with a timescale comparable to lifespan. In particular, the hypothalamus is the principal coordinator of the homeostatic, autonomic and hormonal aspects of the internal milieu \cite{Mesulam1998}. The characteristic timescales of other brain areas would come between, e.g.\ $\approx 1\text{--}100$ ms for sensory areas, $\approx 10 \text{--} 10^3$ ms for association areas (e.g.\ 0.8 s for LIP \cite{Huk2005}), $\approx 10\text{--}10^3$ ms for primary and premotor areas \cite{Poeppel2008}, $\approx 10$s for superior temporal sulcus and precuneus, $\approx 36$s for temporo-parietal junction and frontal eye field \cite{Hasson2008}, $> 10$s for anterior cingulate cortex and lateral prefrontal cortex, and a very long characteristic timescale of perhaps years for the medial temporal lobe (which contains high level objects such as person identities \cite{QuianQuiroga2005}) and orbitofrontal cortex (OFC; which harbors long-lasting affective values)\cite{Kiebel2008}. Some areas near the top level such as lateral prefrontal and/or posterior parietal areas  \cite{Mesulam1998} are particularly interesting because they play the crucial role of mapping internal states to action; this enables living systems to actively counteract exogenous (fatal) perturbations \cite{Friston2010a}. 

The sensory level operates $\approx 100$ times faster than the top (seventh) level, while carrying $\approx 10^7/10^2$ times its information rate (Section \ref{sec:toymodel_EarthSurf}, \cite{Buzsaki2004}). If the number of neurons in each cortex level were proportional to their information content, then there ought to be $\approx 10^3$ times more neurons in the sensory than in the top level, which gives an incremental number of neurons ratio of $\approx 10^{3/7} \approx 2.7$ between consecutive levels. Applying this quotients to a $10^{11}$-neuron brain yields an estimate of $\approx 6 \cdot 10^7$ neurons in the top level, which matches the order of magnitude of the number of hippocampal neurons \cite{West1990}.
However, this hierarchical layout should be taken with a grain of salt. The hippocampus has also been proposed to be a temporal storage of explicit facts \cite{Buzsaki1996} that are transferred into the neocortex in order to improve the generative model of the brain, and a structure that encodes (temporal) ordinal structure without reference to particular events \cite{Friston2016}, which would place it near, but not at the top. Other areas might that might fit in the top level are those harboring long-term declarative memories (temporal cortex) \cite{Thompson1996a}, affective values (OFC) \cite{Kiebel2008} and in general slowly oscillating areas \cite{Hasson2008,Honey2012,Murray2014}.

\subsection{Criticality in the brain \label{sec:toymodel_crit}}

One of the reasons critical systems are interesting is that they may resemble life. Complex dissipative structures (NESS) typically arise in domains perfused by a steady flow of energy. Conversely, creatures persisting long enough in complex environments must be able to counteract the perturbations stirred by the energy flow, which means they can anticipate and act upon these perturbations \cite{Friston2006}. Life on Earth evolved in a critical world \cite{Chialvo2006} that exhibits power law or fractal spatio-temporal spectra across a wide range of magnitudes \cite{Paczuski1996} (e.g.\ natural scenes \cite{Dong1995b,Ruderman1997}). This suggests that living creatures embody critical internal representations that unfold across a multiplicity of temporal scales. In humans, this is reflected for instance in the temporal frequency perception of natural scenes \cite{Billock2001a} and in brain synchronization metrics \cite{Kitzbichler2009}.

Systems that perform inference about the causes of inputs through a variational free energy formulation (e.g.\ systems incorporating DEM) are by design disposed to destabilize their own dynamics (autovitiation); this is because they seek out regions of low free energy that have a low curvature, which leads to the apparition of slow dynamic modes (critical slowing) \cite{Friston2012a}. This is consonant with research suggesting that the brain lies on average in a slightly subcritical state \cite{Priesemann2014,Gollo2017} (but see \cite{Poil2012}) with occasional incursions into a supercritical state \cite{Freeman1987,Bonachela2010,Friston2012a}. This suggests that the purpose itself that defines perception (reducing the mismatch between environmental and internal states) is what induces criticality in perceptive systems \cite{Shew2015}, similarly to SOC or quasi-SOC systems.
In PC, the exogenous input rate is infinitesimal relative to discharge rates (i.e.\ the next input waits for all chain elements to ``decide'' whether to discharge ot not) and energy is bulk conserved (within the chain). 
This entails that, if we neglect threshold fluctuations (FTA, Section \ref{sec:FTA}), PC is analogous to a SOC model operating near criticality  \cite{Vespignani1998}. The finite number of levels $n_l$ together with the boundary (top level) dissipation induce finite-size scaling. True scale invariance would be achieved in the limit of infinite number of levels \cite{Vespignani1998}. Conversely, allowing exogenous input injection before the cascade ends would render the system supercritical (cf. sandpile model with finite driving rate \cite{Vespignani1998}).
When threshold fluctuations are allowed, the system hovers around the critical point, in a stretched critical phase grounded on threshold adaptability, similarly to how neural models with  dynamic synapses manifest a Griffiths phase or stretched criticality range \cite{Moretti2013,Larremore2014,Levina2014,DeAndradeCosta2015}. Finally, introducing surprisal (energy) bulk dissipation would lead to a pseudo-critical phenomenon called self-organized quasi-criticality \cite{Bonachela2009,Bonachela2010}.

Empirically, the hypothesis that the brain operates near criticality \cite{Chialvo1999,Beggs2008} has also gained much (if not conclusive) support \cite{Beggs2012,Shew2013,Cocchi2017}. Power-law temporal correlations have been found in EEG \cite{Novikov1997,Linkenkaer-Hansen2001} and MEG signals \cite{Shriki2013} across a broad range of frequencies \cite{Miller2009,Scott2014}, in functional neuroimaging data \cite{Tagliazucchi2012}, and real neuronal avalanches (stereotyped population-size firing events) have been repeatedly observed in animals in multielectrode arrays, \textit{in vitro} in acute cortical slices and organotypic cultures \cite{Beggs2003}, and \textit{in vivo} \cite{Gireesh2008,Shew2013}, displaying spatio-temporal and neuronal group size scale invariance \cite{Petermann2009}.

The continuous-input Bak-Tang-Wiesenfeld lattice sandpile model \cite{Bak1987,DeLosRios1999} shares many traits with PC: discrete pulse density peaks (which may be related to DSI, see section \ref{sec:toymodel_septs}), energy conservation, and power law spectra for avalanches and stored energy, and Lorentzian temporal spectra \cite{DeLosRios1999}. When energy is injected from one end and dissipation is added, long range correlations are quenched, yet temporal spectra become $\sim \nu^{-1}$. This results from a linear superposition of the (now) uncorrelated independent lattice elements, which are composed of exponentials with decay times that depend on their location in the lattice. This is like the mechanism that underpins the PC power spectrum, where uncorrelated energy (surprisal) accumulates sequentially at increasingly slow levels, setting up independently excitable elements of different intrinsic timescales (also to McWhorter\'s model of shot noise \cite{McWhorter1957}). Their common key features are separation of timescales (achieved by decorrelating dissipation in \cite{DeLosRios1999} and by hierarchical design in PC) and unidirectional propagation of energy.
In another similar model, input is injected into the center of a square lattice, constituted by elements characterized by threshold firing, refractory period, and synaptic plasticity \cite{DeArcangelis2006}. This model displays power law distributions of avalanches (with exponent $\approx -1.2$) and spectrum ($\approx -.8$) presumably because its built-in plasticity in conjunction with random initial conditions carves a complex landscape of synapses with a near-critical branching parameter; however, it has many parameters. Adjusting the ratio of excitation and inhibition can yield similar results \cite{Poil2012}, which furnishes an explanation for the wide range of temporal scaling exponents found in the literature.

\subsection{Neuronal avalanches \label{sec:toymodel_nav}}

Cortical neurons are likely to be segregated into layers by their duties as predictive coding hardware. Specifically, prediction error (pulse) neurons would be allocated to supragranular and state neurons to infragranular cortical layers \cite{Friston2006,Bastos2012}. Thus, we can relate PC pulses to prediction error spikes in supragranular cortical layers.
Neuronal avalanches presumably originate from activity in superficial (supragranular) layers 2 and 3 that propagates across multiple cortical columns in all regions of the cortex \cite{Stewart2006,Gireesh2008,Fagerholm2016}. This can be taken one more step further, by relating pulses to local field potentials (LFPs) \cite{Destexhe1999,Friston2005} (the electric potentials recorded in extracellular space in brain tissue), which are the signals picked up by micro-electrodes and by EEG ---the latter after filtering and diffusion across head tissues.

Somewhat atypically, here avalanche denotes a volley of pulses triggered within a singe time step, and its size is proportional to the sum of its constituting pulse sizes. The mixture density $f_{\mathcal{E}_{mix}}$ (Section \ref{sec:properties_ava}) then corresponds to recordings with high enough temporal (or spatial, if levels were spatially segregated) resolution to tell apart each individual pulse or avalanche within a single time step ---akin to high resolution multielectrode recordings in electrophysiology. In contrast, the sum density $f_{\Sigma\mathcal{E}}$ results from registering all the pulses or avalanches generated in one time step as their linear superposition, i.e. lumping them together \cite{Srinivasan1999,Lombardi2014,Blythe2017}. The sum density is analogous to the linear superposition of temporally overlapping post-synaptic potentials on micro-electrodes or LFPs on surface electrodes \cite{Srinivasan1996}, where temporal subsampling \cite{Priesemann2014,Blythe2017} or limited spatial resolution may blend separate avalanches into a single time-varying medley. However, note that the sequential nature of avalanches is lost in our system due to time discretization, which merges all pulses within a time step, and the lack of spatial extent within levels. At any rate, recall that in our system pulses (and avalanches) cannot be naively interpreted as actual electrical potentials or spike counts (Section \ref{sec:toymodel_ciu}); for example, the analogous to the branching parameter \cite{Beggs2003} would be the gain $g$, but the lack of a within-level spatial dimension precludes sensible comparison.

\begin{figure}
\includegraphics[width=.5\textwidth]{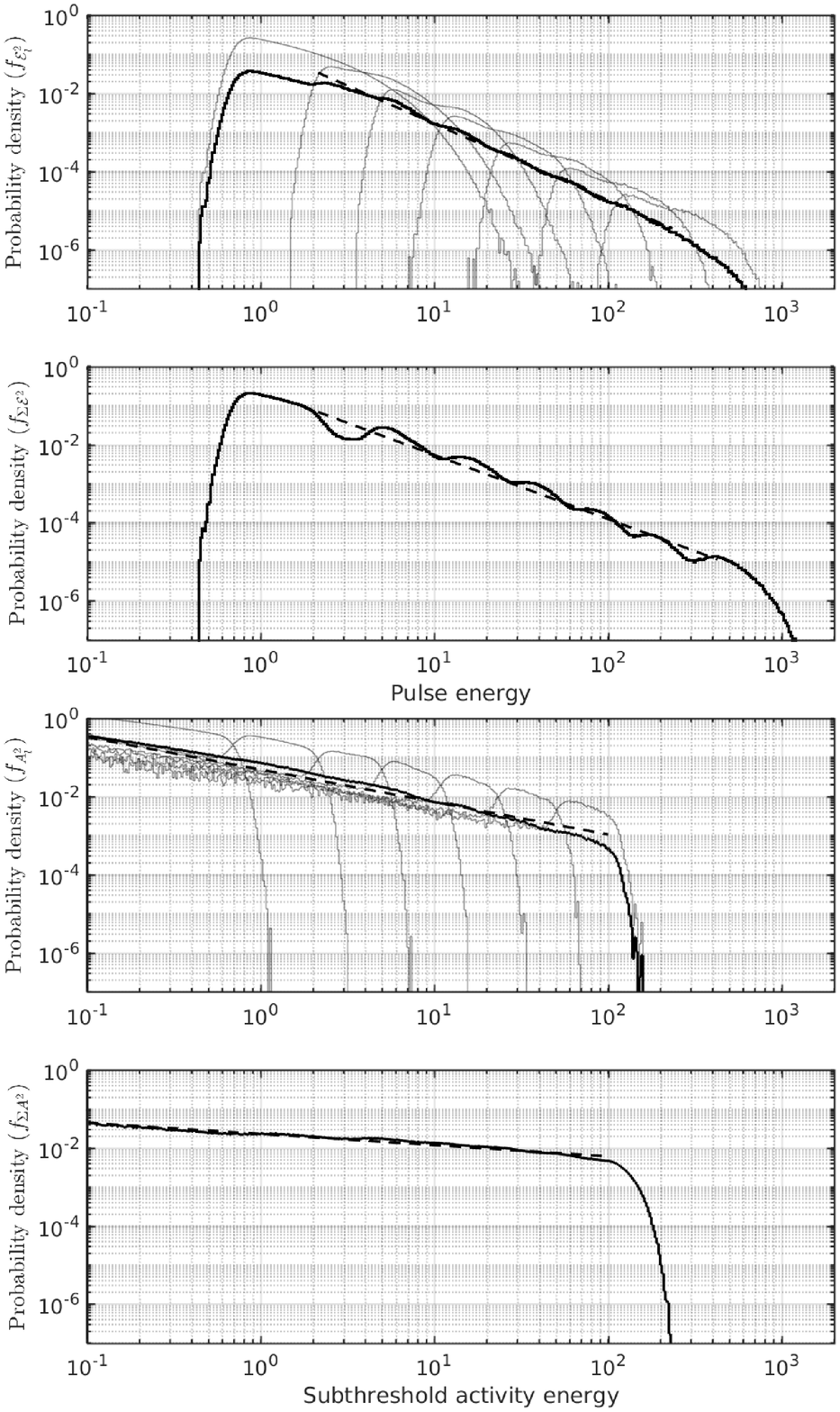}
\caption{Density estimates of pulse and subthreshold activity energies for $n_l=7$ levels. First from above: pulse energy mixture density $f_{\mathcal{E}_{mix}^2}$ (black), and individual levels $f_{\mathcal{E}_l^2}$ (light grey) for all $l=1\ldots 7$. Second: sum of pulse energies $f_{\Sigma \mathcal{E}^2}$. Third: subthreshold activity mixture $f_{\mathcal{A}_{mix}^2}$ and individual level energy $f_{A_l^2}$ densities. Fourth: sum of subthreshold activity energies $f_{\Sigma \mathcal{A}^2}$. Pulse energy slopes (dashed, top -1.95, bottom -1.64) were fitted in the least squares sense in a range that excluded finite-size effects: $[\langle \mathcal{E}_1^2 \rangle,\langle \mathcal{E}_7^2 \rangle]$ for $f_{\mathcal{E}_{mix}^2}$ and $[\langle \mathcal{E}_1^2 \rangle, \sum_{i=1}^7\langle \mathcal{E}_i^2 \rangle]$ for $f_{\Sigma \mathcal{E}^2}$. Subthreshold activity slopes (dashed, top -.83, bottom -.29) were fitted in the range $[10^{-2}, 10^2]$. \label{fig:AE2pdfhist}}
\end{figure}

Empirical estimates from microelectrode arrays suggest that neuronal avalanche size and duration bear scaling exponents 3/2 and 2 respectively \cite{Beggs2003,Stewart2006,Shriki2013} (cf. 3/2 with 1.77 in Fig.\ \ref{fig:Epdfhist}; or with 1.64 in Fig.\ \ref{fig:AE2pdfhist}). The number 3/2 matches the avalanche size scaling exponent in a model of globally coupled (mean-field) threshold elements with random stimulation location \cite{Eurich2002} similar to the Olami-Feder-Christensen earthquake model \cite{Olami1992}, mean-field self-organized branching processes \cite{Zapperi1995}, many flavors of mean-field zero-temperature Ising models \cite{Kuntz2000}, mean-field directed percolation and mean-field dynamical percolation \cite{Munoz1999}. This is notable because the critical behavior of directed percolation belongs to a robust universality class that is widespread in systems exhibiting cascading transitions from active to absorbing phases \cite{Zapperi1995,Munoz1999,Dickman2000,Lubeck2004}. The closely related (but different for low dimensions) Manna class or conserved directed percolation class is particularly relevant to SOC and is associated to conserving dynamics \cite{Lubeck2004,Bonachela2009}. However, the significance of 3/2 is moot because there are similar systems with absorbing-state phase transitions that have different scaling exponents \cite{Munoz1999}, mean-field approximations are typically incorrect for low dimensional systems (e.g.\ 2-dimensional sandpile model \cite{Bak1987,Vespignani1998}, but see \cite{MichielsVanKessenich2016}) and the exponent estimate $\approx 1.5 \textit{--} 2.1$ may depend on the spacing between electrodes in the array \cite{Beggs2008}. Further, avalanches occur in short timescales of 1--100 ms \cite{Plenz2007,Palva2013}, but electrographic long-range temporal correlations timescales are in the range 1--1000 s, so it is plausible that the 3/2 exponent be restricted to spreading of activity within (isofrequency) levels.

The densities of pulse activities and their avalanches (Fig.\ \ref{fig:Epdfhist}) are a priori the variables most related to neural activity, but energy densities allow to visualize the distribution of stored and relayed surprisal across levels (Fig.\  \ref{fig:AE2pdfhist}). Energy is also interesting because it is the only variable conserved in PC, defines discharge rates, and has the physiological interpretation of irreducible uncertainty. The scaling exponents of pulse (-1.95 for mixture, -1.64 for sum density) and subthreshold energy densities (-.83 for mixture, -.29 for sum density) are shown in Fig.\ \ref{fig:AE2pdfhist}. Both variables display power law or DSI behavior within their range, with a right tail roll-off more conspicuous for subthreshold activities caused by the thresholding.

\subsection{Autocorrelated and non-Gaussian inputs \label{sec:toymodel_autcorr}}

For inferential systems that are good enough regulators \cite{Conant1970,Friston2006}, the Hurst exponent of the input stochastic process is $H \approx 0.5$, so $k \approx 1.5$ (Section \ref{sec:RRLC_autocorr}). This reflects that good regulators expect only unbiased, temporally uncorrelated surprisal ---which is why Brownian motion is a central component of standard models of decision making in neuroscience \cite{Gold2007}.
Autocorrelated input may occur when the system is not a good enough model of its environment or input. Then predictions may be biased, which in PC manifests as error signal distribution being biased at all levels. For $H \neq 0.5$, there is a transition at $H \approx 0.75$, beyond which the subthreshold density splits into two disjoint modes (section \ref{sec:RRLC_autocorr}).

The threshold adaptation rule of Eq.\ \ref{eq:thetal} allows PC variables to converge (if there are enough levels) to the rescaled recurrent configuration state (Section \ref{sec:RRLC}; Fig.\ \ref{fig:SFL}) if the input density is not too heavy-tailed: the tail should have a finite first moment (mean). The reason is that iteratively applying Eq.\ \ref{eq:thetal} converges to the mean (Section \ref{sec:toymodel_ciu}). Approximating the tail with a Pareto distribution $\sim x^{-\alpha}$ (Appendix \ref{sec:appPulseGpd}), this is satisfied if $\alpha > 2$. The thresholds diverge if the input density mean is undefined (e.g.\ Cauchy).

\subsection{Subthreshold bistability \label{sec:toymodel_stb}}

\begin{figure}
\includegraphics[width=.5\textwidth]{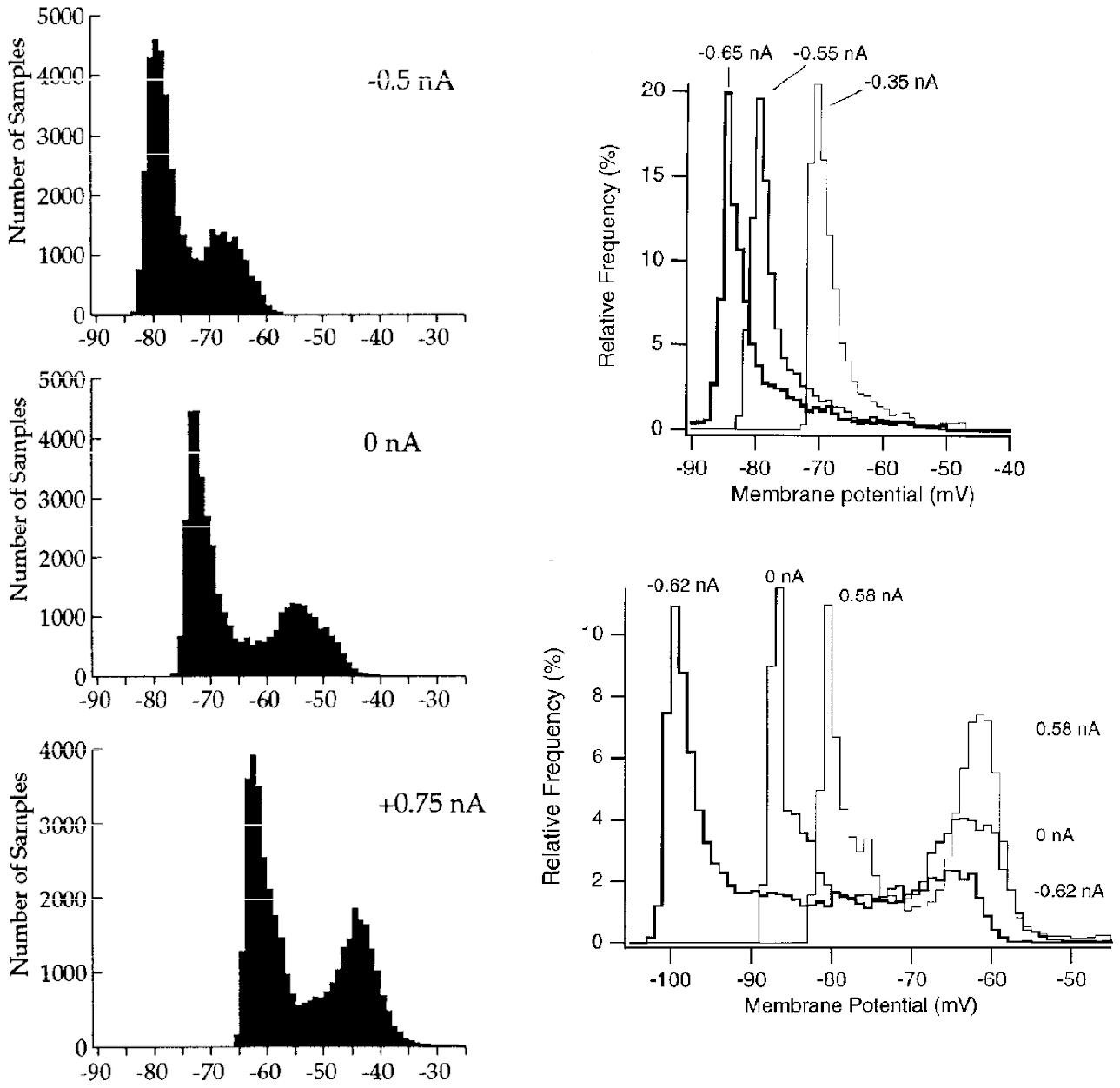}
\caption{Effect of polarizing current on membrane potential density experimental recordings. Compare with Figs.\ \ref{fig:pdfs} and \ref{fig:AncsHists}. Left: \textit{In vivo} intracellular recordings of spontaneous activity of neo-striatal spiny cells displaying Up-Down modes. -0.5 nA denotes a strong hyperpolarizing current, 0 nA no injected current and 0.75 nA a depolarizing current. Reused from \cite{Wilson1996}. Copyright 1996 Society for Neuroscience. Right: Cat deep pyramidal neuron recordings under  barbiturate anesthesia (top) and ketamine-xylazine anesthesia (bottom) under hyperpolarizing and depolarizing currents. Reused with permission from \cite{Pare1998}. \label{fig:WilsonPare}}
\end{figure}

Remarkably, the bimodal PC subthreshold activity densities (Fig.\ \ref{fig:pdfs}, \ref{fig:SFL}, in black, note the probability mass $g$ associated to zero is not shown) are similar to \textit{in vivo} intracellular spontaneous membrane potential shift recordings (anesthetized rat corticostriatal, Fig.\ 5 of \cite{Cowan1994}; neo-striatal, Fig.\ 3 of \cite{Wilson1996}, reproduced in Fig.\ \ref{fig:WilsonPare}, left; cat neocortical pyramidal neurons, Figs.\ 5A, 6A of \cite{Pare1998}, reproduced in Fig.\ \ref{fig:WilsonPare}, right) after equating zero to resting membrane potential and $\langle \Theta \rangle$ to threshold potential. Note that hyperpolarizing currents lead to near-unimodal densities, whereas depolarizing currents lead to bimodal densities (Fig. \ref{fig:WilsonPare}),   in a manner that strongly resembles the PC densities induced by driving input of Hurst parameter $H<0.5$ and $H>0.5$ respectively (Fig.\ \ref{fig:AncsHists}). In agreement with \cite{Mihalas2014}, for PC only the sensory unit lacks a bimodal distribution, and activity densities with positively (negatively) autocorrelated exogenous input (Fig.\ \ref{fig:AncsHists}) resemble membrane potential histograms under depolarizing (hyperpolarizing) currents \cite{Wilson1996,Pare1998}.

The membrane potential of cortical neurons responds to synaptic input in a two-state manner \cite{Cossart2003}. Some neurons toggle between one very hyperpolarized (Down state), and one more depolarized (Up state) membrane potential \cite{Wilson2008a}, where fluctuations around the Up state are of higher amplitude and follow stimulation \cite{Cossart2003}, whereas the Down state is relatively free of noise \cite{Wilson2008a}. Similarly, waiting times between avalanches have a bimodal distribution \cite{Lombardi2014}.
The possible coexistence of both SOC and both continuous and discontinuous phase transitions via a cusp bifurcation has been demonstrated by modeling the electrochemical dynamics of synaptic terminals with a few parameters \cite{Levina2009}, which provides an explanation for Up-Down states. This behavior has been construed as an effect of homeostatic regulation \cite{Lombardi2012}.
Bistability (in alpha rhythms) has also been proposed to ensue from a (fifth-order branch) subcritical bifurcation in corticothalamic activity dynamics, that crucially is noise-driven \cite{Freyer2011}.
The PC bimodal density of subthreshold activity in both feedforward and rescaled recurrent loop configurations provides a simple explanation for Up-Down states as the steady state expression of excitable elements that accumulate synaptic input while adjusting the firing threshold to the incoming pulse sizes, which in turn is a manifestation of the trade-off between metabolic power consumption and performance.

\section{Caveats and prospective research paths}

The bare bones design of PC unavoidably brings about shortfalls. For instance, we only examined one adaptive threshold rule that weights mean squared error, but there are other biologically plausible options (Section \ref{sec:toymodel_ciu}). The simple daisy chain linear topology of PC precludes modeling intra-level effects (e.g.\ lateral connections between neurons or microcolumns in the same level). We likened activity energy or surprisal to irreducible uncertainty, thereby neglecting fluctuations derived from error signals that are successfully suppressed, which are likely to account for a substantial fraction of activity and surprisal in full-blown good regulators. Although simplifications warrant caution in assessing the applicability of modeling results, they may also indicate new appealing research paths. 

Crucially, we left out top-down effects, which are a fundamental aspect of hierarchical Bayesian generative models, where the causal asymmetry between top-down modulatory effects and bottom-up driving effects is pivotal to inference and learning \cite{Dayan1995,Friston2003}. These lead to back and forth volleys between contiguous levels, and the effect of bottom-up signals would be modulated by supraordinate priors at every level. By shifting the reference or zero value of activity, top-down effects by themselves could suppress discharges (by reconfiguring internal states to accurately predict input, e.g.\ eureka moment) or encourage them (e.g.\ under flawed perceptual inference). A top-down signal could set off a surge of activity in subordinate levels, with its size being exponentially related to the rank of the emitting level (Section \ref{sec:toymodel_hie}). 
An expedient course for future work on top-down effects could be to explicitly model top-down signals that suppress uncertainty in subordinate levels and to split the time step into the number of hierarchy levels traversed. Potentially this could inform the presence of local peaks commonly superimposed on M/EEG power spectra \cite{Buzsaki2004} and untangle avalanche clusters due to temporal subsampling \cite{Priesemann2014,Blythe2017} (Section \ref{sec:toymodel_nav}). Finally, allowing PC to accept input before stimulus-evoked cascades end would render it (more) subcritical (Section \ref{sec:toymodel_crit}).

In PC, energy (surprisal) is forced to traverse the whole chain before leaving the system. One interesting accessory feature would be fixing the top level threshold and considering the surprisal quenching at the top level as caused by the action of the system on the environment (recall that top level surprisal cannot be suppressed by supraordinate levels). Then, the PC feedforward configuration would become a metaphor for passive perception under strong priors and occasional stimulus-driven action (akin to mind-wandering). In addition, difficult or surprising situations could be modeled by increasing the input variance, or even its tail shape. 
With only bottom-up signals there are few alternatives, but when top-down signals are included many plausible variations appear. 
For instance, although PC's behavior suggests that the hierarchical coupling of brain levels is essential to its critical behavior, the switching between transient states characteristic of cognitive dynamics \cite{Friston1997a,Friston1997c,Rabinovich2008b} is not explicitly modeled. Allowing to suppress surprisal by acting on the environment (as opposed to only reconfiguring internal states) would provide an explicit mechanism that displays the same sort of itinerant dynamics arising in SOC, synergetics \cite{Haken1983}, and living creatures \cite{Friston2012a}.
The dichotomy between perception and action is a fundamental design principle of living systems that dictates the architecture of hierarchical brain models through their intimate relationship \cite{Friston2010}. However, the implementation of action effectors does not arise from the top level: the timescale of the faster goal-directed motor planning is roughly $.1 \text{--} .5$ Hz for humans, which (not casually) corresponds to the time window of the stream of consciousness \cite{Varela1999,Eagleman2000,Herzog2016}. Hence, the top level associated with action is lower than the top level associated with perception. In motor control and free-energy formulations of behavior, action is computed via an inverse model \cite{Wolpert1998a,Wolpert1998b,Friston2010} that maps sensory prediction errors to actions. Thus, a hierarchy level operating roughly in the theta band, consistent with large theta-cycle avalanches triggering smaller and faster ones \cite{Lombardi2014}, is likely to encompass the top level of the inverse model, which would be akin to a ``sandbox'' used to simulate future courses of action.
A candidate for implementing the highest modules of the inverse model is the default mode network. The resting brain evinces large-amplitude spontaneous $<$0.1 Hz fluctuations in neuromagnetic signals that are temporally correlated across functionally related areas (chiefly medial prefrontal cortex, posterior cingulate cortex, and angular gyrus \cite{Leech2012}) that are posited to constitute a default mode network subserving long term planning, recollection \cite{Biswal1995,Achard2006,Fox2007,Haimovici2013} and mood \cite{Gollo2015}. These notions suggest research pathways for how itinerant dynamics emerge from the exploration of the intertwined action-perception space of possibilities and sensory inputs, and how this relates to the trade-off between metabolic power saving and performance.

The analysis of autocorrelated input (Section \ref{sec:toymodel_autcorr}) hints that multifractal analysis \cite{Mandelbrot1974} could be a useful tool in assessing potential time-varying multiplicative cascading processes underlying dynamics of PC aggregate activity. This is because multifractal analysis can be used to break down a time series of multiplicative interactions that transport information or energy across multiple temporal scales into overlapping contributions derived from a time-varying scaling exponent. Its application to theoretical neuroscience entails ``a local threshold dynamic between adjacent neurons that leads to a dissipation of exogenous or afferent stimuli across multiple spatial scales of the nervous system ... This intermittent spatial dissipation ... defines the sensitivity and adaptability of the cognitive system to exogenous perturbations'' \cite{Ihlen2010}. This would allow examining the degree of incoming pulses autocorrelation ---which are likely to play an important role in the operation of PC-like architectures--- at both level-wise and whole system scales.

\section{Conclusion}

The perfusive cascade is a minimal model of a good regulator \cite{Conant1970} that is wholly stripped of the salient behavior of its generative model, but displays dynamics characteristic of highly optimized systems that model a complex, multi-scale environment. Surprisal transport through its stylized hierarchy is near self-similar \cite{Mandelbrot1974,Ihlen2010}, and leads to scaling laws for all its derived variables. The state units that embody its model of the world are self-dissimilar \cite{Carlson1999} and are not explicitly modeled, although they admit the assumption that activities at all levels reflect the mismatch between incoming pulses (derived from sensory input) and expectations about the sensory input \cite{Friston2003}. 
Its essential features are no adjustable parameters, a universal scaling function in the limit of infinite levels (under weak conditions; Section \ref{sec:toymodel_autcorr}), trade-off between metabolic power consumption and performance, SOC or autovitiation \cite{Friston2012a}, a hierarchy of timescales \cite{Kiebel2008,Hasson2015} where each level represents brain regions characterized by timescale, and bistable subthreshold fluctuations closely resembling intracellular recordings (Section \ref{sec:toymodel_stb}).
The subthreshold activities and pulses have Lorentzian and flat temporal power spectra, respectively; their energy represents surprisal or irreducible uncertainty that is accumulated and relayed across levels. With a biologically plausible refractory period ($T_0=5$ms) and hierarchy depth ($n_l=7$ \cite{Mesulam1998}), the resulting PC scaling range of 0.8--85 Hz roughly matches the physiologically relevant brain rhythms. In contrast to other similar models that predict a wide range of scaling exponents \cite{Poil2012,DeArcangelis2006}, the spectrum scaling exponent ensues from the weighted linear combination of spectra from each level, which can yield any $\beta \in [0,2]$--- as opposed parameter tuning. These weights are proportional to the number of neurons in each level, which decreases exponentially while ascending the hierarchy (Section \ref{sec:toymodel_hie}). Thus, the temporal scaling patterns found in electrophysiological recordings could be explained as weighted linear combinations of subthreshold activities and pulses from different hierarchy levels.
Finally, this scheme suggests some testable conjectures, pertinent to experimental neuroscience: (1) prediction error neurons (presumably in supragranular layers) that exhibit Up-Down states exist all over the brain; (2) power spectra of M/EEG traces can display any scaling exponent in the range $[-2, 0]$ (depending on the task and recorded brain area composition) within the frequency band where circuit level neural dynamics unfold (which corresponds roughly to 0--100 Hz for a refractory period of 5 ms); and 	(3) temporal power spectra scaling exponents are steeper in rest than under stimulation, because (stimuli-driven) error signals have flatter spectra.

\begin{acknowledgments}
The article was supported by the HSE University Basic Research Program.
\end{acknowledgments}

\appendix

\section{\label{sec:appa} Analytic expressions for stationary densities}

Let $w \equiv w^+ = w^-$, the exogenous driving input be $I \sim \mathcal{N}(0,1)$, $f_{A_l}$ be the steady state subthreshold activity density, and post-pulse activity be defined as $\tilde{A}_l = A_l + \mathcal{E}_{l-1}$ (with $\mathcal{E}_0 \equiv I$). Then, for $l=1 \ldots n_l$:
\begin{eqnarray}
f_{\tilde{A}_l} & = & f_{A_l} * f_{\mathcal{E}_{l-1}}, \label{eq:Patl} \\
f_{A_l} & = & \sum^{\infty}_{i=0} f_{A_l^{(i)}} q_l^{(i)} \prod^{i-1}_{j=0} (1-q_l^{(j)}) \label{eq:Pal} \\
& = & f_{A_l^{(0)}} q_l^{(0)} + (1-q_l^{(0)}) (f_{A_l^{(1)}} q_l^{(1)} + \nonumber \\
&& (1-q_l^{(1)})((f_{A_l^{(2)}} q_l^{(2)} + ...)), \nonumber
\end{eqnarray}
where $*$ denotes convolution product \footnote{Alternatively to \ref{eq:Pal}, $f_{A_l}$ could be calculated as $f_{A_l} = B_{\theta_l}[f_{\tilde{A}_l}] q_l^{(0)} + (1-q_l^{(0)}) (B_{\theta_l} [f_{\tilde{A}_l^{(1)}}] q_l^{(1)} + (1-q_l^{(1)})(\ldots))$, with $f_{\tilde{A}_l^{(t)}} = B_{\theta_l}[ f_{\tilde{A}_l}^{(t-1)}] * I$, $f_{\tilde{A}_l^{(0)}} = f_{\mathcal{E}_{l-1}}$, and $f_{\tilde{A}_l^{(0)}} = I$.}. This is reminiscent of a iterative method devised to study irreversible growth processes (Fixed Scale Transformation \cite{Pietronero1988b}). $f_{A_l^{(t)}}$ is the stationary density of $A_l$ at iteration $t$ after the last reset to zero, and $q_l^{(t)} = 1 - \int_{-\theta_l}^{\theta_l} f_{\tilde{A}_l^{(t)}}$. Explicitly, $f_{A_l^{(t)}} = B_{\theta_l}[ f_{A_l^{(t-1)}} * f_{\mathcal{E}_{l-1}}]$; in particular,  $f_{A_l^{(0)}} = \delta(0)$, where $\delta$ is the Dirac delta; $f_{A_l^{(1)}} = B_{\theta_l}[f_{\mathcal{E}_{l-1}}]$, $f_{A_1^{(1)}} = B_{\theta_1}[I]$. For example, for $l=1$:
\begin{eqnarray*}
f_{A_1} & = & \delta(0) q_1^{(0)} + (1-q_1^{(0)})(B_{\theta_1}[I] q_1^{(1)} + \\
&& (1-q_1^{(1)}) (B_{\theta_1} [B_{\theta_1}[I]*I] q_1^{(2)} + \dots)). 
\end{eqnarray*}
$q_l^{(0)}$ is the probability of firing consecutively (twice or more), and in general $q_l^{(t)}$ the probability of firing $t+1$ iterations after the last firing time. In particular, note that $q_l^{(0)} = 1 - \int_{-\theta_l}^{\theta_l} f_{\mathcal{E}_{l-1}}$ would be on average $1/2$ if activities were reset after every iteration (which would lead to $\langle A_l \rangle = \langle \Theta_l \rangle = \langle \mathcal{E}_{l-1} \rangle$).

In the limit $t \to \infty$, the probability $P(\mathcal{E}_l \neq 0) = P(A_l=0) = P(|\tilde{A}_l| > \Theta_l )$ of unit $l$ firing at any time $t$ is
\begin{eqnarray}
P(\mathcal{E}_l \neq 0) & = & 1 - \sum^{\infty}_{i=0} q_l^{(i)} \prod^{i}_{j=0} (1-q_l^{(j)}) \label{eq:Pfiring} \\
& = & \sum^{\infty}_{i=0} [q_l^{(i)}]^2 \prod^{i-1}_{j=0} (1-q_l^{(j)}) \nonumber \\ 
& = & [q_l^{(0)}]^2 + (1-q_l^{(0)}) \cdot \nonumber \\
&& ([q_l^{(1)}]^2 + (1-q_l^{(1)})(...)). \nonumber
\end{eqnarray}

A coarse estimation of $\langle \Theta_1 \rangle$ using Eqs.\  \ref{eq:meanthetal}, \ref{eq:Patl} and \ref{eq:Pal} up to first order approximation ---one iteration (till $i=1$), ignoring $O(A_1^{(2)})$ terms and above--- yields $\langle \Theta_1 \rangle \approx 0.8197$, in reasonable approximation to the numerical simulations (Table \ref{tab:a1}). Plugging into Eq.\ \ref{eq:Pfiring} the numerical estimate for $\langle \Theta_1 \rangle$ and truncating the series at $i=3$, we get $f_{\mathcal{E}_1}(\varepsilon_1 \neq 0) \approx 0.4336$, which face-validates Eq.\ \ref{eq:Pfiring}. Table \ref{tab:a1} displays the average activities, thresholds, and pulses or error signals obtained through numerical simulations with $L=7$ units and $n_t = 9 \cdot 10^5$ iterations and exogenous input distributed as white Gaussian noise.

\begin{table}
\caption{Numerical estimates of mean subthreshold activities, thresholds, and pulse absolute values. In theory $P(A_l = 0) = P(|A_l + \mathcal{E}_{l-1}| > \Theta_l \;|\; \mathcal{E}_{l-1} \neq 0) = P(\mathcal{E}_l \neq 0 \;|\; \mathcal{E}_{l-1} \neq 0) \equiv g_l : \forall l>1$ should hold because, given a uncorrelated exogenous input, $\mathcal{E}_l^{(t)}$ and $\mathcal{E}_l^{(t+1)}$ must also be uncorrelated. The discrepancies are due to sampling error. Parameters as in Table \ref{tab:01}. \label{tab:a1}}
\begin{ruledtabular}
\begin{tabular}{clllc}
Level $l$ & $ \langle |A_l| \rangle$ & $\langle \Theta_l \rangle$ & $\langle |\mathcal{E}_l| \rangle$  &  $P(A_l = 0)$ \\
\hline
1 & 0.2309(2) & 0.8464      & 1.4479(8)    & .4251(5) \\
2 & 0.4782(5) & 1.4769      & 2.2214(14)   & .4501(5) \\
3 & 0.7408(9) & 2.2470      & 3.2759(29)   & .4605(5) \\
4 & 1.0686(13) & 3.3212(2)  & 4.8099(60)   & .4696(5)  \\
5 & 1.5937(19) & 4.8576(2)  & 7.0698(131)  & .4620(5) \\
6 & 2.2914(28) & 7.1114(4)  & 10.3020(282) & .4707(5) \\
7 & 3.4148(42) & 10.4729(6) & 15.2187(606) & .4648(5)
\end{tabular}
\end{ruledtabular}
\end{table}

Solving for $f_{\Theta_l}$ would be laborious ---it requires specifying $w$ to solve simultaneously the interlocked pair of Eqs.\ \ref{eq:Pal} and \ref{eq:thetal}. However, by the central limit theorem, in the steady state  $f_{\Theta_l}$ converges in distribution to a normal function, for $l=1 \ldots n_l$. A larger $w$ induces flatter $f_{\Theta_l}$ and $f_{A_l}$ densities. By Eq.\ \ref{eq:thetal}, the expectation of the threshold equals the expectation of post-pulse activity $\langle \Theta_l \rangle = \langle |\tilde{A}_l| \rangle$. Thus, if only the expectation $\langle \Theta_l \rangle$ is sought, the calculation of joint densities can be eschewed by calculating $\langle \tilde{A}_l\rangle$ or equivalently by finding the value that on average matches the signs of threshold fluctuations, which numerically can be estimated by minimizing the functional
\begin{equation}
\langle \Theta_l \rangle = \langle |\tilde{A}_l| \rangle \approx \arg \min_\theta \left| \langle \tilde{A}_l - \theta \rangle_{\tilde{A}_l > \theta} - \langle \theta - \tilde{A}_l \rangle_{\tilde{A}_l < \theta} \right|, \label{eq:meanthetal}
\end{equation}
where $\langle \tilde{A}_l - \theta \rangle_{\tilde{A}_l > \theta} = \int_{\theta}^{\infty} (x - \theta) f_{\tilde{A}_l}(x)dx$, and $\langle \theta - \tilde{A}_l \rangle_{\tilde{A}_l < \theta} = \int_{0}^{\theta} (\theta - x) f_{\tilde{A}_l}(x)dx$ and $f_{\tilde{A}_l}$ is approximated with its numerical distribution. This approximation becomes exact in the limit $w \to 0$.

\section{\label{sec:appb} Brownian neurons: firing behavior at fine-grained timescales}

We have assumed that integrator units discharge only at discrete time steps. This trait mirrors refractoriness, or the property of being unresponsive to stimuli for a period of time (refractory period). Neurons are excitable elements with refractory period, but it is also interesting to study the case of non-refractory elements or ``Brownian neurons'', which are continuously responsive and may take in continuous input. These elements are a kind of stochastic model of neuron membrane voltage, similar to Stein's or Ornstein-Uhlenbeck neuron models \cite{Stein1965, Ditlevsen2005}. Here we examine excitable behavior as the refractory period $T_0$ approaches zero, which is the scaling limit where a random walk converges to the Wiener process. With the adaptive threshold rule of Eq. \ref{eq:thetal}, a refractory period $T_0=0$ under a continuous-time driving process entails a zero threshold, so we will fix the threshold to a positive value. 

\subsection{Firing rate}
First we derive bounds and approximations for the discharge rates of the sensory unit and for units located far enough from the input. The firing rate of the sensory unit is the probability that it fires at any time $P(A_1=0)=P(\mathcal{E}_1 \neq 0)$, given by Eq.\ \ref{eq:Pfiring}. As mentioned in Appendix \ref{sec:appa}, the probability of discharging at iteration $t$ is the exceedance probability $q_1^{(t)} = 1 - \int_{-\theta_1}^{\theta_1} f_{\tilde{A}_1^{(t)}}$. Since $f_{\tilde{A}_1}$ falls exponentially-like, $\theta_1$ is guaranteed to lie not far from the median of $f_{\tilde{A}_1}$, so the firing rate is guaranteed to remain close to $1/2$, which would not occur if $f_{\tilde{A}_1}$ decayed as a power law \cite{Sornette2004}. Because of diffusion, this probability would be lower (higher) if the activity $\tilde{A}_1$ were distributed as white Gaussian noise $\mathcal{N}(0,1) = W^{(1)}$ (Wiener process $W^{(t)}$). A tighter upper bound can be achieved by approximating $\tilde{A}_1^{(t)}$ with the sum of Gaussian input with the activity attained in the limit of infinitely many iterations without firing, namely $\tilde{A}_1^{(t)} \sim \mathcal{N}(0,1) + \tilde{A}_1^{(\infty)}$, where $\tilde{A}_1^{(\infty)} \sim \mathcal{U}_{[-\theta_1,\theta_1]}$ in $[-\theta,\theta]$ as a result of iteratively applying convolution and truncation. For fixed $t$, this random variable has lower variance than Brownian motion, but higher than $\tilde{A}_1$. In particular, Eq.\ \ref{eq:Pfiring} yields for the sensory unit $P_{\mathcal{N}(0,1)}(\mathcal{E}_1 \neq 0) = .3973 < P_{\tilde{A}_1^{(t)}}(\mathcal{E}_1 \neq 0) = .4251 < P_{\tilde{A}_1^{(\infty)}+\mathcal{N}(0,1)}(\mathcal{E}_1 \neq 0) = .4608 < P_{B^{(t)}}(\mathcal{E}_1 \neq 0) = .5157$.   

\subsection{Hitting time density \label{sec:appb_HT}}
The hitting time is the random time when a stochastic process first reaches some target subset of the codomain. For Brownian motion on the real line starting at zero, the hitting time for the interval $[\theta, \infty)$ follows \cite{Taylor1998} a L\'evy distribution with scale parameter $\theta^2$, as $\tau_{\theta_1} = \frac{\theta}{\sqrt{2\pi}} e^{-\theta^2/(2t)} t^{-3/2}$. Its heavy tail falls off as $\sim t^{-3/2}$, and its mean is undefined. For fractional Brownian motion with Hurst parameter $H$, it can be shown that it follows $\sim t^{2-H}$ \cite{Ding1995}. However, the problem is essentially changed if the target subset is defined by two flanking thresholds as $\mathbb{R} \setminus [-\theta, \theta]$. Then $\tau_\theta = \min{\{t \geq 0 \, : \, W^{(t)} = \theta \}}$, which is the same as the first exit time from $[-\theta, \theta]$. If we assume no reset, the first exit time of $A_1$ from $[-\theta_1,\theta_1]$ (or first hit time of its complement interval) would be the same as the hitting time of $\theta_1$ for reflected Brownian motion $|W|$ (RBM). The expected value of RBM is $\langle |W^{(t)}| \rangle = \sqrt{2t/\pi}$ (from the Wiener process transition density function with origin at zero $f_{W^{(t)}}(x) = \frac{1}{\sqrt{2 \pi t}} e^{-\frac{x^2}{2t}}$ \cite{Taylor1998}), so the threshold $\langle |W^{(t_1)}| \rangle = \theta_1$ should be attained at time $t_1=\frac{\pi}{2}\theta_1^2$. Thus, if $\tilde{A}_1 \sim |W|$, the threshold $\theta_1 = \sqrt{2/\pi}$ would be on average reached at $t_1=1$. However, the actual values are $t_1 = 1.125$ and $\theta_1 = .8464$ because leftover activity from quiescent trials and the threshold updating rule (Eq.\ \ref{eq:Tae}) push the threshold outwards.

The first exit time of $\tilde{A}_1$ can be calculated as the weighted sum of each time step by its firing probability. Simulations yield $\langle \tau_{\pm \theta_1} \rangle = 1.655$, but the expected hitting time or residence time for RBM is 
\begin{equation}
\langle \tau_{\pm \theta} \rangle = \theta^2,  \label{eq:rbmhte}
\end{equation}
which can be derived from the knowledge that $(W^{(t)})^2-t$ is a martingale \cite{Durrett2019}. $\langle \tau_{\pm \theta_1} \rangle =.7164$ if $\theta_1 = .8464$. This discrepancy (a factor of $\approx 2.31$) is due to the discretization of allowed firing times imposed by the existence of a refractory period $T_0= 1 > \langle \tau_{\pm \theta_1} \rangle$. The discrepancy  is roughly the fraction of energy ``skipping the threshold during the refractory period'', which coincides with the inverse of the gain $g_1^{-1} \approx 2.35$ (see Table \ref{tab:01} to look up $g_1$).

We could stop here, but in pursuance of getting a comprehensive picture of the dynamics of integrator units in the limit of non-refractoriness, let us derive the RBM hitting time density $f_{\tau_{\pm \theta}}$. The reflection principle \cite{Durrett2019}, which stands on the observation that the distribution of future paths is symmetric at any given time, relates the running maximum of a Wiener process up to time $t$ to its tail distribution at time $t$ \footnote{Notice this implies that for fixed $t$, the distribution of RBM coincides with the running maximum distribution for Brownian motion $f_{|B^{(t)}|}(x) = f_{\max_{0 \leq u \leq t} W^{(u)}}(x)$}. In turn, the running maximum is equal to the cumulative distribution of the hitting time $F_{\tau_{\pm \theta}}(t)=P(\tau_{\pm \theta} \leq t)$ \cite{Taylor1998}, so
\[
P(\tau_{\pm\theta} \leq t) = P (\max_{0 \leq u \leq t} W^{(u)} \geq \theta) = 2 P(W^{(t)} > \theta). \label{eq:RBMmeanHT}
\]
This can also be applied to calculate the hitting time of RBM, but some caution is called for. The idea is applying the reflection principle iteratively to the thresholds $-\theta$ and $\theta$. The  RBM hitting time is the first exit time from the $[-\theta,\theta]$ stripe: $\tau_{\pm \theta} = \min{\{t\, : \, W^{(t)} \notin [-\theta,\theta] \}}$. As in Brownian motion, when a particle first reaches the threshold $\theta$, it carries on outwards or reverts with the same probability (by symmetry, the same reasoning applies to particles attaining first $-\theta$). Hence, $P(\max_{0 \leq u \leq t} |B^{(u)}| \geq \theta) = 2 P(|W^{(t)}| > \theta)$ as long as paths do not travel further than $2\theta$ from the first hit, i.e., if $|B^{(t)}| < 3\theta$. Beyond this point, some of the reverting paths will attain the opposite threshold $-\theta$. To exclude the contribution from virtual paths that hit thresholds more than once, we have to remove the path densities located in $[(4k-1)\theta,(4k+1)\theta]$ for $k \in \mathbb{Z}$ ---these are the odd-numbered interleaved stripes, resulting from iteratively reflecting the $[-\theta,\theta]$ stripe over the thresholds \footnote{This is a common procedure in finance for pricing double barrier options}. This ensures we count only paths that hit at least one of the thresholds, and not paths hitting multiple times. To sum up, 
\begin{eqnarray}
F_{\tau_{\pm \theta}}(t) &=& P(\tau_{\pm\theta} \leq t) = P(\max_{0 \leq u \leq t} |W^{(u)}| \geq \theta) \nonumber  \\ 
&=& \int_{-\theta}^{\theta} P(|W^{(t)}|>\theta) dy \nonumber  \\
&=& \int_{-\theta}^{\theta} \sum_{k \in \mathbb{Z}} 2 f_{W^{(t)}}(y +(4k+2)\theta)dy \nonumber \\
&=& \sum_{k \in \mathbb{Z}} 2 \int_{-\theta}^{\theta} \frac{1}{\sqrt{2\pi t}} e^{-\frac{(y+(4k+2)\theta)^2}{2t}}dy \nonumber \\
&=& \sqrt{\frac{2}{\pi t}} \sum_{k \in \mathbb{Z}} \int_{(4k+1)\theta}^{(4k+3)\theta}  e^{-\frac{x^2}{2t}}dx \label{eq:Ftau} \\
&=& \sum_{k \in \mathbb{Z}} \erf(\frac{(4k+3)\theta}{\sqrt{2t}}) - \erf(\frac{(4k+1)\theta}{\sqrt{2t}}), \nonumber
\end{eqnarray}
where we performed the change of variables $x = y + (4k+2)\theta$, and the error function $\erf$ takes $\frac{x}{\sqrt{2t}}$ as variable. This is equivalent to the generalized reflection principle formula \cite{Faulhaber02}: $P(\max_{0 \leq u \leq t} |W^{(u)}| \leq \theta) = \sum_{k \in \mathbb{Z}} f_{W^{(t)}}(y +4k\theta) - f_{W^{(t)}}(y +(4k+2)\theta) = 1 - P(\max_{0 \leq u \leq t}|W^{(u)}|>\theta)$. Finally, differentiating $P(\tau_{\pm\theta} \leq t)$ returns the hitting time density:
\begin{eqnarray}
f_{\tau_{\pm\theta}}(t) &=& \frac{d}{dt} P(\tau_{\pm\theta} \leq t) \nonumber \\
& = & \sum_{k \in \mathbb{Z}} \frac{d}{dt} (\erf(\frac{(4k+3)\theta}{\sqrt{2t}}) - \erf(\frac{(4k+1)\theta}{\sqrt{2t}})) \nonumber \\
& = & \frac{\theta}{\sqrt{2 \pi t^3}} \sum_{k \in \mathbb{Z}} (4k+1)e^{-\frac{\theta^2(4k+1)^2}{2t}} - (4k-1)e^{-\frac{\theta^2(4k-1)^2}{2t}} \nonumber \\
& = & \frac{\theta}{\sqrt{2 \pi t^3}} \sum_{k \in \mathbb{Z}} (-1)^k (2k+1)e^{-\frac{\theta^2(2k+1)^2}{2t}} \nonumber \\
& = & \frac{\theta\sqrt{2}}{\sqrt{\pi t^3}} \sum_{k=0}^\infty (-1)^k (2k+1)e^{-\frac{\theta^2(2k+1)^2}{2t}} \label{eq:rbmtau1}
\end{eqnarray}

Alternatively to Eq.\ \ref{eq:rbmtau1}, $f_{\tau_{\pm\theta}}$ can be calculated from the moment-generating function of $f_{W^{(t)}}$, which can be derived from the properties of martingales. $e^{\sigma W^{(t)} - \frac{\sigma^2}{2} t}$ is a martingale \cite{Durrett2019}, so for a Wiener process starting at the origin, it holds that $\langle e^{0}\rangle = \langle \exp{(\sigma W^{(\min(\tau_{+\theta}, \tau_{-\theta}))} - \frac{\sigma^2}{2}\min(\tau_{+\theta}, \tau_{-\theta}}))\rangle$, so $1 = \frac{1}{2}\langle \exp{(\sigma \theta - \frac{\sigma^2}{2}\tau_{\pm\theta})} + \exp{(-\sigma \theta - \frac{\sigma^2}{2}\tau_{\pm\theta})} \rangle$, and rearranging $\langle e^{-\frac{\sigma^2}{2} \tau_{\pm\theta}} \rangle \frac{e^{\sigma \theta} + e^{-\sigma \theta}}{2}= 1$. After the change of variables $\sigma = \sqrt{2\theta}$, we get an expression for the Laplace transform of the hitting time density \cite{Durrett2019}:
\[
\langle e^{-s \, \tau_{\pm\theta}} \rangle = \sech{\theta\sqrt{2s}}.
\]
Its inverse Laplace transform $\mathcal{L}^{-1}\{\sech{\theta\sqrt{2s}}\}(t) = \frac{1}{2 \pi i} \lim_{R \to \infty} \int_{c - iR}^{c + iR} \frac{e^{st} ds}{\cosh{\theta\sqrt{2s}}}$ is the hitting time density $f_{\tau_{\pm\theta}}(t)$, with $c \in \mathbb{R}$ such that the real part of any of the singularities of the integrand is less than $c$. Cauchy's residue theorem allows to compute the line integral of an analytic function $g$ over a closed rectifiable curve $\Gamma$ in terms of the function's residues \cite{Markushevich1950}. Let $g(s) = \frac{e^{st}}{\cosh{\theta\sqrt{2s}}}$ be the integrand, and $\Gamma = C \cup \gamma$, where $\gamma = [c -iR, c +iR]$ is the line integral we wish to calculate, and $C= c+Re^{i\varphi} : \varphi \in [\frac{\pi}{2},\frac{3\pi}{2}]$ is the arc of the semicircle with center $c$ and radius $R$. Then, the residue theorem states that $\oint_{\Gamma} g = \int_{\gamma} g + \int_{C} g = 2\pi i \sum_k \Res{\left(g, s_k \right)}$, where $s_k$ are the poles of $g$ located inside $\Gamma$. Hence, letting the radius $R$ of the semicircle $\Gamma$ grow to infinity,
\begin{equation}
\frac{1}{2 \pi i} \lim_{R \to \infty} \int_{c - iR}^{c + iR} g =  \sum_k \Res{\left(g, s_k \right)} -\frac{1}{2 \pi i} \lim_{R \to \infty} \int_C g. \label{eq:CRT}
\end{equation}
Since $e^{st}$ is an entire function, the poles $s_k$ of $g$ are the zeroes of $\cosh{\theta\sqrt{2s}}$, which are defined by $\cosh{\theta \sqrt{2s_k}} = 0$; this yields the distinct poles $s_k = -\frac{\pi^2}{8 \theta^2}(2k+1)^2$, with $k \in \mathbb{N}_0$; seeing that all poles satisfy $s_k < 0$, we can take $c=0$. We can evaluate the second term of the r.h.s.\ of Eq.\ \ref{eq:CRT} through an approach akin to that of Jordan's lemma \cite{Markushevich1950}, i.e.\ by finding an upper bound to the contour integral modulus 
\begin{eqnarray}
\left| \int_C g \right| &\leq& \int_C |g| = \int_{\frac{\pi}{2}}^{\frac{3\pi}{2}} \left| \frac{e^{tR(\cos{\varphi}+i\sin{\varphi})} Rie^{i\varphi}}{\cosh{(\theta\sqrt{2R}e^{i\varphi/2}})}  \right| d\varphi \nonumber \\ 
&=& R \int_{\frac{\pi}{2}}^{\frac{3\pi}{2}} \frac{e^{tR\cos{\varphi}}}{|\cosh{(\theta\sqrt{2R}e^{i\varphi/2}})|} d\varphi \nonumber \\
& \leq &  2R \int_{\frac{\pi}{2}}^{\frac{3\pi}{2}} e^{tR\cos{\varphi}} d\varphi, \nonumber
\end{eqnarray}
which has limit zero as $R \to \infty$ because $\cos{\varphi} \leq 0$ for $\varphi \in [\frac{\pi}{2},\frac{3\pi}{2}]$. The last step requires calculating $\max_{\varphi \in [\frac{\pi}{2},\frac{3\pi}{2}]} |\cosh{(\theta\sqrt{2R}e^{i\varphi/2}})|^{-1} = \max_{\varphi \in [\frac{\pi}{2},\frac{3\pi}{2}]} 2|\cosh{(2\theta\sqrt{2R}\cos{\frac{\varphi}{2}})} + \cos{(2\theta\sqrt{2R}\cos{\frac{\varphi}{2}})}|^{-2} = 2$, which is achieved for $\varphi=\pi$. Consequently, 
\[
f_{\tau_{\pm\theta}} =  \frac{1}{2 \pi i} \lim_{R \to \infty} \int_{-iR}^{iR}g =  \sum_k \Res{\left(g, s_k \right)}.
\]
Since $\frac{d}{ds}\cosh{\theta \sqrt{2s}} \rvert_{s=s_k} = \frac{\theta}{\sqrt{2s_k}} \sinh{\theta \sqrt{2s_k}} = \frac{\theta^2}{\frac{\pi}{2} (2k+1)} \sin{\frac{\pi}{2}(2k + 1)} \neq 0$, all poles ($\forall k \in \mathbb{N}^0$) are simple and isolated. Hence, $\frac{e^{st}}{\cosh{\theta\sqrt{2s}}}$ is a meromorphic function (except at infinity), and it can be expanded (by Mittag-Leffler's theorem) around each of its poles $s_k$ as a sum of partial fractions, where each partial fraction expansion is the principal part of $\sech{\theta\sqrt{2s}}$ at the pole $s_k$ \cite{Markushevich1950}. The residue at pole $s_k$ is thus
\begin{eqnarray}
\Res{\left(g, s_k \right)} &=& \lim_{s \to s_k} (s-s_k)g(s) \nonumber \\
&=& \frac{e^{s_kt}}{\frac{\theta^2}{\frac{\pi}{2} (2k+1)} \sinh{\theta\sqrt{2s_k}}} \nonumber \\
&=& \frac{\frac{\pi}{2}(2k+1)}{\theta^2}  \frac{e^{-\frac{\pi^2}{8\theta^2}(2k+1)^2t}}{\sin{\frac{\pi}{2}(2k + 1)}} \nonumber \\
&=& \frac{\pi}{2\theta^2}(2k+1)(-1)^k e^{-\frac{\pi^2}{8\theta^2}(2k+1)^2t}. \nonumber
\end{eqnarray}
Finally, by the residue theorem
\begin{eqnarray}
f_{\tau_{\pm\theta}}(t) &=& \frac{\pi}{2\theta^2} \sum_{k=0}^\infty (-1)^k(2k+1) e^{-\frac{\pi^2}{8\theta^2}(2k+1)^2t}. \label{eq:rbmtau2}
\end{eqnarray}
Numerical simulations show that Eq.\ \ref{eq:rbmtau1} converges much faster than Eq.\ \ref{eq:rbmtau2} for $t \ll 1$, and vice versa for $t \gg 1$. We can use Eq.\ \ref{eq:rbmtau2} to calculate the expected hitting time $\tau_W$ of a Brownian neuron, with threshold $\theta_1$, where firing of sensory units is possible only at specific times separated by $T_0=1$, as $\langle \tau_W \rangle = \sum_{k \in \mathbb{Z}^+} k\frac{f_{\tau_{\pm\theta}}(k)}{\sum_{i \in \mathbb{Z}^+}  f_{\tau_{\pm\theta}}(i)} = 1.21756$, which lies between the values of $\langle \tau_{\theta_1} \rangle$ for RBM and the sensory unit. Importantly, this shows that the hitting time density of Brownian neurons has an exponentially decaying tail.

\subsection{Brownian neurons mimicking refractory units: reduced variance and noisy relaxation processes}
How to tinker with Brownian neurons so they reproduce the firing rates and mean hitting times of discrete (sensory level) neurons? For mean hitting times, perhaps the simplest way is reducing the variance of the Brownian neuron. Reusing Eq.\ \ref{eq:RBMmeanHT} gives for the standard deviation of the Brownian neuron that equates its mean hitting time to that of the sensory unit $\sigma_\text{red} = \theta_1/\sqrt{\langle \tau_1 \rangle} = .6579$. 
For firing rates, we can counterpoise the discrepancy between Brownian and sensory neurons by adding a zero-reverting term to the Brownian neuron transition density, thus rendering it a noisy relaxation or (driftless) ``Ornstein-Uhlenbeck neuron''. This mimics the activity resetting effect of discharges (Eq.\ \ref{eq:a}). The problem now is to find the mean-reverting coefficient $\rho$ of the stochastic differential equation $dx = -\rho x dt + dW$ obeyed by the Ornstein-Uhlenbeck neuron that matches the firing rate of the sensory unit $P_{\tilde{A}_1^{(t)}}(\mathcal{E}_1 \neq 0)$. Numerical optimization of $\rho$ with respect to its transition density \cite{Taylor1998} yields $\rho = .2819$, which compared to $0$ shows the influence of discharge-induced suppression on the dynamics of $\tilde{A}_1$.

\section{\label{sec:appc} Dynamics of thresholds, pulses, and activities}
Threshold dynamics can be described with an AR(1) model. Recalling Eq.\ \ref{eq:thetal}, and assuming $w=w^+=w^-$:
\begin{eqnarray}
\theta^{(t+1)} &=& (1-w) \theta^{(t)} + w |\tilde{a}^{(t)}| \nonumber \\
&=&  w\langle \Theta \rangle + (1-w)\theta^{(t)} + \tilde{\epsilon}^{(t)} \label{eq:ar1}
\end{eqnarray}
where $w\langle \Theta \rangle$ is the average, $1-w$ is the regressor parameter, and $\tilde{\epsilon}^{(t)} = w(|\tilde{a}^{(t)}| - \langle \Theta \rangle)$ is a white noise process with zero mean and variance $\sigma_{\tilde{\epsilon}}^2 = w^2 \var{|\tilde{A}|}$. The $l$ subscripts have been omitted for clarity. For $1-w \approx 1$, this constitutes a wide-sense stationary process and the central limit theorem ensures that (in the steady state) $\Theta$ is approximately Gaussian-distributed, with mean $\langle \Theta \rangle$ and covariance $C_{\theta,\theta}(n) \equiv \cov{[\theta^{(t+n)]},\theta^{(t)}]}= \frac{\sigma_{\tilde{\epsilon}}^2}{1-(1-w)^2}(1-w)^{|n|} = \frac{w^2\var{|\tilde{A}|}}{1-(1-w)^2}(1-w)^{|n|} \approx \frac{\var{|\tilde{A}|}}{2}w(1-w)^{|n|} = \frac{\sigma_{\tilde{\epsilon}}^2}{2w}(1-w)^{|n|}$, whence
\begin{equation}
C_{\theta,\theta}(n) \approx \frac{\sigma_{\tilde{\epsilon}}^2}{2w}(1-w)^{|n|}, \label{eq:Ctt}
\end{equation}
where the approximation is valid for small $w$ \cite{Shumway2006}. $C_{\theta,\theta}(n)$ decays with a time constant $-T_s/\log{(1-w)} \approx T_s/w$ ($T_s = T_l$ sampling period) \footnote{The expected mean $\langle \theta^{(t+n)} | \theta^{(t)} \rangle$ and variance $\var{(\theta^{(t+n)} | \theta^{(t)})}$ after $n$ iterations can be derived explicitly by iterated application of Eq.\ \ref{eq:ar1} $\langle \theta^{(t+n)} | \theta^{(t)} \rangle = \langle \Theta \rangle (1-(1-w)^n)+ \theta^{(t)}(1-w)^n$ and $\var{(\theta^{(t+n)} | \theta^{(t)})} \approx (1-(1-w)^{2n}) \frac{w}{2}\var{|\tilde{a}^{(t)}|}$}. Notice that we set different sampling periods $T_s=T_l= \nu_{l-1}^{-1}$ corresponding to each level $l=1 \ldots 7$ (with $\nu_0^{-1} = T_0$), because thresholds can be updated only upon arrival of pulses, so the firing rate of the preceding unit determines the threshold updating rate at each unit. Since $\tau \gg T_l$ for small $w$, we can calculate the power spectrum through the Fourier transform \footnote{Specifically, the non-unitary Fourier transform, in ordinary frequency units.} of the (continuum approximation) covariance:
\begin{eqnarray}
S_{\theta,\theta} (\nu | l) &=& \frac{\sigma_{\tilde{\epsilon}}^2}{1-(1-w)^2}\frac{2\tau^{-1}}{\tau^{-2}+(2\pi\nu)^2} \nonumber \\
&\approx& \frac{w^2 \var{|\tilde{A}|}}{2w}\frac{2w/T_l}{w^2/T_l^2+(2\pi\nu)^2} \nonumber  \\
&=& \var{|\tilde{A}|} \frac{T_l}{1+\left( \frac{T_l}{w}2\pi\nu \right)^2}, \label{eq:Stt1}
\end{eqnarray}
which is a Lorentzian function, with maximum $S_{\theta,\theta}(0) = T_l\var{|\tilde{A}|}$ at $\nu=0$ and corner frequency $\gamma_\theta = \frac{w}{2\pi T_l}$. The corner frequencies for each unit firing period are shown in Table \ref{tab:c1}. Note that the time resolution a each level $T_s = T_l$ decreases with the level number, reflecting the slower dynamics of later units.
 
The continuous-time analogue of the AR(1) process of Eq.\ \ref{eq:ar1} is a Ornstein-Uhlenbeck process (OUP) with mean $\langle \Theta_l \rangle$, mean-reversal term $-w$, and noise variance $\sigma_{\tilde{\epsilon}}$. This can be deduced by rearranging Eq.\ \ref{eq:ar1} as $\theta_l^{(t)} = \theta_l^{(t-1)} -w(\theta_l^{(t-1)} - \langle \Theta_l \rangle)  + \tilde{\epsilon}^{(t)}$ and taking the limit as the step size approaches zero:
\begin{equation}
d\theta = -w(\theta-\langle \Theta \rangle)dt + \sigma_{\tilde{\epsilon}}dW. \label{eq:Toup1}
\end{equation}
Donsker's invariance principle guarantees that the random walk series of $\tilde{\epsilon}^{(t)}$ increments $\lim_{n \to \infty}\frac{1}{\sqrt{n}} \sum_{i=1}^{nt} \tilde{\epsilon}^{(i)}$ converges in distribution to $\sigma_{\tilde{\epsilon}_l}W(t)$ for $t \in [0,1]$, where $W(t)$ is the standard Wiener process. This corresponds to the scaling limit $T_s \to 0$ of the discrete AR(1) process in Eq.\ \ref{eq:ar1}. When $T_s$ is not negligible, Eq.\ \ref{eq:Toup1} is no longer a good approximation and it has to be corrected with additional factors that depend on the step size. In the steady state, the OUP covariance function \cite{Taylor1998} is $\cov{[\theta^{(t+u)},\theta^{(t)}]} = \frac{\var{|\tilde{A}|}}{2}we^{-wu} = \frac{\sigma_{\tilde{\epsilon}}^2}{2w}e^{-wu}$, which for $w \to 0$ approaches the covariance function of the discrete AR(1) process (Eq.\ \ref{eq:Ctt}). Numerical simulations with $w=0.01$ yielded $\sigma_{|\tilde{A}|} \equiv \sqrt{\var{|\tilde{A}|}} \approx 0.53 \langle \Theta_l \rangle$ for $l \geq 3$ (Table \ref{tab:c1}) \footnote{cf. variance of pulses $\sigma_{\mathcal{E}} = 0.35 \langle \Theta_l \rangle$.}. This gives the following reasonable continuum approximation to the threshold dynamics for $l \geq 3$:
\begin{equation}
\frac{1}{w}d\theta \approx -(\theta - \langle \Theta \rangle) dt + 0.53 \langle \Theta \rangle dW. \label{eq:Toup2}
\end{equation}
The solid curves in Fig.\ \ref{fig:psdT} are derived from Eq.\ \ref{eq:Toup2}, except the noise variances from Table \ref{tab:c1} were used instead of 0.53.

\begin{table}
\caption{Numerical estimates of threshold noise variance components $\sigma_{|\tilde{A}_l|}$ normalized by the average threshold, and corner frequencies $\gamma_{\theta,l}$ and $\gamma_{a,l}$ (time series of length $n_t = 9 \cdot 10^6$).\label{tab:c1}}
\begin{ruledtabular}
\begin{tabular}{cllll}
Level $l$ & $\frac{\sigma_{|\tilde{A}_l|}}{\langle \Theta_l \rangle}$ & $\frac{\sigma_{|\mathcal{E}_l|}}{\langle |\mathcal{E}_l| \rangle}$ & $\gamma_\theta(l)$ (Hz) & $\gamma_a(l)$ (Hz) \\  \hline
1 & .7557 & .3493 & .318 & 59.2 \\
2 & .5664 & .2576 & .135 & 19.5 \\
3 & .5393 & .2491 & .061 & 8.41 \\
4 & .5301 & .2421 & .028 & 3.85 \\
5 & .5395 & .2419 & .013 & 1.80 \\
6 & .5253 & .2425 & .006 & 0.82 \\
7 & .5452 & .2411 & .003 & 0.38
\end{tabular}
\end{ruledtabular}
\end{table}

Computing the power spectral density (PSD) using the discrete Fourier transform on windowed finite time series entails unavoidably an estimation bias. This is because using a finite-length signal is equivalent to multiplying  the original infinite-length signal by a window function, and the frequency response of a finite window (Fej\'{e}r's kernel) is not just a Dirac delta, but a distribution with a finite-sized mainlobe and many tailing off sidelobes that spread out to cover the whole frequency range \cite{Percival1994}. As a result, the estimate of a windowed signal has power leaked into all frequencies around any true signal frequency. This effect is called spectral leakage, and it affects the periodogram and its derivatives such as Welch's method, typically biasing PSD estimates upwards \cite{Percival1994} (but flat spectrum or white noise signals are not affected). In particular, the high frequency power of AR(1) models does not decay as a power law ---as it does for the continuous OUP--- because these frequencies correspond to trial-by-trial changes, that cannot vanish in discrete implementations. In general, on top of estimation bias, there is always a trade-off between variance reduction and resolution when estimating PSDs. Using Welch's method with a (rectangular) window length $n_w = 10^5$, the squared Fej\'{e}r's kernel has a peak mainlobe of $10^{10}$ and its largest sidelobes peak at about 27dB below it (at a distance of $f_s/n_w = \approx 10^{-3}$Hz, the lowest resolvable frequency). In Fig.\ \ref{fig:psdT}, observe that spectral leakage inflates the PSD at the lowest and highest frequencies; the PSD cannot be estimated at the lowest frequencies ($\approx 10^{-3}$Hz) due to the loss of resolution attending the improvement in variance afforded by Welch's method. To verify that the bias is due to spectral leakage, we also estimated PSDs using the computationally expensive Thompson's multitaper method (not shown), which uses a bank of optimal bandpass filters called Slepian sequences (instead of a bank of rectangular bandpass filters as in and Welch's method) that reduces Fej\'{e}r's kernel's sidelobes. For a time-bandwidth parameter $3$, the bias is substantially reduced at both lowest and highest frequencies, although at the expense of higher variance.

\begin{figure}
\includegraphics[width=1\textwidth]{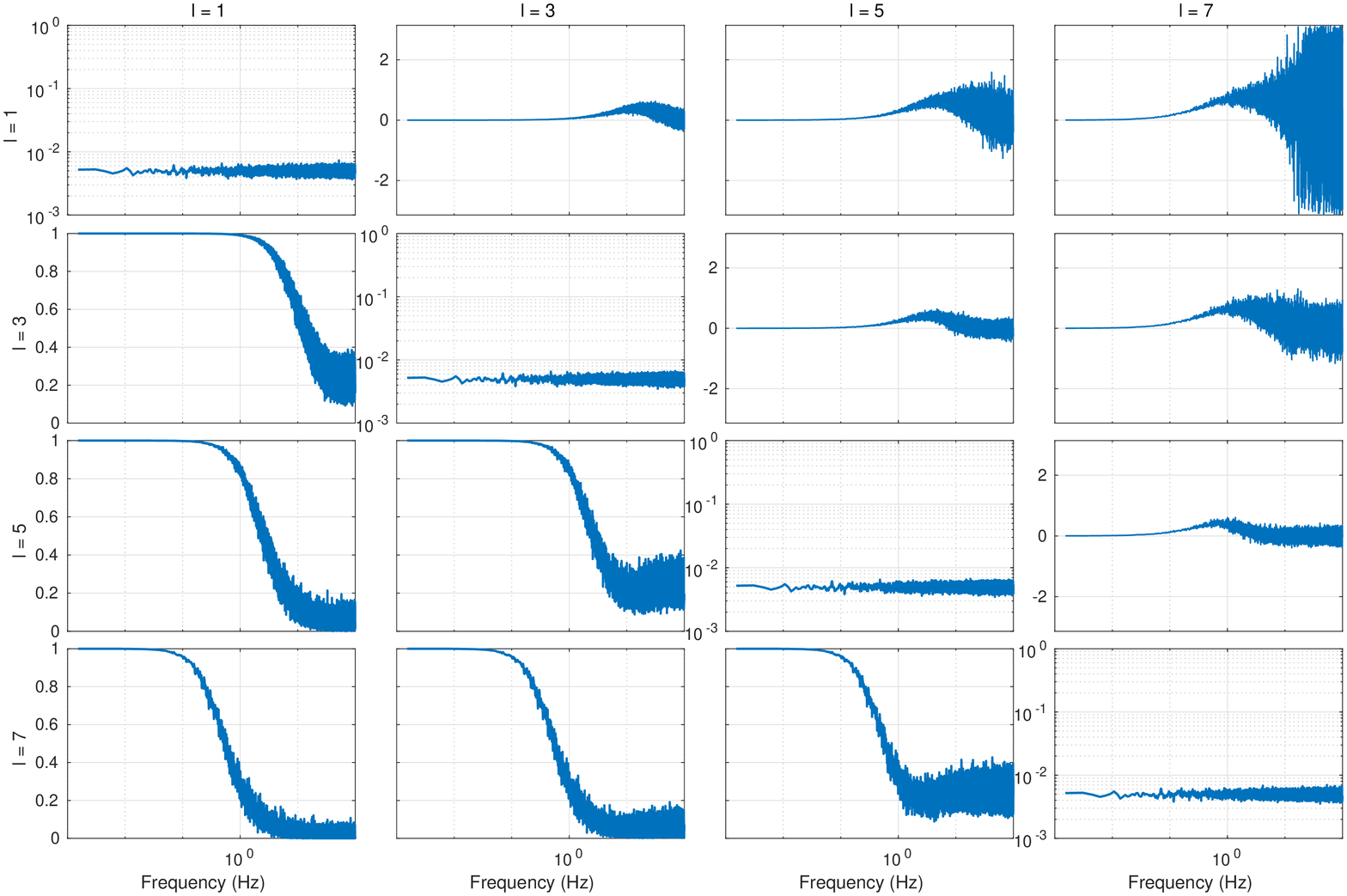}
\caption{Cross-spectral phase (cells above the main diagonal) and magnitude-squared coherence (below the main diagonal) pair-wise estimates between pulses $\varepsilon_l^{(t)}$ for $l=1,3,5,7$. The main diagonal is populated by PSD plots, as in Fig.\ \ref{fig:epsactcoh}. The cross-spectral phase was defined as the argument of the cross-spectral density $S_{\varepsilon_i,\varepsilon_j}$; and the magnitude-square coherence was defined as $\frac{|S_{\varepsilon_i,\varepsilon_j}|^2}{S_{\varepsilon_i} S_{\varepsilon_j}}$, with PSD estimation parameters set as in Fig.\ \ref{fig:psdT}.\label{fig:epscoh}}
\end{figure}

The PSD of a sum of signals is the sum of the signals' PSDs only if the signals are uncorrelated. Otherwise, $S_{\theta_i + \theta_j}(\nu) = S_{\theta_i,\theta_i}(\nu) + S_{\theta_j,\theta_j}(\nu) + 2 \operatorname{Re}({S_{\theta_i,\theta_j}(\nu)})$, where $S_{\theta_i,\theta_j}(\nu)$ is the cross power spectral density (CPSD). Hence, estimating the PSD of $\sum_{i=1}^L \theta_l^{(t)}$ in general requires computing numerically  $L(L-1)/2 = 21$ pairwise CPSD functions for $L=7$, with no closed-form expression. However, a reasonable approximation can be achieved computing only the CPSD function that contributes the highest power: $S_{\sum_{i=1}^7 \theta_l}(\nu) \approx \sum_{i=1}^7 S_{\theta_i,\theta_i}(\nu) + 2 \operatorname{Re}({S_{\theta_6,\theta_7}(\nu)})$. Inspection of the PSD of the sum of variables in Figs.\ \ref{fig:psdT} and \ref{fig:psdE} reveals that pulses $\varepsilon_l$ have pairwise CPSD components that contribute to the CPSD of the sum only  at low frequencies. This is confirmed by their close to 1 coherence estimates and zero-lag cross-phase (Fig.\ \ref{fig:epscoh}).

The activities $a_l^{(t)}$ follow a random walk with stationary and independent increments distributed as $\varepsilon_l^{(t)}$ that resets to zero upon exiting $[-\theta_l, \theta_l]$. Rewriting Eq.\ \ref{eq:a}:
\begin{equation}
a_l^{(t)} = (a_l^{(t-1)} + \varepsilon_{l-1}^{(t)}) \mathbb{1}_{a_l^{(t-1)} + \varepsilon_{l-1}^{(t)} \in [-\theta_l,\theta_l]}\, , \label{eq:a1}
\end{equation}
where $\mathbb{1}_x$ is the indicator function, which equals zero when its argument $x$ is false and one when true, and $\theta_l = \theta_l^{(t)}$ is also a function of $t$. In general, subthreshold activity can be modeled as a Markov renewal process with discrete discharge times $t$, inter-event times $\nu_{a,l-1}^{-1}$ that equal the inverse of the firing rate of the preceding unit, and activity as a continuously valued state $A_l$, where the jumps are the incoming pulses from the preceding unit $\mathcal{E}_{l-1}$. Alternatively, it can be construed as a discrete version of Brownian motion with stochastic resetting \cite{Evans2011a}: since $\varepsilon^{(t)}$ has a well defined variance, in the scaling limit $a_l^{(t)}$ behaves as a Brownian particle that resets to zero as soon as it exits $[-\theta_l, \theta_l]$, so Eq.\ \ref{eq:a1} becomes:
\begin{equation}
da = \sigma_{\mathcal{E}}dW \quad \text{if} \; a\in[-\theta,\theta] \label{eq:da}
\end{equation}
and $a=0$ otherwise, with $\sigma_{\mathcal{E}}$ the standard deviation of $\mathcal{E}$.

The covariance function $C_{a,a}(t_1,t_2)$ of the process $a^{(t)}$ with stochastic reset can be expressed \cite{Majumdar2018a} in terms of the covariance function of the same process without reset ---which is a Wiener process scaled by $\sigma_{\mathcal{E}_0}$--- $\sigma_{\mathcal{E}_0}^2C_{W,W}(t_1,t_2)$, with $t_2>t_1$, as $C_{a,a}(t_1,t_2) = P(\tau > t_2-t_1) \left[ \int_0^{t_1} P(\tau<t')P(t' < \tau < t'+dt') \sigma_{\mathcal{E}_0}^2C_{W,W}(t',t_2-t_1+t') + P(\tau > t_1) \sigma_{\mathcal{E}_0}^2C_{W,W}(t_1,t_2)  \right]$, where $\tau$ is used to denote $\tau_{\pm\theta}$ for clarity. Since the activity process without reset becomes a Wiener process in the scaling limit, $C_{W,W}(t_1,t_2)=\min(t_1,t_2) = t_1$, and in the steady state $P(\tau > t_1) C_{W,W}(t_1,t_2) = \bar{F}_\tau(t_1) t_1 \to 0$ as $t_1 \to \infty$, so we obtain
\begin{equation}
C_{a,a}(t_1,t_2) = \bar{F}_\tau(t_2-t_1) \int_0^{t_1}F_{\tau}(t') f_{\tau}(t') \sigma_{\mathcal{E}_0}^2 t' dt', \label{eq:Caa1}
\end{equation}
where $\bar{F}_\tau(t) = 1 - F_\tau(t)$ is the tail distribution of $\tau$, and $\sigma_{\mathcal{E}_0}^2=2D$ is just twice the diffusivity parameter $D$ used in diffusion problems. There is no closed-form expression for this equation that we know of, so we shall approximate it.

Repeated firing or discharging can be modeled as a renewal process, which involves ignoring the activity values and focusing only on the accumulated number of discharges $N(t)$ at each iteration $t$, where the jumps are resets to zero. The inter-event times (the duration between successive discharges) are distributed as the first exit time $f_{\tau}$ (Eq.\ \ref{eq:rbmtau1} or \ref{eq:rbmtau2}). Although the probability of firing in a given time interval depends on the last discharge time, this dependency vanishes when averaging over long time spans. By the elementary renewal theorem \cite{Taylor1998}, the number of discharges is asymptotically linear, i.e.\ in the long run  $t\to\infty$ the averaged firing frequency approaches the inverse of the hitting time mean
\begin{equation}
\lim_{t \to \infty} \langle \nu \rangle \equiv \lim_{t \to \infty} \frac{1}{t}\langle N(t) \rangle = \langle \tau \rangle^{-1} = \theta^{-2}, \label{eq:nuev}
\end{equation}
where the last equality follows from Eq.\ \ref{eq:rbmhte}. Since the inverse is a convex function, by Jensen's inequality the expectation of the inverse is larger or equal than the inverse of the expectation:  $\langle \nu \rangle = \langle \tau^{-1} \rangle \geq  \langle \tau \rangle^{-1}$, which indicates that $\theta^{-2}$ is also a lower bound of the average firing rate. For large $t$ the asymptotic variance of $N(t)$ behaves \cite{Taylor1998} according to
\begin{equation}
\lim_{t \to\ \infty} \frac{\var{[N(t)]}}{t} = \frac{\var{[\tau]}}{\langle \tau \rangle^3} = \frac{2}{3}\theta^{-2}, \label{eq:nuvar}
\end{equation}
where in the rightmost equality we used $\var{[\tau_{\pm\theta}]} = \frac{2}{3}\theta^4$, which can be derived from the properties of martingales \cite{Durrett2019} in a similar manner to Eq.\ \ref{eq:rbmhte}. Together with Eq.\ \ref{eq:nuev}, this determines the magnitude of the fluctuations of inter-firing times about their expectation $\theta^{-2}$; these fluctuations are approximately normally distributed \cite{Taylor1998}.  

From the first exit time density, we can deduce the probability that a unit fires with rate $r=t^{-1}$ from $f_\tau(t)dt = f_\nu(r)dr$. Thus, the sequence of firing times can be interpreted as a inhomogeneous Poisson point process over $t$ with rate intensity function $rf_{\tau}(t)$ (cf. Eq.\ \ref{eq:rbmtau2}). Hence the probability that the activity fires or reaches the threshold $n$ times within a segment $[0,t]$ starting from the previous reset at zero time is
\[
P(\text{firing }n\text{ times}) = \frac{[\Lambda(t)]^n}{n!}e^{-\Lambda(t)},  \label{eq:poisfire}
\]
where $\Lambda(t) = \int_0^t \tau^{-1} f_\tau(\tau) d\tau =  \langle t^{-1} \rangle$ before the first discharge, but as $t \to \infty$, by the elementary renewal theorem $\Lambda(t) \to \langle \tau \rangle^{-1} t = rt$.

Thus, by neglecting short time correlations, the elementary renewal theorem furnishes a simple approximation to the firing rate. This effectively amounts to construing repeated firing as a Poisson point process with rate $r = \theta^{-2}$. By considering the inter-event times of $a^{(t)}$ as a (homogeneous) Poisson process with constant resetting rate, we can derive again (Eq.\ \ref{eq:Caa1}) the covariance function of $a^{(t)}$ in terms of the covariance function of Brownian motion \cite{Majumdar2018a} by substituting $F_\tau$ for its homogeneous counterpart as
\begin{eqnarray}
C_{a,a}(t_1,t_2) &\approx& e^{-r(t_2-t_1)}\int_0^{t_1} r e^{-rt'} \sigma_{\mathcal{E}}^2 t'dt' \nonumber \\
&=& r\sigma_{\mathcal{E}_0}^2 e^{-r(t_2-t_1)} \int_0^{t_1} t'e^{-rt'} dt' \nonumber \\
&=& \frac{\sigma_{\mathcal{E}_0}^2}{r} e^{-r(t_2-t_1)}[1 - e^{-rt_1}(rt_1+1)] \nonumber \\
&\underset{t_1 \to \infty}{=}& \frac{\sigma_{\mathcal{E}_0}^2}{r}e^{-r(t_2-t_1)}, \label{eq:Caa2}
\end{eqnarray}
where, in the first line, the first factor is the probability that there is no reset between $t_1$ and $t_2$ (Eq.\ \ref{eq:poisfire} with $n=0$); and the integrand denotes the probability that the last reset before $t_1$ occurs between $t'$ and $t'+dt'$ \cite{Majumdar2018a}. Note that this approximated covariance function is stationary only if fluctuations around the expectation are ignored (from the elementary renewal theorem). These fluctuations introduce a bias, but we can alleviate it by making use of Eq.\ \ref{eq:nuvar} and taking the expectation of Eq.\ \ref{eq:Caa2} with respect to the approximately normally distributed $r$:
\begin{equation}
C_{a,a}(t) \approx \sigma_{\mathcal{E}_0}^2 \langle \frac{e^{-r|t|}}{r} \rangle \approx \theta^2\sigma_{\mathcal{E}_0}^2 e^{-\frac{4}{3}\theta^{-2}|t|}, \label{eq:Caa3}
\end{equation}
where $t=t_2-t_1$ (when $t_2>t_1$); we approximated $\langle r^{-1} \rangle\approx \langle r \rangle^{-1}$; we used $r= \langle  \nu \rangle \approx \mathcal{N}(\theta^{-2},\frac{2}{3}\theta^{-2})$ from Eqs.\ \ref{eq:nuev} and \ref{eq:nuvar}; and $\langle e^{\mathcal{N}(\theta^{-2},\frac{2}{3}\theta^{-2})} \rangle =  e^{\frac{4}{3}\theta^{-2}}$ by the properties of the lognormal distribution. $C_{a,a}$ decays with time constant $\frac{3}{4}\theta^2$. Finally, the PSD of $a^{(t)}$ can be computed by Fourier-transforming Eq.\ \ref{eq:Caa3}:
\begin{equation}
S_{a,a}(\nu|l) = \theta_l^2\sigma_{\mathcal{E}_0}^2 \frac{2\frac{4}{3}\theta_l^{-2}T_0^{-1}}{(2\pi\nu)^2 + (\frac{4}{3}\theta_l^{-2}T_0^{-1})^2} = \frac{\sigma_{\mathcal{E}_0}^2 \theta_l^2 \gamma_a}{\pi(\nu^2 + \gamma_a^2)}, \label{eq:Saa1}
\end{equation}
which is again a Lorentzian function ---like the PSD of $\theta^{(t)}$--- with corner frequency $\gamma_a = (2\pi\frac{3}{4}\theta_l^2T_0)^{-1}$ and maximum $S_{a,a}(0) = 2\sigma_{\mathcal{E}_0}^2T_0(\frac{3}{4})^2\theta_l^4$. Notice that, unlike for $S_{\theta,\theta}(\nu|l)$, the ``sampling rate'' is fixed to $T_0$ across all levels. This is because the Brownian motion approximation to activities (Eq.\ \ref{eq:da}) and its subsequent assumptions are not valid at the large timescales of higher levels: they incorporate large and sparse fluctuations induced by $\varepsilon^{(t)}_l$ that can only be modeled as Brownian motion by breaking up the $\varepsilon^{(t)}_l$ into small fluctuations across many small $T_0$ iterations.

\begin{figure}
\includegraphics[width=0.5\textwidth]{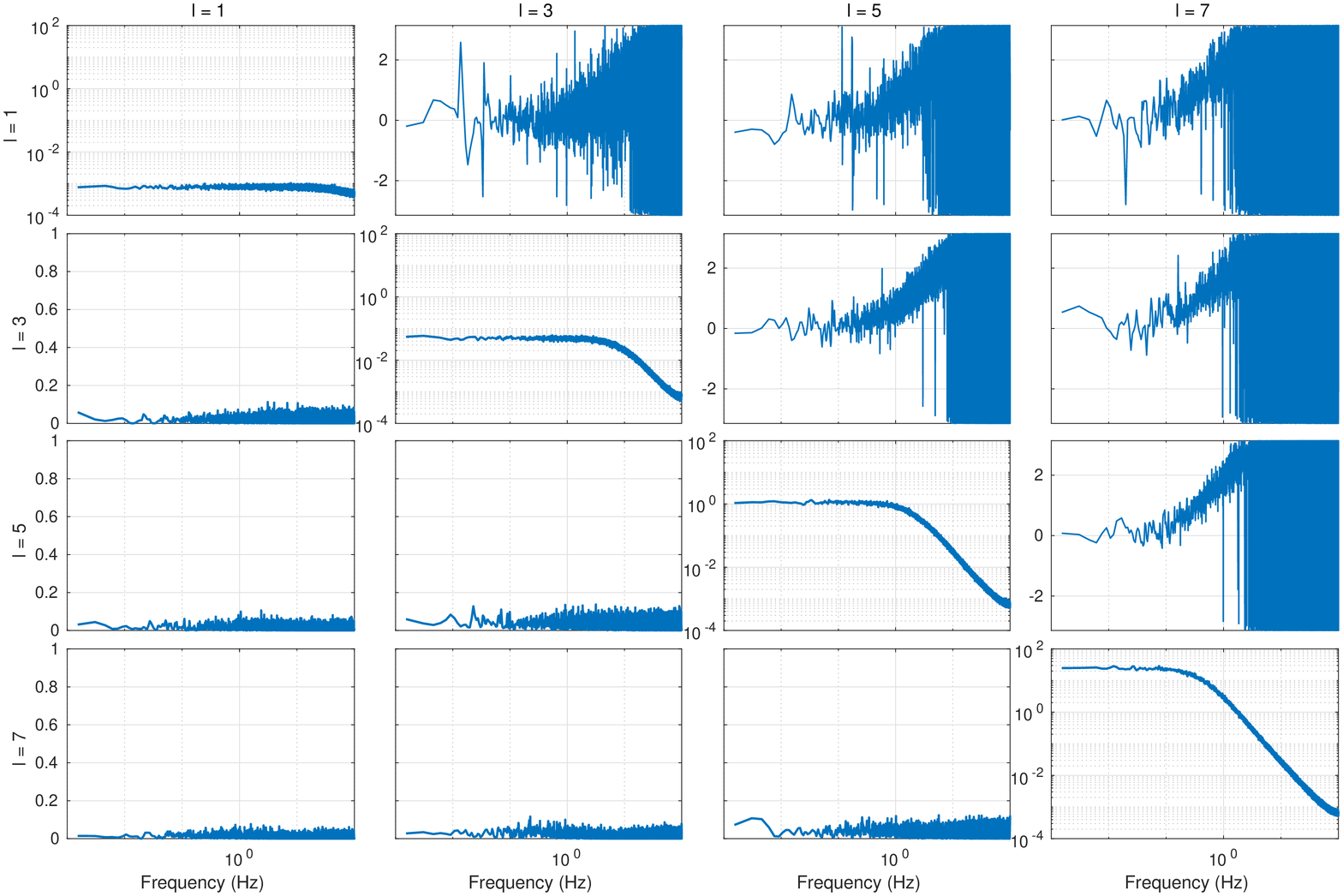}
\caption{Cross-spectral phase and magnitude-squared coherence pair-wise estimates of subthreshold activities $a_l^{(t)}$ for $l=1,3,5,7$. Layout and estimation parameters as in Fig.\ \ref{fig:epscoh}.\label{fig:actcoh}}
\end{figure}

This still does not yield a good estimate of the numerically computed PSD because it overestimates power across the whole frequency range with respect to the numerical simulation output. The reason lies in its discretization and recording schemes: since activities are reset at every iteration when a discharge occurs, $P(A_l=0) = g_l$, so the amplitude of the simulated time series of activities is a fraction $1-g_l$ of the amplitude of the continuous-time analog, upon which we derived the PSD estimate. With this last modification, the PSD estimate of the discretized time series becomes
\begin{equation}
S_{a,a}(\nu|l) \approx \frac{\sigma_{\mathcal{E}_0}^2 \theta_l^2 (1-g_l)^2 \gamma_a}{\pi(\nu^2 + \gamma_a^2)}. \label{eq:Saa2}
\end{equation}
Unlike pulses, activities are pair-wise incoherent over the whole frequency range (Fig.\ \ref{fig:actcoh}), so a specific frequency power in a given level is not transmuted into power in the subsequent level. As a corollary, the PSD of the sum of activities can be approximated by the sum of their PSDs (Fig. \ref{fig:psdA}). 

Regardless of the specific parameters, the takeaway is that activity dynamics, characterized by interleaved periods of diffusion and resetting forces, are well described by the OUP, a similar stochastic process where mean-reverting forces replace jump resets. 

\section{\label{sec:app} Some relationships among activity variables' statistics}

From the definition of variance:

$\langle |\tilde{A}_l|^2 \rangle = \langle \tilde{A}_l^2 \rangle = \var{|\tilde{A}_l|} + \langle |\tilde{A}_l| \rangle^2 = \var{|\tilde{A}_l|} + \langle \Theta_l \rangle^2 = \langle A_l^2 \rangle + \langle \mathcal{E}_{l-1}^2 \rangle = \langle A_l^2 \rangle + \langle \mathcal{E}_l^2 \rangle g_l$

$\langle |\tilde{A}_l|^2 \rangle = \langle |A_l+\mathcal{E}_{l-1}|^2 \rangle = \langle (A_l+\mathcal{E}_{l-1})^2 \rangle = \langle A_l^2 +\mathcal{E}_{l-1}^2 + 2A_l\mathcal{E}_{l-1} \rangle = \langle A_l^2 \rangle + \langle \mathcal{E}_{l-1}^2 \rangle + 2 \langle A_l\mathcal{E}_{l-1} \rangle = \langle |A_l|^2 \rangle + \langle |\mathcal{E}_{l-1}|^2 \rangle$

From the law of total variance applied to $|A|$, $|A^+|$, and $|\tilde{A}|$:

$\var{|\tilde{A}_l|} = (1-g_l)\var{|A_l^+|} + g_l \var{|\mathcal{E}_l|} + g_l(1-g_l)(\langle |\mathcal{E}_l| \rangle^2 + \langle |A_l^+| \rangle^2) - 2g_l(1-g_l)\langle |\mathcal{E}_l| \rangle\langle |A_l^+| \rangle$

$\var{|A_l|} = \var{A_l^+}(1-g_l) - \langle |A_l^+| \rangle^2 (1-g_l)^2$

$\var{|A_l^+|} = \var{A_l}(1-g_l)^{-1} - \langle |A_l| \rangle^2 (1-g_l)^{-2}$

From the definition of variance:

$\var{|\tilde{A}_l|} = (1-g_l)\var{|A_l^+|} + g_l \var{|\mathcal{E}_l|} + g_l(1-g_l)(\langle |\mathcal{E}_l| \rangle - \langle |A_l^+| \rangle)^2 = \langle A_l^2 \rangle - \langle |A_l| \rangle^2(1-g_l)^{-1} + g_l \var{|\mathcal{E}_l|} + g_l(1-g_l)\langle |\mathcal{E}_l| \rangle^2 + g_l(1-g_l)^{-1}\langle |A_l| \rangle^2 - 2g_l\langle |\mathcal{E}_l| \rangle\langle |A_l| \rangle  = \langle A_l^2 \rangle + g_l \var{|\mathcal{E}_l|} + g_l(1-g_l)\langle |\mathcal{E}_l| \rangle^2 - \langle |A_l| \rangle^2 - 2g_l\langle |\mathcal{E}_l| \rangle\langle |A_l| \rangle = \var{|A_l|} + g_l \var{|\mathcal{E}_l|} + g_l(1-g_l)\langle |\mathcal{E}_l| \rangle^2 - 2g_l\langle |\mathcal{E}_l| \rangle\langle |A_l| \rangle$

$\langle |\tilde{A}_l|^2 \rangle = \var{|\tilde{A}_l|} + \langle \Theta_l \rangle^2 = \var{|A_l|} + g_l \var{|\mathcal{E}_l|} + g_l(1-g_l)\langle |\mathcal{E}_l| \rangle^2 - 2g_l\langle |\mathcal{E}_l| \rangle\langle |A_l| \rangle + g_l^2 \langle |\mathcal{E}_l| \rangle^2 + \langle |A_l| \rangle^2 + 2g_l \langle |\mathcal{E}_l| \rangle\langle |A_l| \rangle = \var{|A_l|} + g_l \var{|\mathcal{E}_l|} + g_l(1-g_l)\langle |\mathcal{E}_l| \rangle^2 + g_l^2 \langle |\mathcal{E}_l| \rangle^2 + \langle |A_l| \rangle^2 = \var{|A_l|} + g_l \var{|\mathcal{E}_l|} + g_l\langle |\mathcal{E}_l| \rangle^2 + \langle |A_l| \rangle^2 = \langle A_l^2 \rangle + g_l\langle \mathcal{E}_l^2 \rangle$

$\langle \Theta_l \rangle^2  = \langle A_l^2 \rangle + \langle \mathcal{E}_l^2 \rangle g_l - \var{|A_l|} - g_l \var{|\mathcal{E}_l|} - g_l(1-g_l)\langle |\mathcal{E}_l| \rangle^2 + 2g_l\langle |\mathcal{E}_l| \rangle\langle |A_l| \rangle = \langle |A_l| \rangle^2 + g_l^2\langle |\mathcal{E}_l| \rangle^2 + 2g_l\langle |\mathcal{E}_l| \rangle\langle |A_l| \rangle  $

The relation $\langle A^2 \rangle \approx \langle (\theta_*-|A^+|)^2 \rangle$ was verified to reasonably hold in numerical simulations. It equates the average subthreshold activity energy to the average post-pulse subthreshold energy deficit below the threshold energy in quiescent trials, which is a manifestation of energy conservation.

\begin{figure*}
\includegraphics[width=\textwidth]{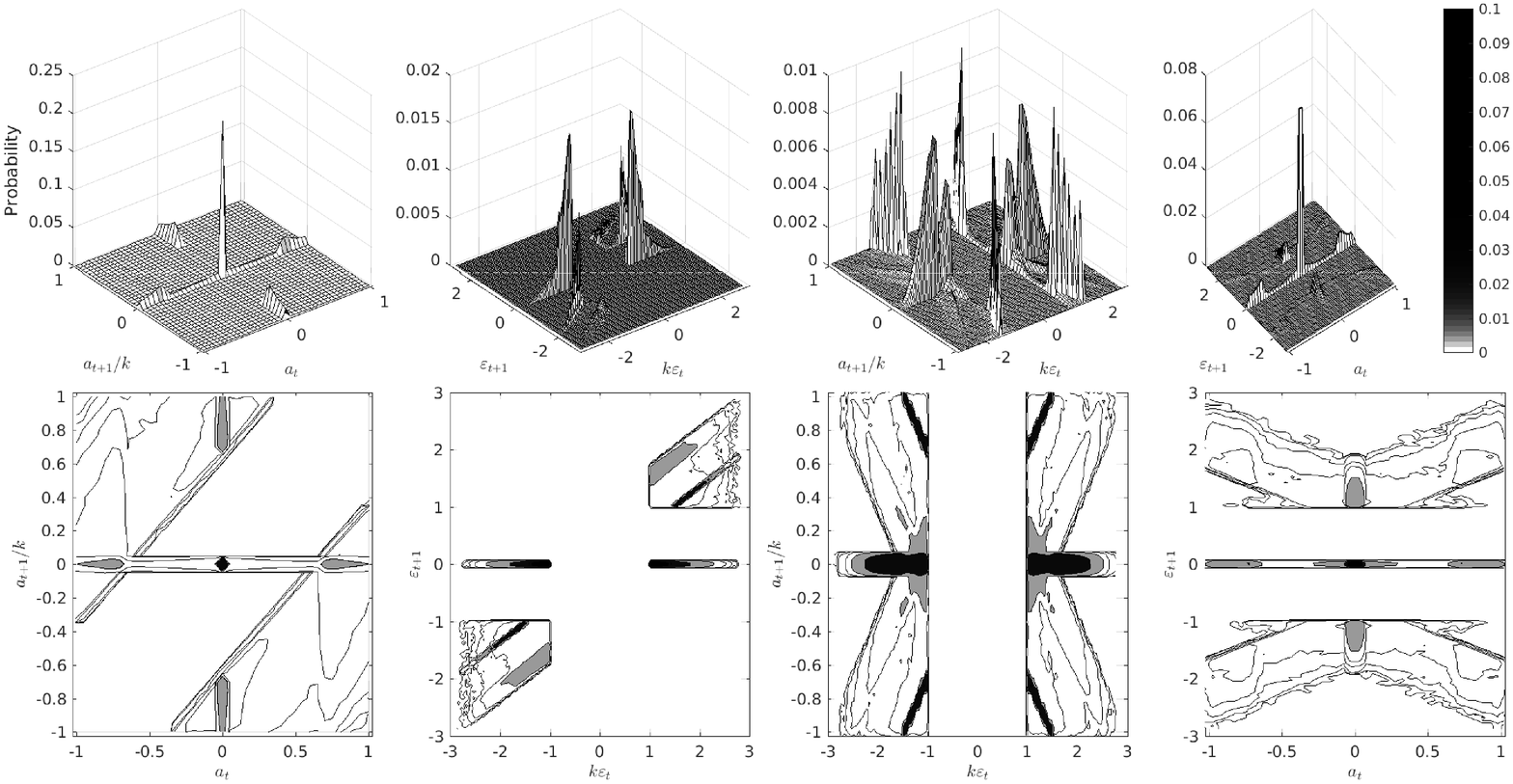}
\caption{Surface (top) and contour (bottom) plots of the joint densities $P(a_{t+1}/k, a_t)$ in left, $P(\varepsilon_{t+1}, k\varepsilon_t)$ in middle left, $P(a_{t+1}/k, k\varepsilon_t)$ in middle right, and $P(\varepsilon_{t+1}, a_t)$ in right. Variables have been scaled by $k$ or $1/k$ to facilitate comparison. The color bar at the right of the top right plot is the color legend of the bottom row contour plots, it represents probability. \label{fig:TransMatJoint}}
\end{figure*}

\section{\label{sec:RecLoopMcmc} Numerical analysis of the rescaled recurrent loop configuration}
The recurrence relation specified by Eqs.\ \ref{eq:hc1}--\ref{eq:hc4} in Section \ref{sec:RRLC} implements a hybrid of bootstrap resampling and discrete-time dynamical system that enables estimating the densities of PC variables, in rescaled recurrent loop configuration, a sort of ``perfusive rescaled loop'' that bites its own tail. This can be programmed as an iterative loop.

One pool of $N=10^5$ data points was defined for each variable. Each pool was initialized with values randomly picked out from the time series ($9 \cdot 10^6$ points) of the homologous variable at level $l=1$ (see Fig.\ \ref{fig:pdfs}, second row of left column) in the PC feedforward configuration described in Section \ref{sec:gausim}. At each iteration $t$, pulses $\varepsilon_i$ were resampled randomly from the pulse pool $\varepsilon_{1:N}$, whereas the gains $k_t$, thresholds $\theta_t$ and subthreshold activities $a_t$ took the values they had in the previous iteration. Then each variable was allocated a new value at iteration $t+1$ according to Eqs.\ $\ref{eq:hc1}$--$\ref{eq:hc4}$. This routine was looped $10^6$ times. Instead of using the last trial value $a_t$, subthreshold activities could also be resampled randomly from the pool $a_{1:N}$ (just as pulses were resampled from $\varepsilon_{1:N}$) to achieve the same results, because same-trial subthreshold activities and incoming pulses are independent. The same holds for $\theta_t$ and $k_t$ (in Eq.\ \ref{eq:hc3}). This was verified in simulations. This scheme generated the data underlying Figs.\ \ref{fig:SFL}, \ref{fig:TransMat}, \ref{fig:TransMatJoint}. Some histogram functions (for plotting transition matrices) are susceptible to binning artifacts; this can be often averted by adding a randomly signed machine epsilon to ``jitter'' the samples.

Unlike in the feedforward configuration, here we need to estimate $k_t$ to iterate the map. Since $k_t$ is a function of the pool of pulses $\varepsilon_{1:N}$ (Eq.\ \ref{eq:hc3}), it is susceptible to estimation errors (through pulses), which in turn propagate to the other variables at each iteration. This feature compounds the estimation uncertainty of all variables, and precludes simulating the recurrent loop with fixed threshold and scaling factor (i.e.\ $\theta_*$ and $k_*$ in FTA) ---except in the limit of infinitely large bootstrapping pools. Another caveat is that a recurrent loop embodies a positive feedback that tends to magnify skewness: if left unchecked, the densities will become  increasingly lopsided, to the point that the resulting densities will correspond to the case of perfectly correlated driving input ($H=1$, see below). To avoid this effect, every few iterations the pools must be re-symmetrized, by e.g.\ randomly flipping the sign of all pulse samples.
By definition, FTA is achieved in the limit of vanishing $\theta_t$ fluctuations: by calculating $\theta$ with a pool of subthreshold activity samples $\tilde{a}_{1:N}$, i.e.\ using Eq.\ \ref{eq:hc3mean} instead of Eq.\ \ref{eq:hc3}, and setting a large enough $N$ ---by the law of large numbers, threshold fluctuations wane along with $N$. However, the larger is the pool, the more ``inertia'' it has, so in practice it is expedient to start with a small (``light'') pool, and only when it reaches a steady state fill it up with new samples; a large biased pool will remain so for many iterations, which is computationally wasteful.

The above scheme can accommodate autocorrelated driving input with little modification. First, the pulse $\varepsilon_{1:N}$ and subthreshold activity $a_{1:N}$ pools should take in samples after applying the absolute value function, so they contain only positive or zero values; since all densities are symmetrical about the origin, no information is thereby lost. Second, the incoming pulse magnitude is defined by the pool sample, whereas its sign is pseudo-randomly generated as $\sgn{(\mathcal{U}_{[0,1]} < H)}$, where $\mathcal{U}_{[0,1]}$ the uniform density with support $[0,1]$, and $H \in [0, 1]$ is a parameter describing the autocorrelation of driving input, which  for fractional Gaussian noise is the Hurst exponent. Thus, the larger is $H$, the higher is the probability that two consecutive pulses have the same sign, with $H=0.5$ characterizing white noise. The duplets used to estimate transition matrices in Figs.\ \ref{fig:TransMat}, \ref{fig:TransMatJoint} were stored at every iteration in separate variables.

\begin{figure}
\includegraphics[width=0.5\textwidth]{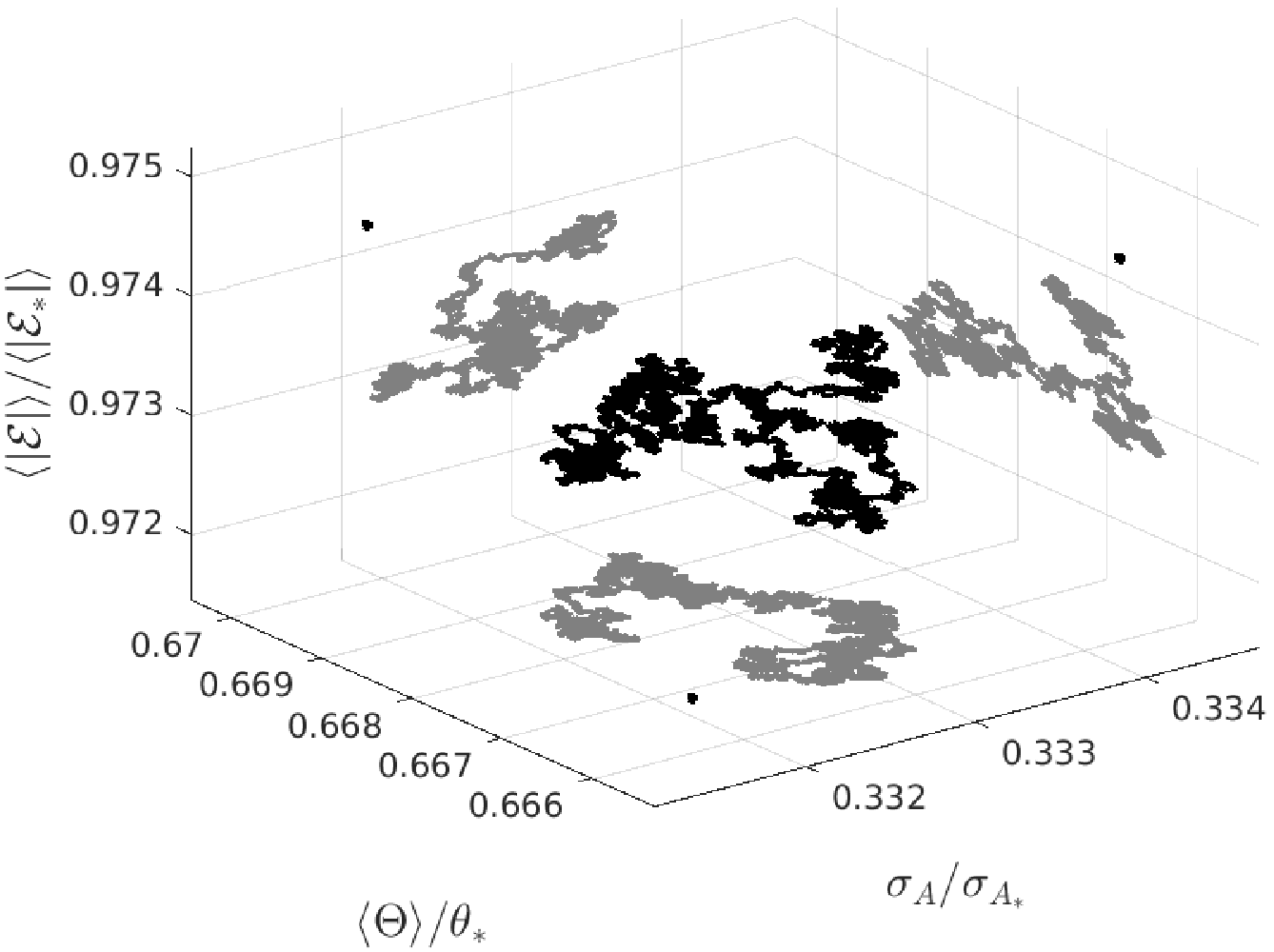}\caption{Simulated trajectory of states for $10^5$ iterations in rescaled recurrent loop configuration (Eqs.\ \ref{eq:hc1}, \ref{eq:hc2}, \ref{eq:hc3mean}, \ref{eq:hc4}), embedded in the three-dimensional manifold of the normalized variables $\sigma_A / \sigma_{A_*}$, $\langle \Theta \rangle / \theta_*$ and $\langle |\mathcal{E}|\rangle / \langle |\mathcal{E}_*|\rangle$. The dot indicates the fixed threshold approximation (FTA) estimate location. This and the trajectory are also shown projected (in gray) onto the three planes defined pair-wise by the FTA estimates. Parameters: Pool size $N=10^5$, Gaussian noise input Hurst exponent $H=0.5$, $10^6$-iteration burn-in. \label{fig:3dBmATE}}
\end{figure}

A numerically simulated trajectory of the triplet  $[\sigma_A / \sigma_{A_*}, \langle \Theta \rangle / \theta_*, \langle |\mathcal{E}|\rangle / \langle |\mathcal{E}_*|\rangle]$ in phase space for a $10^5$ iterations fragment is shown in Fig.\ \ref{fig:3dBmATE}.
The system was let settle in the steady state beforehand for $10^6$ iterations, so the trajectory wiggles in an attractor located nearby the FTA point (although visually it is similar to a 3-dimensional Brownian particle) while displaying a small bias respect to the FTA estimate (Fig. \ref{fig:ErrorbarGaetResc}). In general, reducing the pool size $N$ leads to larger fluctuations, but increasing it too much hinders the trajectory from returning to the vicinity of the expected value after a (random) fluctuation. Increasing the learning rate while using the delta learning rule (Eq.\ \ref{eq:hc3} instead of Eq.\ \ref{eq:hc3mean}) results in larger oscillations in the $\langle \Theta \rangle / \theta_*$ dimension (not shown).

\begin{figure}
\includegraphics[width=0.5\textwidth]{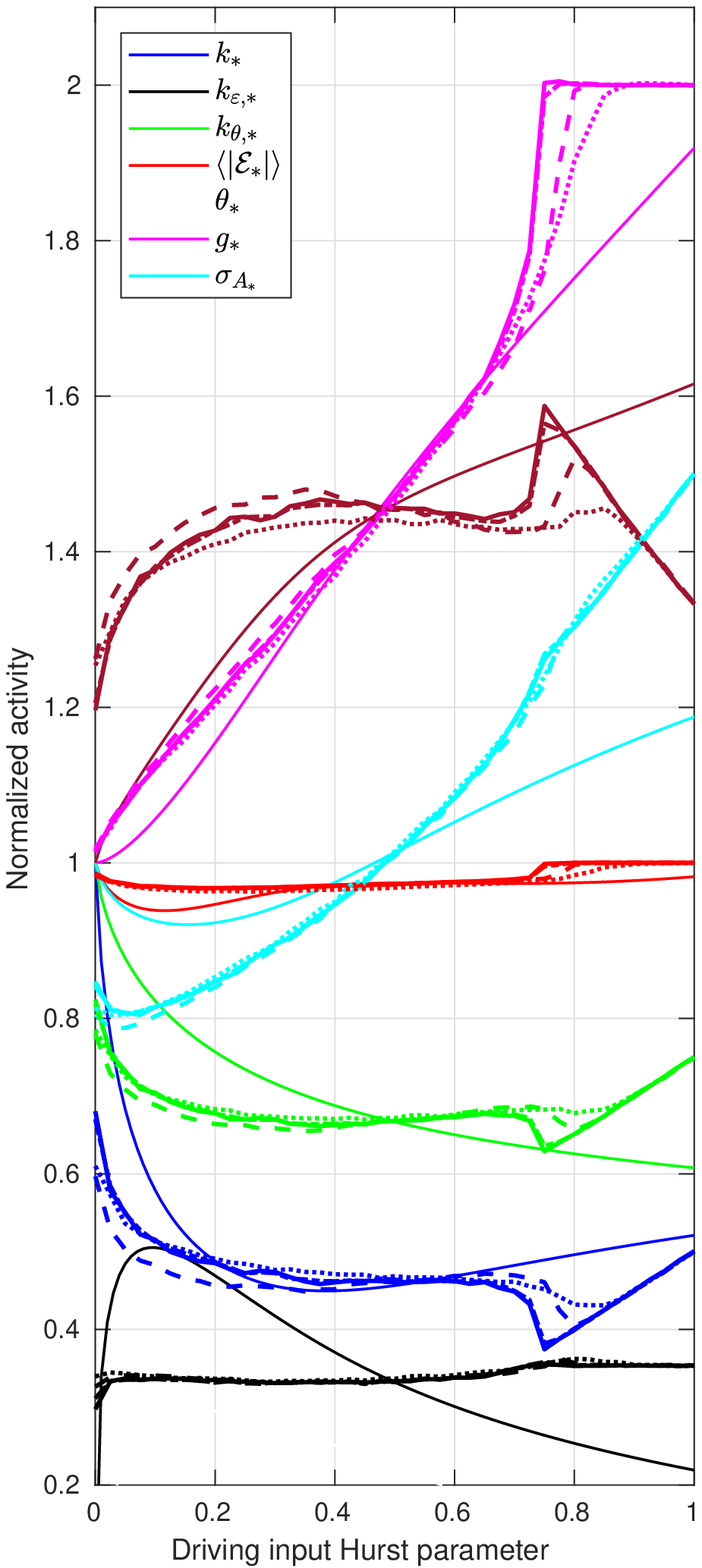}
\caption{Perfusive cascade variables' (see legend) estimates with delta learning rule, as a function of driving input autocorrelation, in recurrent rescaled loop configuration. $A$, $\mathcal{E}$ are scaled consistently with $\langle \mathcal{E}^2 \rangle = 1$. Line style indicates the estimation procedure: closed form expression (Eqs.\ \ref{eq:Hc1}--\ref{eq:Hc4}) derived through the fixed threshold approximation (solid thin); simulation with learning rate $w=10^{-1}$ (dotted), $N=10^{-2}$ (dash-dotted), $N=10^{-3}$ (solid thick), $N=2\cdot 10^{-4}$ (dashed). Dashed lines largely overlap with solid thick lines. \label{fig:AncsH1}}
\end{figure}

Fig.\ \ref{fig:AncsH1} displays variable estimates using the delta learning rule for thresholds in recurrent configuration as a function of input Hurst parameter. Although it conveys similar information with respect to its homologous Fig.\ \ref{fig:AncsH2} with explicit threshold averaging, we can observe that with the smallest learning rate of $w=10^{-4}$ (dashed line), the hill peak at $H \approx .74$ is shifted to $H \approx .8$. The reason seems to lie in that under superdiffusion, this configuration greatly overestimates the discharge rate $g$ and underestimates the threshold with respect to the fixed threshold approximation, as can be seen in Fig.\ \ref{fig:AncsH1}. Finally, note that in the feedforward configuration simulations in Section \ref{sec:properties}, we used a learning rate $w=0.01$, which corresponds to the dashed-dotted line in Fig.\ \ref{fig:AncsH1}.

\section{\label{sec:appPulseGpd} Modeling of pulse size densities with the generalized Pareto distribution}

We suggested in Section \ref{sec:gausim} that pulse density profiles follow approximately a generalized Pareto distribution (GPD), which is a family of densities, often used to model the shape of the tail of other distributions \cite{Coles2001}. Here we use the GPD as a parametric approximation to the tails of $f_{\tilde{A}}$, or equivalently to the pulse sizes $f_{|\mathcal{E}|}$. The GPD density is defined as
\begin{equation}
f_{GPD}(x | \mu,\sigma,\xi) = \frac{1}{\sigma}\left( 1+ \xi \frac{x-\mu}{\sigma} \right)^{-1-\frac{1}{\xi}}, \nonumber
\end{equation}
with parameters location $\mu$ (which defines the minimum of $\mathcal{E}$), scale $\sigma$ (note this runs counter to its meaning of standard deviation in other sections), and shape $\xi$ (which characterizes how steeply the tail falls off). Its support is $x \geq \mu$ when $\xi \geq 0$ and $\mu \leq x \leq \mu-\sigma/\xi \equiv M$ when $\xi < 0$. Its mean is $\mu + \frac{\sigma}{1-\xi}$ when $\xi < 1$, its median is $\mu + \sigma\frac{2^\xi-1}{\xi}$, and its mode is $\mu$.

By definition, $f_{|\mathcal{E}_l|}$ is the tail distribution of post-pulse activity absolute value $|\tilde{A}_l|$. Its cumulative distribution function $F_{|\mathcal{E}_l|}(x) \equiv P( |\mathcal{E}_l| \leq x)$ is also the excess distribution of post-pulse activities above the firing threshold $P(|\tilde{A}_l| - \theta_l \leq x \; | \; |\tilde{A}_l| > \theta_l) = \frac{F_{|\mathcal{E}_l|}(\theta_l + x) - F_{|\mathcal{E}_l|}(\theta_l)}{1-F_{|\mathcal{E}_l|}(\theta_l)}$. For larger enough $l$, this can be interpreted as the tail of the cumulative distribution of a sequence of identically distributed and independent random variables conditioned to be larger than some threshold: each $\varepsilon_l$ can be decomposed as the sum of inputs $\varepsilon_0$'s by recursively substituting each $\varepsilon_l$ by its summands $\varepsilon_{l-1} + a_{l-1}$, which by construction are already conditioned on their absolute value exceeding $\theta_{l-1}$. This supplies the conditions of applicability of the Gnedenko-Pickands-Balkema-de Haan theorem, which states that the tail of the excess distribution is asymptotically given by a GPD as the variable approaches the end of the tail of the distribution \cite{Sornette2000a}. This implies that in the steady state, and for large enough $l$, the tail distribution $F_{\mathcal{E}_l}$ can be asymptotically characterized by a GPD.

This justifies using the GPD to model $f_{|\mathcal{E}_l|}$ for large enough $l$. Maximum likelihood fits (through MATLAB's fitdist function) of the GDP parameters to the numerically obtained histograms of $f_{|\mathcal{E}_l|}$ yielded the estimates displayed in Table \ref{tab:GpdParHistFits}.

\begin{table}
\caption{Maximum likelihood estimates of GPD parameters fit to the numerical pulse size distributions of the perfusive cascade levels $l=1 \ldots 7$. Location parameters $\mu_l$ were fixed to their corresponding level threshold means (cf. Table \ref{tab:a1}); scale parameters are scaled by their corresponding location parameters. Only significant digits are shown. Numerical simulation parameters are as in Fig.\ \ref{fig:pdfs}.\label{tab:GpdParHistFits}}
\begin{ruledtabular}
\begin{tabular}{clll}
Level $l$ & $\mu_l$ & $\sigma_l / \mu_l$ & $\xi_l$ \\ 
\hline
1 & $\langle \Theta_1 \rangle$ & .832 & -.151 \\
2 & $\langle \Theta_2 \rangle$ & .596 & -.169 \\
3 & $\langle \Theta_3 \rangle$ & .571 & -.214 \\
4 & $\langle \Theta_4 \rangle$ & .564 & -.239  \\
5 & $\langle \Theta_5 \rangle$ & .606 & -.304  \\
6 & $\langle \Theta_6 \rangle$ & .606 & -.319  \\
7 & $\langle \Theta_7 \rangle$ & .609 & -.315 
\end{tabular}
\end{ruledtabular}
\end{table}

The GPD parameters corresponding to the fixed threshold approximation can be obtained through the estimates for $\theta$ and $\sigma_{|\mathcal{E}|}$ given at the end of Section \ref{sec:FTA}; thereby we can establish the equalities $\mu_* = \theta_*$ and $\mu_* + \frac{\sigma_*}{1-\xi_*} = \sigma_{|\mathcal{E}_*|}$. This yields
\begin{eqnarray}
\frac{\sigma_*}{\mu_*} &\approx& 0.6579 \nonumber \\
\xi_* &\approx& \nonumber -0.4274. 
\end{eqnarray}
It follows that $-1-\frac{1}{\xi_*} \approx 1.3396$ is the exponent of $f_{GPD}$,  the mean is $\frac{\sigma_{|\mathcal{E}_*|}}{\mu_*} \approx 1.4609$ (which entails a gain of $.4349$), the median $1 + \frac{\sigma_*}{\mu_*}\frac{2^\xi_*-1}{\xi_*} \approx 1.3947$ (gain $.5$), and the maximum of the support $\frac{M_*}{\mu_*} \approx 2.5393$ (notice all these values have been scaled by the threshold $\mu_* = \theta_*$). Finally, the derivative of the density at the threshold is $f_{GPD}'(\mu_*) = -\frac{1+\xi_*}{\sigma_*} \approx -0.8703$.

\bibliography{perfcasc}

\end{document}